%% file: main.tex
\documentclass[pdftex]{article}
\usepackage{amsmath}
\usepackage{amssymb}
\usepackage{latexsym}
\usepackage{cmll}
\usepackage{url}
\usepackage{bussproofs}

\usepackage{tikz-inet}

\usepackage{alltt}

\input local

\input notation
\input tikznotation

\title{An introduction to Differential
  Linear Logic: proof-nets,  models and antiderivatives}
\author{Thomas Ehrhard
  \hfill \\
  CNRS, IRIF, UMR 8243 \hfill \\
  Univ Paris Diderot, Sorbonne Paris Cit\'e F-75205 Paris, France }

\date{February 2016}

\begin{document}
\maketitle

\label{firstpage}

\begin{abstract}
  Differential Linear Logic enriches Linear Logic with additional logical rules
  for the exponential connectives, dual to the usual rules of dereliction,
  weakening and contraction. We present a proof-net syntax for Differential
  Linear Logic and a categorical axiomatization of its denotational models. We
  also introduce a simple categorical condition on these models under which a
  general antiderivative operation becomes available. Last we briefly describe
  the model of sets and relations and give a more detailed account of the model
  of finiteness spaces and linear and continuous functions.
\end{abstract}

\section*{Introduction}

Extending Linear Logic (\LL) with differential constructs has been considered
by Girard at a very early stage of the design of this system. This option
appears at various places in the conclusion of~\cite{Girard86}, entitled
\emph{Two years of linear logic: selection from the garbage collector}. In
Section~\emph{V.2~The quantitative attempt} of that conclusion, the idea of a
syntactic Taylor expansion is explicitly mentioned as a syntactic counterpart
of the quantitative semantics of the
$\lambda$-calculus~\cite{Girard88c}. However it is contemplated there as a
reduction process rather than as a transformation on terms. In
Section~\emph{V.5~The exponentials}, the idea of reducing $\lambda$-calculus
substitution to a more elementary linear operation explicitly viewed as
differentiation is presented as one of the basic intuitions behind the
exponential of \LL. The connection of this idea with Krivine's
Machine~\cite{Krivine85,Krivine05} and its \emph{linear head reduction}
mechanism~\cite{DanosRegnier99} is explicitly mentioned. In this mechanism,
first considered by De~Bruijn and called \emph{mini-reduction}
in~\cite{DeBruijn87}, it is only the head occurrence of a variable which is
substituted during reduction. This restriction is very meaningful in \LL: the
head occurrence is the only occurrence of a variable in a term which is linear.

\LL{} is based on the distinction of particular proofs among all proofs,
that are linear wrt.~their hypotheses. The word \emph{linear} has here two
deeply related meanings.
\begin{itemize}
\item An algebraic meaning: a linear morphism is a function which preserves
  sums, linear combinations, joins, unions (depending on the context). In most
  denotational models of \LL, linear proofs are interpreted as
  functions which are linear in that sense.
\item An operational meaning: a proof is linear wrt.~an hypothesis if the
  corresponding argument is used exactly once (neither erased nor duplicated)
  during cut-elimination.
\end{itemize}
\LL{} has an essential operation, called \emph{dereliction}, which allows one
to turn a linear proof into a non linear one, or, more precisely, to forget the
linearity of a proof. Differentiation, which in some sense is the converse of
dereliction, since it turns a non linear morphism (proof) into a linear one,
has not been included in \LL{} at an early stage of its development.


We think that there are two deep reasons for that omission.
\begin{itemize}
\item First, differentiation seems fundamentally incompatible with
  \emph{totality}, a denotational analogue of normalization usually considered
  as an essential feature of any reasonable logical system. Indeed, turning a
  non-linear proof into a linear one necessarily leads to a loss of information
  and to the production of \emph{partial} linear proofs. This is typically what
  happens when one takes the derivative of a constant proof, which must produce
  a zero proof.
%
%
\item Second, they seem incompatible with determinism because, when one
  linearizes a proof obtained by contracting two linear inputs of a proof, one
  has to choose between these two inputs, and there is no canonical way of
  doing so: we take the non-deterministic superposition of the two
  possibilities. Syntactically, this means that one must accept the possibility
  of adding proofs of the same formula, which is standard in mathematics, but
  hard to accept as a primitive logical operation on proofs (although it is
  present, in a tamed version, in the additive rules of \LL{}).
\end{itemize}
The lack of totality is compatible with most mathematical interpretations of
proofs and with most denotational models of \LL: Scott domains (or more
precisely, prime algebraic complete lattices,
see~\cite{Huth93,Winskel04,Ehrhard12a}), dI-domains, concrete data structures,
coherence spaces, games, hypercoherence spaces etc. Moreover, computer
scientists are acquainted with the use of syntactic partial objects (fix-point
operators in programming languages, B\"ohm trees of the $\lambda$-calculus etc.)
and various modern proof formalisms, such as Girard's Ludics, also incorporate
partiality for enlarging the world of ``proof-objects'' so as to allow the
simultaneous existence of ``proofs'' and ``counter-proofs'' in order to obtain
a rich duality theory on top of which a notion of totality discriminating
genuine proofs from partial proof-objects can be developed.

It is only when we observed that the differential extension of \LL{} is
the mirror image of the structural (and dereliction) rules of \LL{} that
we considered this extension as logically meaningful and worth being studied
more deeply. The price to pay was the necessity of accepting an intrinsic non
determinism and partiality in logic (these two extensions being related:
failure is the neutral element of non-determinism), but the gain was a new
viewpoint on the exponentials, related to the Taylor Formula of calculus.

In \LL, the exponential is usually thought of as the modality of duplicable
information. Linear functions are not allowed to copy their arguments and are
therefore very limited in terms of computational expressive power, the
exponential allows one to define non linear functions which can duplicate and
erase their arguments and are therefore much more powerful. This duplication
and erasure capability seems to be due to the presence of the rules of
contraction and weakening in \LL{}, but this is not quite true: the genuinely
infinite rule of \LL{} is promotion which makes a proof duplicable an arbitrary
number of times, and erasable. This fact could not be observed in \LL{} because
promotion is the only rule of \LL{} which allows one to introduce the ``$!$''
modality: without promotion, it is impossible to build a proof object that can
be cut on a contraction or a weakening rule.

In Differential \LL{} (\DILL), there are two new rules to introduce the ``$!$''
modality: \emph{coweakening} and \emph{codereliction}. The first of these rules
allows one to introduce an empty proof of type $\Excl A$ and the second one
allows one to turn a proof of type $A$ into a proof of type $\Excl A$,
\emph{without making it duplicable} in sharp contrast with the promotion
rule. The last new rule, called \emph{cocontraction}, allows one to merge two
proofs of type $\Excl A$ for creating a new proof of type $\Excl A$. This
latter rule is similar to the \emph{tensor} rule of ordinary \LL{} with the
difference that the two proofs glued together by a cocontraction must have the
same type and cannot be distinguished anymore deterministically, whereas the
two proofs glued by a tensor can be separated again by cutting the resulting
proof against a \emph{par} rule. These new rules are called \emph{costructural}
rules to stress the symmetry with the usual structural rules of \LL.

\DILL{} has therefore a \emph{finite} fragment which contains the standard
``$\wn$'' rules (weakening, contraction and dereliction) as well as the new
``$\oc$'' ones (coweakening, cocontraction and codereliction), but not the
promotion rule. Cut elimination in this system generates sums of proofs, and
therefore it is natural to endow proofs with a vector space (or module)
structure over a field (or more generally over a semi-ring\footnote{This
  general setting allows us to cover also ``qualitative'' situations where
  sums of proofs are lubs in a poset.}). This fragment has the following
pleasant properties:
\begin{itemize}
\item It enjoys strong normalization, even in the untyped case, as long as one
  considers only proof-nets which satisfy a correctness criterion similar to
  the standard Danos-Regnier criterion for multiplicative \LL{} (\MLL).
\item In this fragment, all proofs are linear combinations of ``simple proofs''
  which do not contain linear combinations: this is possible because all the
  syntactic constructions of this fragment are multilinear. So proofs are
  similar to polynomials or power series, simple proofs playing the role of
  monomials in this algebraic analogy which is strongly suggested by the
  denotational models of \DILL.
\end{itemize}
Moreover, it is possible to transform any instance of the promotion rule (which
is applied to a sub-proof $\pi$) into an infinite linear combination of proofs
containing copies of $\pi$: this is the Taylor expansion of promotion. This
operation can be applied hereditarily to all instances of the promotion rule in
a proof, giving rise to an infinite linear combinations of promotion-free
\DILL{} simple proofs with positive rational coefficients.

\paragraph*{Outline.}
We start with a syntactic presentation of \DILL{}, in a proof-net formalism
which uses terms instead of graphs (in the spirit of Abramsky's linear chemical
abstract machine~\cite{Abramsky93} or of the formalisms studied by Fernandez
and Mackie, see for instance~\cite{FernandezMackie99,MackieSato08}) and we
present a categorical formalism which allows us to describe denotational models
of \DILL. We define the interpretation of proof-nets in such categories.

Then we shortly describe a differential $\lambda$-calculus
formalism and we summarize some results, giving bibliographical references.

The end of the paper is devoted to concrete models of \DILL{}. We briefly
review the relational model, which is based on the $*$-autonomous category of
sets and relations (with the usual cartesian product of sets as tensor product)
because it underlies most denotational models of (differential) \LL{}.
Then we describe the \emph{finiteness space} model which was one of our main
motivations for introducing \DILL{}. We provide a thorough description of this
model, insisting on various aspects which were not covered by our initial
presentation in~\cite{Ehrhard00b} such as \emph{linear boundedness} (whose
relevance in this semantical setting has been pointed out by Tasson
in~\cite{Tasson09,Tasson09b}), or the fact that function spaces in the Kleisli
category admit an intrinsic description.

\medbreak

One important step in our presentation of the categorical setting for
interpreting differential \LL{} is the notion of an \emph{exponential
  structure}. It is the categorical counterpart of the finitary fragment of
\DILL{}, that is, the fragment \DILLZ{} where the promotion rule is not
required to hold. 

An exponential structure consists of a preadditive\footnote{This means that the
  monoidal category is enriched over commutative monoids. Actually, we assume
  more generally that it is enriched over the category of $\Field$-modules,
  where $\Field$ is a given semi-ring.}  $*$-autonomous category $\cL$ together
with an operation which maps any object $X$ of $\cL$ to an object $\Excl X$ of
$\cL$ equipped with a structure of $\otimes$-bialgebra (representing the
structural and costructural rules) as well as a ``dereliction'' morphism in
$\cL(\Excl X,X)$ and a ``codereliction'' morphism $\cL(X,\Excl X)$. The
important point here is that the operation $X\mapsto\Excl X$ is not assumed to
be functorial (it has nevertheless to be a functor on isomorphisms). Using this
simple structure, we define in particular morphisms $\Coderc
X\in\cL(\Tens{\Excl X}{X},\Excl X)$ and $\Derc X\in\cL(\Excl X,\Tens{\Excl
  X}{X})$.

An element of $\cL(\Excl X,Y)$ can be considered as a non-linear morphism from
$X$ to $Y$ (some kind of generalized polynomial, or analytical function), but
these morphisms cannot be composed. It is nevertheless possible to define a
notion of polynomial such morphism, and these polynomial morphisms can be
composed, giving rise to a category which is cartesian if $\cL$ is cartesian.

By composition with $\Coderc X\in\cL(\Tens{\Excl X}{X},\Excl X)$, any element
$f$ of $\cL(\Excl X,Y)$ can be differentiated, giving rise to an element $f'$
of $\cL(\Tens{\Excl X}X,Y)$ that we consider as its derivative\footnote{Or
  differential, or Jacobian: by monoidal closedness, $f'$ can be seen as an
  element of $\cL(\Excl X,\Limpl XY)$ where $\Limpl XY$ is the object of
  morphisms from $X$ to $Y$ in $\cL$, that is, of linear morphisms from $X$ to
  $Y$, and the operation $f\mapsto f'$ satisfies all the ordinary properties of
  differentiation.}. This operation can be performed again, giving rise to
$f''\in\cL(\Tens{\Excl X}{\Tens XX},Y)$ and, assuming that cocontraction is
commutative, this morphism is symmetric in its two last linear parameters (a
property usually known as \emph{Shwarz Lemma}).

In this general context, a very natural question arises.  Given a morphism
$g\in\cL(\Tens{\Excl X}X,Y)$ whose derivative $g'\in\cL(\Tens{\Excl X}{\Tens
  XX},Y)$ is symmetric, can one always find a morphism $f\in\cL(\Excl X,Y)$
such that $g=f'$? Inspired by the usual proof of \emph{Poincar\'e's Lemma}, we
show that such an \emph{antiderivative} is always available as soon as the
natural morphism $\Id_{\Excl X}+(\Coderc X\Compl\Derc X)\in\cL(\Excl X,\Excl
X)$ is an isomorphism for each object $X$. We explain how this property is
related to a particular case of integration by parts. We also describe briefly
a syntactic version of antiderivatives in a promotion-free differential
$\lambda$-calculus.

To interpret the whole of \DILL, including the promotion rule, one has to
assume that $\Excl\_$ is an endofunctor on $\cL$ and that this functor is
endowed with a structure of comonad and a monoidal structure; all these data
have to satisfy some coherence conditions wrt.~the exponential structure. These
conditions are essential to prove that the interpretation of proof-nets is
invariant under the various reduction rules, among which the most complicated
one is an \LL{} version of the usual \emph{chain rule} of calculus. Our main
references here are the work of Bierman~\cite{Bierman95},
Melli\`es~\cite{Mellies09} and, for the commutations involving costructural
logical rules, our concrete models~\cite{Ehrhard00b,Ehrhard00c}, the
categorical setting developed by Blute, Cockett and
Seely~\cite{BluteCockettSeely06} and, very importantly, the work of
Fiore~\cite{Fiore07}.
\medbreak

One major \emph{a priori} methodological principle applied in this paper is to
stick to \emph{Classical Linear Logic}, meaning in particular that the
categorical models we consider are $*$-autonomous categories. This is justified
by the fact that most of the concrete models we have considered so far satisfy
this hypothesis (with the noticeable exception
of~\cite{BluteEhrhardTasson10}) and it is only in this
setting that the new symmetries introduced by the differential and costructural
rules appear clearly. A lot of material presented in this paper could probably
be carried to a more general intuitionistic Linear Logic setting.
\medbreak

Some aspects of \DILL{} are only alluded to in this presentation, the most
significant one being certainly the Taylor expansion formula and its connection
with linear head reduction. On this topic, we refer
to~\cite{EhrhardRegnier06a,EhrhardRegnier06b,Ehrhard10b}.

\paragraph*{Note on this version.}
A final version of this paper will appear in \emph{Mathematical Structures in
  Computer Science}. The first version of this survey (already containing the
material on antiderivatives) has been written in 2011 and has been improved and
enriched since that time.

\section*{Notations}

In this paper, a set of coefficients is needed, which has to be a commutative
semi-ring. This set will be denoted as $\Field$. In
Section~\ref{sec:finiteness-spaces}, $\Field$ will be assumed to be a field but
this assumption is not needed before that section.

\section{Syntax for \DILL{} proof-structures}
We adopt a presentation of proof-structures and proof-nets which is based on
terms and not on graphs. We believe that this presentation is more suitable to
formalizable mathematical developments, although it sometimes gives rise to
heavy notations, especially when one has to deal with the promotion rule
(Section~\ref{sec:promotion-syntax}). We try to provide graphical intuitions on
proof-structures by means of figures.  

\subsection{General constructions}
\label{sec:gen-proof-structures}

\paragraph{Simple proof-structures.}
Let $\Varset$ be an infinite countable set of variables. This set is equipped
with an involution $x\mapsto\Covar x$ such that $x\not=\Covar x$ for each
$x\in\Varset$. 

Let $u\subseteq\Varset$. An element $x$ of $u$ is \emph{bound} in $u$ if $\Covar
x\in u$. One says that $u$ is \emph{closed} if all the elements of $u$ are
bound in $u$. If $x$ is not bound in $u$, one says that $x$ is \emph{free} in
$u$.

Let $\Conset$ be a set of \emph{tree constructors}, given together with an
arity map $\Arity:\Conset\to\Nat$.

\emph{Proof trees} are defined as follows, together with their associated set of
variables:
\begin{itemize}
\item if $x\in\Varset$ then $x$ is a tree and $\Varnet(x)=\{x\}$;
\item if $\phi\in\Conset_n$ (that is $\phi\in\Sigma$ and $\Arity(\phi)=n$)
  and if $\List t1n$ are trees with $\Varnet(t_i)\cap\Varnet(t_j)=\emptyset$
  for $i\not=j$, then $t=\phi(\List t1n)$ is a tree with
  $\Varnet(t)=\Varnet(t_i)\cup\cdots\cup\Varnet(t_n)$. As usual, when $\phi$ is
  binary, we often use the infix notation $t_1\Rel\phi t_2$ rather than
  $\phi(t_1,t_2)$.
\end{itemize}

A \emph{cut} is an expression $\Cut{t}{t'}$ where $t$ and $t'$ are trees
such that $\Varnet(t)\cap\Varnet(t')=\emptyset$. We set
$\Varnet(c)=\Varnet(t)\cup\Varnet(t')$. 

A \emph{simple proof-structure} is a pair $p=\Net{\Vect t}{\Vect c}$ where
$\Vect t$ is a list of proof trees and $\Vect c$ is a list of cuts, whose sets
of variables are pairwise disjoint.

\begin{remark}
  The order of the elements of $\Vect c$ does not matter; we could have used
  multisets instead of sequences. In the sequel, we consider these sequences of
  cuts up to permutation.
\end{remark}

Bound variables of $\Varnet(p)$ can be renamed in the obvious way in $p$
(rename simultaneously $x$ and $\Covar x$ avoiding clashes with other variables
which occur in $p$) and simple proof-structures are considered up to such
renamings: this is $\alpha$-conversion. Let $\Fvarnet(p)$ be the set of free
variables of $p$. We say $p$ is closed if $\Fvarnet(p)=\emptyset$.

The simplest simple proof-structure is of course $\Net{}{}$. A less trivial
closed simple proof-structure is $\Net{}{\Cut{x}{\Covar x}}$ which is a loop.

\paragraph{\LL{} types.}
Let $\Atoms$ be a set of type atoms ranged over by $\alpha,\beta,\dots$,
together with an involution $\alpha\mapsto\Coatom\alpha$ such that
$\Coatom\alpha\not=\alpha$. Types are defined as follows.
\begin{itemize}
\item if $\alpha\in\Atoms$ then $\alpha$ is a type;
\item if $A$ and $B$ are types then $\Tens AB$ and $\Par AB$ are
  types;
\item if $A$ is a type then $\Excl A$ and $\Int A$ are types.
\end{itemize}
The linear negation $\Orth A$ of a type $A$ is given by the following inductive
definition: $\Orth\alpha=\Coatom\alpha$, $\Orth{(\Tens AB)}=\Par{\Orth A}{\Orth
  B}$; $\Orth{(\Par AB)}=\Tens{\Orth A}{\Orth B}$; $\Orth{(\Excl A)}=\Int{\Orth
  A}$ and $\Orth{(\Int A)}=\Excl{\Orth A}$.

An \MLL{} type is a type built using only the $\ITens$ and $\IPar$
constructions\footnote{We do not consider the multiplicative constants $\One$
  and $\Bot$ because they are not essential for our purpose.}.

\subsection{Proof-structures for \MLL}
Assume that $\Conset_2=\{\mathord{\ITens},\mathord{\IPar}\}$ and that
$\Arity(\mathord{\ITens})=\Arity(\mathord{\IPar})=2$.

A \emph{typing context} is a finite partial function $\Phi$ (of domain
$\Dom\Phi$) from $\Varset$ to formulas such that $\Phi(\Covar
x)=\Orth{(\Phi(x))}$ whenever $x,\Covar x\in\Dom\Phi$. 

\paragraph{Typing rules.}
We first explain how to type \MLL{} proof trees. The corresponding typing
judgments of the form $\Typing\Phi tA$ where $\Phi$ is a typing context, $t$ is
a proof tree and $A$ is a formula.

The rules are
\begin{center}
  \AxiomC{}
  \UnaryInfC{$\Typing{\Phi,x:A}{x}{A}$}
  \DisplayProof
\end{center}
\begin{center}
  \AxiomC{$\Typing{\Phi}{s}{A}$}
  \AxiomC{$\Typing{\Phi}{t}{B}$}
  \BinaryInfC{$\Typing{\Phi}{\Tens st}{\Tens AB}$}
  \DisplayProof
  \quad
  \AxiomC{$\Typing{\Phi}{s}{A}$}
  \AxiomC{$\Typing{\Phi}{t}{B}$}
  \BinaryInfC{$\Typing{\Phi}{\Par st}{\Par AB}$}
  \DisplayProof
\end{center}

Given a cut $c=\Cut s{s'}$ and a typing context $\Phi$, one writes
$\TypingCut\Phi c$ if there is a type $A$ such that
$\Typing\Phi sA$ and $\Typing{\Phi}{s'}{\Orth A}$.

Last, given a simple proof-structure $p=\Net{\vec s}{\vec c}$ with $\vec
s=(\List s1n)$ and $\vec c=(\List c{1}k)$, a sequence $\Gamma=(\List A1l)$ of
formulas and a typing context $\Phi$, one writes $\Typing\Phi p\Gamma$ if $l=n$
and $\Typing{\Phi}{s_i}{A_i}$ for $1\leq i\leq n$ and $\TypingCut{\Phi}{c_i}$
for $1\leq i\leq k$.




\paragraph{Logical judgments.}
A logical judgment is an expression $\Logic\Phi p{\Gamma}$ where $\Phi$ is
a typing context, $p$ is a simple proof-structure and $\Gamma$ is a list of
formulas. 

If one can infer that $\Logic\Phi p\Gamma$, this means that the proof-structure
$p$ represents a proof of $\Gamma$. Observe that the inference rules coincide
with the rules of the \MLL{} sequent calculus.

We give now these logical rules.

\begin{center}
  \AxiomC{}
  \RightLabel{\quad\AXIOM}
  \UnaryInfC{$\Logic{\Phi,x:A,\Covar x:\Orth A}{\Net{x,\Covar x}{}}{A,\Orth A}$}
  \DisplayProof
\end{center}
\begin{center}
  \AxiomC{$\Logic{\Phi}{\Net{\List t1n}{\vec c}}{\List A1n}$}
  \RightLabel{\quad\PERMRULE{}, $\sigma\in\Symgrp n$}
  \UnaryInfC{$\Logic{\Phi}{\Net{t_{\sigma(1)},\dots,t_{\sigma(n)}}{\vec c}}
    {A_{\sigma(1)},\dots,A_{\sigma(n)}}$}
  \DisplayProof
\end{center}
\begin{center}
  \AxiomC{$\Logic\Phi{\Net{\Vect s,s}{\Vect c}}{\Gamma,A}$}
  \AxiomC{$\Logic\Phi{\Net{\Vect t,t}{\Vect d}}{\Delta,\Orth A}$}
  \RightLabel{\quad\CUTRULE}
  \BinaryInfC{$\Logic{\Phi}
    {\Net{\Vect s,\Vect t}
      {\Vect c,\Vect d,\Cut st}}{\Gamma,\Delta}$}
  \DisplayProof
\end{center}
\begin{center}
  \AxiomC{$\Logic\Phi{\Net{\Vect t,s,t}{\Vect c}}{\Gamma,A,B}$}
  \RightLabel{\quad\PARRULE}
  \UnaryInfC{$\Logic\Phi{\Net{\Vect t,\Par st}{\Vect c}}{\Gamma,\Par AB}$}
  \DisplayProof
\end{center}
\begin{center}
  \AxiomC{$\Logic\Phi{\Net{\Vect s,s}{\Vect c}}{\Gamma,A}$}
  \AxiomC{$\Logic\Phi{\Net{\Vect t,t}{\Vect d}}{\Delta,B}$}
  \RightLabel{\quad\TENSRULE}
  \BinaryInfC{$\Logic{\Phi}
    {\Net{\Vect s,\Vect t,\Tens st}
      {\Vect c,\Vect d}}{\Gamma,\Delta,\Tens AB}$}
  \DisplayProof
\end{center}
We add the mix rule for completeness because it is quite natural
denotationally. Notice however that it is not necessary. In particular,
mix-free proof-nets are closed under cut elimination.
\begin{center}
  \AxiomC{$\Logic\Phi{\Net{\vec s}{\vec c}}\Gamma$}
  \AxiomC{$\Logic\Phi{\Net{\vec t}{\vec d}}\Delta$}
  \RightLabel{\quad\MIXRULE}
  \BinaryInfC{$\Logic{\Phi}{\Net{\vec s,\vec t}{\vec c,\vec d}}
    {\Gamma,\Delta}$}
  \DisplayProof
\end{center}

\begin{lemma}
  If $\Logic\Phi p\Gamma$ then $\Typing\Phi p\Gamma$ and
  $\Varnet(p)$ is closed.
\end{lemma}
\Beginproof
Straightforward induction on derivations.
\Endproof


\subsection{Reducing proof-structures}
The basic reductions concern cuts, and are of the form
\[
c\Rel\Redcut\Net{\Vect t}{\Vect d}
\]
where $c$ is a cut, $\Vect d=(\List d1n)$ is a sequence of cuts and $\Vect
t=(\List t1k)$ is a sequence fo trees.

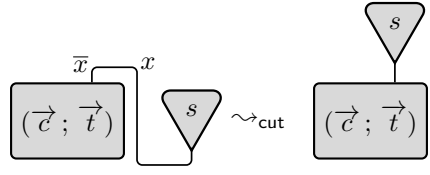
\begin{figure}[t]
  \centering
  \begin{tikzpicture}
    \Somenet{1.5}{1}{cont}{$\Net{\Vect t}{\Vect c}$}
    \node[right=0.2 of cont.north](pos){};
    \node[above=0 of pos](axn){};
    \node[left=-0.2 of axn]{$\Covar x$};
    \node[right=0.4 of axn]{$x$};
    \node[above=-0.5 of cont](ref){};
    \Sometree[right=1.2 of ref]{0.8}{s}{$s$}
    \node[right=1.3 of cont]{$\Redcut$};
    \draw[inetwire](pos.center)|-++(0.2,0.2)-|++(0.4,-1)|-++(0.2,-0.3)-|(s.pal);
    \Somenet[right=2.5 of cont]{1.5}{1}{contr}{$\Net{\Vect t}{\Vect c}$}
    \node[right=0.2 of contr.north](posr){};
    \Sometree[above=0.1 of posr]{0.8}{sr}{$s$}
    \draw[inetwire](sr.pal)--(posr.center);
  \end{tikzpicture}
  \caption{The axiom/cut reduction}
  \label{fig:ax-cut-red}
\end{figure}

With similar notational conventions, here are the deduction rules for the
reduction of \MLL{} proof-structures.
\begin{center}
  \AxiomC{$c\Rel\Redcut\Net{\Vect t}{\Vect d}$}
  \RightLabel{\quad\CONTEXT}
  \UnaryInfC{$\Net{\Vect s}{c,\Vect
      b}\Rel\Redcut\Net{\Vect s,\Vect t}{\Vect d,\Vect b}$} \DisplayProof
\end{center}
\begin{center}
  \AxiomC{$\Covar x\notin\Varnet(s)$}
  \RightLabel{\quad\VARCUT}
  \UnaryInfC{$\Net{\Vect t}{\Cut{x}{s},\Vect c}
    \Rel\Redcut\Subst{\Net{\Vect t}{\Vect c}}{s}{\Covar x}$} \DisplayProof
\end{center}
For applying the latter rule (see Figure~\ref{fig:ax-cut-red}), we assume that
$\Covar x\notin\Varnet(s)$.  Without this restriction, we would reduce the
cyclic proof-structure $\Net{}{\Cut x{\Covar x}}$ to $\Net{}{}$ and erase the
cycle which is certainly not acceptable from a semantic viewpoint. For
instance, in a model of proof-structures based on finite dimension vector
spaces, the semantics of $\Net{}{\Cut x{\Covar x}}$ would be the dimension of
the space interpreting the type of $x$ (trace of the identity).

\begin{remark}
  We provide some pictures to help understand the reduction rules on proof
  structures. In these pictures, logical proof-net constructors (such as
  \emph{tensor}, \emph{par} etc.) are represented as white triangles labeled by
  the corresponding symbol --~they correspond to the \emph{cells} of
  interaction nets or to the \emph{links} of proof-nets~-- and subtrees are
  represented as gray triangles. 

  Wires represent the edges of a proof tree. We also represent axioms and cuts
  as wires: an axiom looks like
  \begin{tikzpicture}
    \draw[inetwire](0,0)|-++(0.2,0.15)-|(0.4,0);
  \end{tikzpicture}
  and a cut looks like
  \begin{tikzpicture}
    \draw[inetwire](0,0)|-++(0.2,-0.15)-|(0.4,0);
  \end{tikzpicture}\,.  In Figure~\ref{fig:ax-cut-red}, we indicate the
  variables associated with the axiom, but in the next pictures, this
  information will be kept implicit.

  Figure~\ref{fig:simple-proof-structure} represents the simple proof-structure
  \[
  p=\Net{\List t1k}{\Cut{s_1}{s'_1},\dots,\Cut{s_n}{s'_n}}\,.
  \]
  with free variables $\List x1l$. The box named \emph{axiom links} contains
  axioms connecting variables occurring in the trees $\List
  s1n,\List{s'}1n,\List t1k$. When we do not want to be specific about its
  content, we represent such a simple proof-structure as in
  Figure~\ref{fig:proof-struct-synth} by a gray box with indices $1,\dots,k$ on
  its border for locating the roots of the trees of $p$. The same kind of
  notation will be used also for proof-structures which are not necessarily
  simple, see the beginning of Paragraph~\ref{par:red-rules} for this notion.
\end{remark}

\begin{figure}[t]
  \centering
  \begin{tikzpicture}
    \Axioms{8}{axioms};

    \node[left=0.7cm of axioms.west](peqp){};
    \node[below=0.2 of peqp]{$p=$};

    \node[left=2.9 of axioms.south](pltc1){};
    \node[below=0.6 of pltc1.center](ltc1){};
    \Sometree[left=0 of ltc1]{0.9}{s1}{$s_1$}
    \node[above=0 of s1.pax]{$\,\dots$};
    \Sometree[right=0 of ltc1]{0.9}{sp1}{$s'_1$}
    \node[above=0 of sp1.pax]{$\,\dots$};
    \Inetcut{s1.pal}{sp1.pal};
    \draw[inetwire](s1.right pax)--(s1.right pax|-axioms.south);
    \draw[inetwire](s1.left pax)--(s1.left pax|-axioms.south);
    \draw[inetwire](sp1.right pax)--(sp1.right pax|-axioms.south);
    \draw[inetwire](sp1.left pax)--(sp1.left pax|-axioms.south);

    \node[left=0.7 of axioms.south](pltcn){};
    \node[below=0.6 of pltcn.center](ltcn){};
    \Sometree[left=0 of ltcn]{0.9}{sn}{$s_n$}
    \node[above=0 of sn.pax]{$\,\dots$};
    \Sometree[right=0 of ltcn]{0.9}{spn}{$s'_n$}
    \node[above=0 of spn.pax]{$\,\dots$};
    \Inetcut{sn.pal}{spn.pal};
    \draw[inetwire](sn.right pax)--(sn.right pax|-axioms.south);
    \draw[inetwire](sn.left pax)--(sn.left pax|-axioms.south);
    \draw[inetwire](spn.right pax)--(spn.right pax|-axioms.south);
    \draw[inetwire](spn.left pax)--(spn.left pax|-axioms.south);

    \node[left=1.78 of axioms.south](pldots){};
    \node[below=0.9 of pldots.center]{$\dots$};

    \node[right=2.6 of axioms.south](prdots){};
    \node[below=0.6 of prdots.center](rtcenter){};
    \Sometree[left=0.4 of rtcenter.center]{0.9}{t1}{$t_1$}
    \Sometree[right=0.4 of rtcenter.center]{0.9}{tk}{$t_k$}

    \draw[inetwire](t1.right pax)--(t1.right pax|-axioms.south);
    \draw[inetwire](t1.left pax)--(t1.left pax|-axioms.south);
    \draw[inetwire](tk.right pax)--(tk.right pax|-axioms.south);
    \draw[inetwire](tk.left pax)--(tk.left pax|-axioms.south);

    \node[above=0 of t1.pax]{$\,\dots$};
    \node[above=0 of tk.pax]{$\,\dots$};

    \node[below=0.9 of prdots.center]{$\dots$};

    \inetwirefree(t1.pal)
    \inetwirefree(tk.pal)

    \node[left=0.5 of axioms.north](lfree){};
    \node[above=0.3 of lfree.center](lvar){$x_1$};
    \draw[inetwire](lfree.center)--(lvar.south);
    \node[right=0.5 of axioms.north](rfree){};
    \node[above=0.3 of rfree.center](rvar){$x_l$};
    \draw[inetwire](rfree.center)--(rvar.south);
    \node[above=0 of axioms.north]{$\dots$};
  \end{tikzpicture}
  \caption{A simple proof-structure}
  \label{fig:simple-proof-structure}
\end{figure}
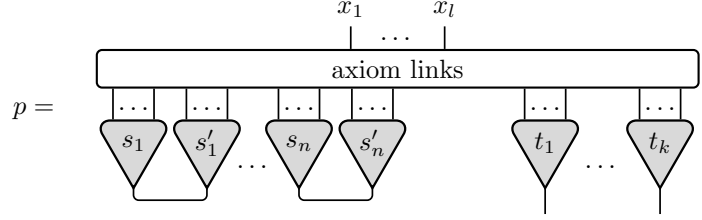

\begin{figure}[t]
  \centering
  \begin{tikzpicture}
    \Somenet{2}{1}{cont}{$p$}

    \node[left=0.5 of cont.south](c1){};
    \node[above=0 of c1.center]{$\scriptstyle{1}$};
    \node[below=0.3 of c1.center](bc1){};
    \draw[inetwire](c1.center)--(bc1.north);

    \node[right=0.5 of cont.south](ck){};
    \node[above=0 of ck.center]{$\scriptstyle{k}$};
    \node[below=0.3 of ck.center](bck){};
    \draw[inetwire](ck.center)--(bck.north);

    \node[below=0 of cont.south]{$\dots$};

    \node[left=0.5 of cont.north](f1){};
    \node[above=0.3 of f1.center](x1){$x_1$};
    \draw[inetwire](f1.center)--(x1.south);

    \node[right=0.5 of cont.north](fl){};
    \node[above=0.3 of fl.center](xl){$x_l$};
    \draw[inetwire](fl.center)--(xl.south);

    \node[above=0 of cont.north]{$\dots$};
  \end{tikzpicture}
  \caption{A synthetic representation of the proof-structure of Figure~\ref{fig:simple-proof-structure}}
  \label{fig:proof-struct-synth}
\end{figure}
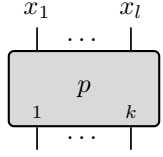

\begin{figure}[t]
  \centering
\begin{tikzpicture}
  \inetcell(tens){$\IPar$}
  \inetcell[right=1.7cm of tens](par){$\ITens$}

  \draw[inetwire](tens.pal)|-++(1,-0.2)-|(par.pal);

  \node[above=0.7cm of tens.pax](tensup){};

  \Sometree[left=0 of tensup]{0.8}{s1}{$s_1$}
  \Sometree[right=0 of tensup]{0.8}{s2}{$s_2$}
  \inetwire(s1.pal)(tens.right pax)
  \inetwire(s2.pal)(tens.left pax)

  \node[above=0.7cm of par.pax](parup){};

  \Sometree[left=0 of parup]{0.8}{t1}{$t_1$}
  \Sometree[right=0 of parup]{0.8}{t2}{$t_2$}
  \inetwire(t1.pal)(par.right pax)
  \inetwire(t2.pal)(par.left pax)

  \node[right=0.6 of par](red){$\Redcut$};
  \node[right=3.1 of par](refp){};
  \node[above=0.3 of refp](ref){};

  \Sometree[left=1cm of ref]{0.8}{sr1}{$s_1$}
  \Sometree[left=0.1cm of ref]{0.8}{sr2}{$s_2$}
  \Sometree[right=0.1cm of ref]{0.8}{tr1}{$t_1$}
  \Sometree[right=1cm of ref]{0.8}{tr2}{$t_2$}
  \draw[inetwire](sr1.pal)|-++(0.5,-0.2)-|(tr1.pal);
  \draw[inetwire](sr2.pal)|-++(0.5,-0.3)-|(tr2.pal);
\end{tikzpicture}
  
  \caption{The tensor/par reduction}
  \label{fig:tens-par-red}
\end{figure}
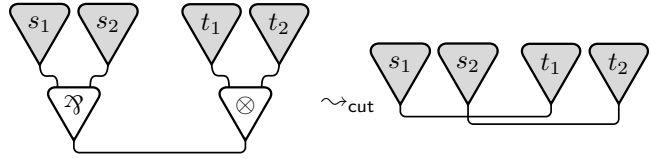

In \MLL, we have only one basic reduction (see Figure~\ref{fig:tens-par-red}):
\begin{equation*}
  \Cut{\Par{s_1}{s_2}}{\Tens{t_1}{t_2}}
  \Rel\Redcut
  \Net{}{\Cut{s_1}{t_1},\Cut{s_2}{t_2}}
\end{equation*}

\subsection{\DILLZ}\label{sec:DILLZ}
This is the promotion-free fragment of differential \LL{}. In \DILLZ{},
one extends the signature of \MLL{} with new constructors:
\begin{itemize}
\item $\Conset_0=\{\WEAK,\COWEAK\}$, called respectively \emph{weakening} and
  \emph{coweakening}.
\item $\Conset_1=\{\DER,\CODER\}$, called respectively \emph{dereliction} and
  \emph{codereliction}.
\item $\Conset_2=\{\IPar,\ITens,\CONTR,\COCONTR\}$, the two new constructors
  being called respectively \emph{contraction} and \emph{cocontraction}.
\item $\Conset_n=\emptyset$ for $n>2$.
\end{itemize}

\paragraph{Typing rules.}
The typing rules for the four first constructors are similar to those of \MLL.
\begin{center}
  \AxiomC{}
  \UnaryInfC{$\Typing\Phi\WEAK{\Int A}$}
  \DisplayProof
  \quad
  \AxiomC{}
  \UnaryInfC{$\Typing\Phi\COWEAK{\Excl A}$}
  \DisplayProof
\end{center}
\begin{center}
  \AxiomC{$\Typing\Phi tA$}
  \UnaryInfC{$\Typing\Phi{\DER(t)}{\Int A}$}
  \DisplayProof
  \quad
  \AxiomC{$\Typing\Phi tA$}
  \UnaryInfC{$\Typing\Phi{\CODER(t)}{\Excl A}$}
  \DisplayProof
\end{center}

The two last rules require the subtrees to have the same type.
\begin{center}
  \AxiomC{$\Typing\Phi{s_1}{\Int A}$}
  \AxiomC{$\Typing\Phi{s_2}{\Int A}$}
  \BinaryInfC{$\Typing\Phi{\CONTR(s_1,s_2)}{\Int A}$}
  \DisplayProof
  \quad
  \AxiomC{$\Typing\Phi{s_1}{\Excl A}$}
  \AxiomC{$\Typing\Phi{s_2}{\Excl A}$}
  \BinaryInfC{$\Typing\Phi{\COCONTR(s_1,s_2)}{\Excl A}$}
  \DisplayProof
\end{center}

\paragraph{Logical rules.}
The additional logical rules are as follows.
\begin{center}
  \AxiomC{$\Logic{\Phi}{\Net{\Vect s}{\Vect c}}{\Gamma}$}
  \RightLabel{\quad\WEAKENING}
  \UnaryInfC{$\Logic{\Phi}{\Net{\Vect s,\WEAK}{\Vect c}}{\Gamma,\Int A}$}
  \DisplayProof
  \quad
  \AxiomC{}
  \RightLabel{\quad\COWEAKENING}
  \UnaryInfC{$\Logic{\Phi}{\Net{\COWEAK}{}}{\Excl A}$}
  \DisplayProof
\end{center}
\begin{center}
  \AxiomC{$\Logic{\Phi}{\Net{\Vect s,s}{\Vect c}}{\Gamma,A}$}
  \RightLabel{\quad\DERELICTION}
  \UnaryInfC{$\Logic{\Phi}{\Net{\Vect s,\DER(s)}{\Vect c}}{\Gamma,\Int A}$}
  \DisplayProof
\end{center}
\begin{center}
  \AxiomC{$\Logic{\Phi}{\Net{\Vect s,s}{\Vect c}}{\Gamma,A}$}
  \RightLabel{\quad\CODERELICTION}
  \UnaryInfC{$\Logic{\Phi}{\Net{\Vect s,\CODER(s)}{\Vect c}}{\Gamma,\Excl A}$}
  \DisplayProof
\end{center}
\begin{center}
  \AxiomC{$\Logic{\Phi}{\Net{\Vect s,s_1,s_2}{\Vect c}}{\Gamma,\Int A,\Int A}$}
  \RightLabel{\quad\CONTRACTION}
  \UnaryInfC{$\Logic{\Phi}{\Net{\Vect s,\CONTR(s_1,s_2)}
      {\Vect c}}{\Gamma,\Int A}$}
  \DisplayProof
\end{center}
\begin{center}
  \AxiomC{$\Logic{\Phi}{\Net{\Vect{s},s}{\Vect{c}}}{\Gamma,\Excl A}$}  
  \AxiomC{$\Logic{\Phi}{\Net{\Vect{t},t}{\Vect{d}}}{\Delta,\Excl A}$}  
  \RightLabel{\quad\COCONTRACTION}
  \BinaryInfC{$\Logic{\Phi}{\Net{\Vect s,\Vect{t},\COCONTR(s,t)}
      {\Vect c,\Vect{d}}}{\Gamma,\Delta,\Excl A}$}  
  \DisplayProof
\end{center}

\paragraph{Reduction rules.}
\label{par:red-rules}
To describe the reduction rules associated with these new constructions, we
need to introduce formal sums (or more generally $\Field$-linear combinations)
of simple proof-structures called \emph{proof-structures} in the sequel, and
denoted with capital letters $P,Q,\dots$\, Such an extension by linearity of
the syntax was already present in~\cite{EhrhardRegnier02}.  

The empty linear combination $0$ is a particular proof-structure which plays an
important role. 
As linear combinations, proof-structures can be linearly combined. 

The typing rule for linear combinations is

\begin{center}
  \AxiomC{$\forall i\in\{1,\dots,n\}\quad\Logic{\Phi}{p_i}{\Gamma}\text{ and
    }\mu_i\in\Field$} \RightLabel{\quad\SUMRULE}
  \UnaryInfC{$\Logic\Phi{\sum_{i=1}^n\mu_i\,p_i}\Gamma$} \DisplayProof
\end{center}

  \begin{figure}[t]
    \centering
  \begin{tikzpicture}
    \inetcell(w){$\wn$}
    \inetcell[right=0.2 of w](cow){$\oc$}
    \Inetcut{w.pal}{cow.pal}
    \node[right=0.3 of cow](){$\Redcut$};
  \end{tikzpicture}    
    \caption{Weakening/coweakening reduction}
    \label{fig:weak-coweak-red}
  \end{figure}
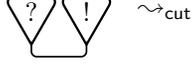

  \begin{figure}[t]
    \centering
  \begin{tikzpicture}
    \inetcell(der){$\wn$}
    \Sometree[above=0.2 of der]{0.8}{s}{$s$}
    \inetcell[right=0.3 of der](cow){$\oc$}
    \inetwire(der.pax)(s.pal)
    \Inetcut{der.pal}{cow.pal}
    \node[right=0.2 of cow](){$\Redcut\ 0$};
  \end{tikzpicture}    
\quad
  \begin{tikzpicture}
    \inetcell(w){$\wn$}
    \inetcell[right=0.3 of w](coder){$\oc$}
    \Sometree[above=0.2 of coder]{0.8}{t}{$t$}
    \inetwire(coder.pax)(t.pal)
    \Inetcut{w.pal}{coder.pal}
    \node[right=0.2 of coder](){$\Redcut\ 0$};
  \end{tikzpicture}
    \caption{Dereliction/coweakening and weakening/codereliction reductions}
    \label{fig:der-coweak-red}
  \end{figure}
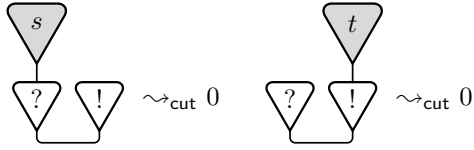

  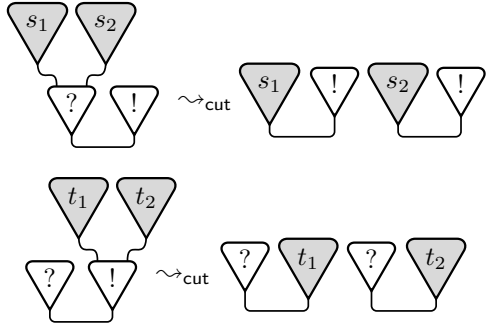
\begin{figure}[t]
    \centering
  \begin{tikzpicture}
    \inetcell(ctr){$\wn$}
    \node[above=0.7 of ctr](ref){};
    \Sometree[left=0 of ref]{0.8}{s1}{$s_1$}
    \Sometree[right=0 of ref]{0.8}{s2}{$s_2$}
    \inetcell[right=0.3 of ctr](cow){$\oc$}
    \inetwire(ctr.right pax)(s1.pal)
    \inetwire(ctr.left pax)(s2.pal)
    \Inetcut{ctr.pal}{cow.pal}
    \node[right=0.2 of cow](red){$\Redcut$};
    \node[right=1.5 of red](refp){};
    \node[above=0 of refp](ref){};
    \Sometree[left=0.8 of ref]{0.8}{sr1}{$s_1$}
    \inetcell[left=0 of ref](cow1){$\oc$}
    \Inetcut{sr1.pal}{cow1.pal}
    \Sometree[right=0 of ref]{0.8}{sr2}{$s_2$}
    \inetcell[right=0.9 of ref](cow2){$\oc$}
    \Inetcut{sr2.pal}{cow2.pal}
  \end{tikzpicture}
  \\[1em]
  \begin{tikzpicture}
    \inetcell(w){$\wn$}
    \inetcell[right=0.3 of w](coctr){$\oc$}
    \node[above=0.7 of coctr](ref){};
    \Sometree[left=0 of ref]{0.8}{t1}{$t_1$}
    \Sometree[right=0 of ref]{0.8}{t2}{$t_2$}
    \inetwire(coctr.right pax)(t1.pal)
    \inetwire(coctr.left pax)(t2.pal)
    \Inetcut{w.pal}{coctr.pal}
    \node[right=0.2 of cow](red){$\Redcut$};
    \node[right=1.5 of red](refp){};
    \node[above=0 of refp](ref){};
    \Sometree[left=0 of ref]{0.8}{tr1}{$t_1$}
    \inetcell[left=0.9 of ref](w1){$\wn$}
    \Inetcutr{tr1.pal}{w1.pal}
    \Sometree[right=0.8 of ref]{0.8}{tr2}{$t_2$}
    \inetcell[right=0 of ref](w2){$\wn$}
    \Inetcutr{tr2.pal}{w2.pal}
  \end{tikzpicture}    
    \caption{Contraction/coweakening and weakening/cocontraction reductions}
    \label{fig:contr-coweak-red}
  \end{figure}

  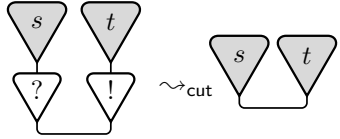
\begin{figure}[t]
    \centering
  \begin{tikzpicture}
  \node(refl){};
    \inetcell[left=0.1 of refl](der){$\wn$}
    \inetcell[right=0.1 of refl](coder){$\oc$}
    \Sometree[above=0.1 of der]{0.8}{s}{$s$}
    \Sometree[above=0.1 of coder]{0.8}{t}{$t$}
    \inetwire(s.pal)(der.pax)
    \inetwire(t.pal)(coder.pax)
    \draw[inetwire](der.pal)|-++(0.5,-0.2)-|(coder.pal);
    \node[right=0.9 of refl](red){$\Redcut$};

    \node[right=2.4 of refl](refrp){};
    \node[above=0.2 of refrp](refr){};

    \Sometree[left=0 of refr]{0.8}{sr}{$s$}
    \Sometree[right=0 of refr]{0.8}{tr}{$t$}
    \draw[inetwire](sr.pal)|-++(0.5,-0.2)-|(tr.pal);
  \end{tikzpicture}    
    \caption{Dereliction/codereliction reduction}
    \label{fig:der-coder-red}
  \end{figure}

  \begin{figure}[t]
    \centering
  \begin{tikzpicture}
    \inetcell(ctr){$\wn$}
    \inetcell[right=1.2cm of ctr](coder){$\oc$}

    \draw[inetwire](ctr.pal)|-++(1,-0.2)-|(coder.pal);

    \node[above=0.7cm of ctr](ctrup){};

    \Sometree[left=0 of ctrup]{0.8}{s1}{$s_1$}
    \Sometree[right=0 of ctrup]{0.8}{s2}{$s_2$}
    \inetwire(s1.pal)(ctr.right pax)
    \inetwire(s2.pal)(ctr.left pax)

    \Sometree[above=0.2cm of coder]{0.8}{t}{$t$}
    \inetwire(t.pal)(coder.pax)

    \node[right=0.4cm of coder](red){$\Redcut$};
    \node[right=3.9cm of red](plus){$+$};
    \node[above=0.5 of plus](refr){};

    \Sometree[left=3.1 of refr]{0.8}{sl1}{$s_1$}
    \Sometree[left=2.2 of refr]{0.8}{tl}{$t$}
    \inetcell[below=0.2cm of tl.pal](coderl){$\oc$}
    \inetwire(tl.pal)(coderl.pax)
    \draw[inetwire](coderl.pal)|-++(-0.5,-0.2)-|(sl1.pal);

    \Sometree[left=1.1 of refr]{0.8}{sl2}{$s_2$}
    \node[left=0.5 of refr](wl){};
    \inetcell[below=0.2cm of wl](cowl){$\oc$}
    \draw[inetwire](cowl.pal)|-++(-0.5,-0.2)-|(sl2.pal);

    \Sometree[right=0.2 of refr]{0.8}{sr1}{$s_1$}
    \node[right=1.2 of refr](wr){};
    \inetcell[below=0.2cm of wr](cowr){$\oc$}
    \draw[inetwire](cowr.pal)|-++(-0.5,-0.2)-|(sr1.pal);

    \Sometree[right=2 of refr]{0.8}{sr2}{$s_2$}
    \Sometree[right=2.9 of refr]{0.8}{tr}{$t$}
    \inetcell[below=0.2cm of tr.pal](coderr){$\oc$}
    \inetwire(tr.pal)(coderr.pax)
    \draw[inetwire](coderr.pal)|-++(-0.5,-0.2)-|(sr2.pal);
  \end{tikzpicture}
  \\[1em]
  \begin{tikzpicture}
    \inetcell(der){$\wn$}
    \inetcell[right=1.2cm of ctr](coctr){$\oc$}

    \draw[inetwire](coctr.pal)|-++(-1,-0.2)-|(der.pal);

    \node[above=0.7cm of coctr](coctrup){};

    \Sometree[left=0 of coctrup]{0.8}{s1}{$t_1$}
    \Sometree[right=0 of coctrup]{0.8}{s2}{$t_2$}
    \inetwire(s1.pal)(coctr.right pax)
    \inetwire(s2.pal)(coctr.left pax)

    \Sometree[above=0.1cm of der]{0.8}{t}{$s$}
    \inetwire(t.pal)(der.pax)

    \node[right=0.7cm of coctr](red){$\Redcut$};
    \node[right=4.1cm of red](plus){$+$};
    \node[above=0.5 of plus](refr){};

    \Sometree[left=3.1 of refr]{0.8}{tl}{$s$}
    \Sometree[left=2.2 of refr]{0.8}{sl1}{$t_1$}
    \inetcell[below=0.2cm of tl.pal](derl){$\wn$}
    \inetwire(tl.pal)(derl.pax)
    \draw[inetwire](derl.pal)|-++(0.5,-0.2)-|(sl1.pal);

    \node[left=1.4 of refr](wl){};
    \inetcell[below=0.2cm of wl](wl){$\wn$}
    \Sometree[left=0.4 of refr]{0.8}{sl2}{$t_2$}
    \draw[inetwire](wl.pal)|-++(0.5,-0.2)-|(sl2.pal);

    \node[right=0.5 of refr](wr){};
    \Sometree[right=1.2 of refr]{0.8}{sr1}{$t_1$}
    \inetcell[below=0.2cm of wr](wr){$\wn$}
    \draw[inetwire](wr.pal)|-++(0.5,-0.2)-|(sr1.pal);

    \Sometree[right=2.2 of refr]{0.8}{tr}{$s$}
    \Sometree[right=3.1 of refr]{0.8}{sr2}{$t_2$}
    \inetcell[below=0.2cm of tr.pal](derr){$\wn$}
    \inetwire(tr.pal)(derr.pax)
    \draw[inetwire](derr.pal)|-++(0.5,-0.2)-|(sr2.pal);
  \end{tikzpicture}    
    \caption{Contraction/codereliction and dereliction/cocontraction reductions}
    \label{fig:contr-coder-red}
  \end{figure}
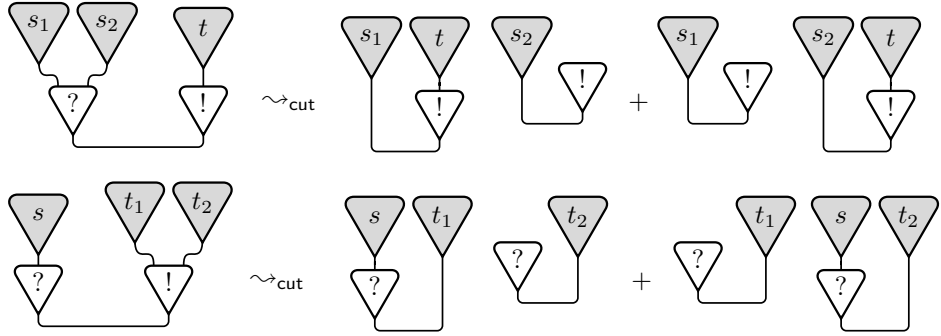

  \begin{figure}[t]
    \centering
  \begin{tikzpicture}
    \inetcell(ctr){$\wn$}
    \inetcell[right=1.7cm of ctr](coctr){$\oc$}

    \draw[inetwire](ctr.pal)|-++(1,-0.2)-|(coctr.pal);

    \node[above=0.7cm of ctr.pax](ctrup){};

    \Sometree[left=0 of ctrup]{0.8}{s1}{$s_1$}
    \Sometree[right=0 of ctrup]{0.8}{s2}{$s_2$}
    \inetwire(s1.pal)(ctr.right pax)
    \inetwire(s2.pal)(ctr.left pax)

    \node[above=0.7cm of coctr.pax](coctrup){};

    \Sometree[left=0 of coctrup]{0.8}{t1}{$t_1$}
    \Sometree[right=0 of coctrup]{0.8}{t2}{$t_2$}
    \inetwire(t1.pal)(coctr.right pax)
    \inetwire(t2.pal)(coctr.left pax)

    \node[right=0.6 of coctr](red){$\Redcut$};

    \node[right=2.3cm of coctrup](resp){};
    \node[below=0.4cm of resp](res){};

    \Sometree[left=0 of res]{0.8}{sr1}{$s_1$}
    \Sometree[right=0 of res]{0.8}{sr2}{$s_2$}

    \node[right=2.6cm of res](mid){};

    \node[right=5.4cm of res](resr){};

    \Sometree[left=0 of resr]{0.8}{tr1}{$t_1$}
    \Sometree[right=0 of resr]{0.8}{tr2}{$t_2$}

    \node[below=0.7cm of res](intcut1){};

    \inetcell[left=0cm of mid](coctr1){$\oc$}
    \inetcell[left=0.8cm of mid](coctr2){$\oc$}
    \inetcell[right=0cm of mid](ctr1){$\wn$}
    \inetcell[right=0.8cm of mid](ctr2){$\wn$}

    \draw[inetwire](sr1.pal)|-++(1,-0.2)-|(coctr2.pal);
    \draw[inetwire](sr2.pal)|-++(1,-0.3)-|(coctr1.pal);
    \draw[inetwire](tr1.pal)|-++(-1,-0.3)-|(ctr1.pal);
    \draw[inetwire](tr2.pal)|-++(-1,-0.2)-|(ctr2.pal);
    \draw[inetwire](coctr2.right pax)|-++(1,0.2)-|(ctr1.right pax);
    \draw[inetwire](coctr2.left pax)|-++(1,0.3)-|(ctr2.right pax);
    \draw[inetwire](coctr1.right pax)|-++(1,0.4)-|(ctr1.left pax);
    \draw[inetwire](coctr1.left pax)|-++(1,0.5)-|(ctr2.left pax);
  \end{tikzpicture}    
    \caption{Contraction/cocontraction reduction}
    \label{fig:contr-cocontr-red}
  \end{figure}
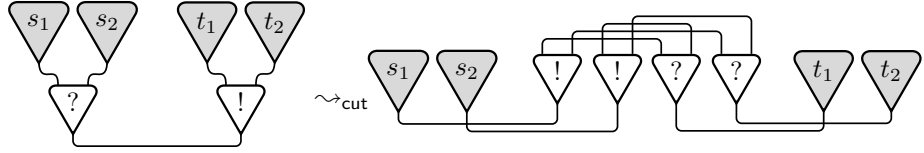

The new basic reduction rules are:
\begin{align*}
  \Cut{\WEAK}{\COWEAK} & \Rel\Redcut\Net{}{}\text{\quad see
    Figure~\ref{fig:weak-coweak-red}}.\\ 
  \Cut{\DER(s)}{\COWEAK} & \Rel\Redcut 0\\
  \Cut{\WEAK}{\CODER(t)} & \Rel\Redcut 0\text{\quad see
    Figure~\ref{fig:der-coweak-red}}.\\
  \Cut{\CONTR(s_1,s_2)}{\COWEAK} &
    \Rel\Redcut \Net{}{\Cut{s_1}{\COWEAK},\Cut{s_2}{\COWEAK}}\\
  \Cut{\WEAK}{\COCONTR(t_1,t_2)} &
    \Rel\Redcut\Net{}{\Cut{\WEAK}{t_1},\Cut{\WEAK}{t_2}}\text{\quad see
    Figure~\ref{fig:contr-coweak-red}}.\\
  \Cut{\DER(s)}{\CODER(t)} & \Rel\Redcut\Net{}{\Cut st}\text{\quad see
    Figure~\ref{fig:der-coder-red}}.\\
  \Cut{\CONTR(s_1,s_2)}{\CODER(t)} &  \Rel\Redcut
    \Net{}{\Cut{s_1}{\CODER(t)},\Cut{s_2}{\COWEAK}}
    + \Net{}{\Cut{s_1}{\COWEAK},\Cut{s_2}{\CODER(t)}}\\
  \Cut{\DER(s)}{\COCONTR(t_1,t_2)} & \Rel\Redcut
    \Net{}{\Cut{\DER(s)}{t_1},\Cut{\WEAK}{t_2}}
    + \Net{}{\Cut{\WEAK}{t_1},\Cut{\DER(s)}{t_2}}\\
    &\hspace{0.5\textwidth}\text{\quad see
    Figure~\ref{fig:contr-coder-red}}.\\
  \Cut{\CONTR(s_1,s_2)}{\COCONTR(t_1,t_2)} &\\
  &\hspace{-7em}\Rel\Redcut\Net{}{
    \Cut{s_1}{\COCONTR(x_{11},x_{12})},
    \Cut{s_2}{\COCONTR(x_{21},x_{22})},
    \Cut{\CONTR(\Covar{x_{11}},\Covar{x_{21}})}{t_1},
    \Cut{\CONTR(\Covar{x_{12}},\Covar{x_{22}})}{t_2}}\\
    &\hspace{0.5\textwidth}\text{\quad see
    Figure~\ref{fig:contr-cocontr-red}}.
\end{align*}
In the last reduction rule, the four variables that we introduce are pairwise
distinct and fresh. Up to $\alpha$-conversion, the choice of these variables is
not relevant.

The contextual rule must be extended, in order to take sums into account.
\begin{center}
  \AxiomC{$c\Rel\Redcut P$}
  \RightLabel{\quad\CONTEXT}
  \UnaryInfC{$\Net{\Vect s}{c,\Vect
      b}\Rel\Redcut\sum_{p=\Net{\Vect t}{\Vect c}}
    P_p\Scalmult\Net{\Vect s,\Vect t}{\Vect{c},\Vect b}$}
  \DisplayProof
\end{center}

\begin{remark}
  In the premise of this rule, $P$ is a linear combination of proof-structures,
  so that for a given proof-structure $p=\Net{\Vect t}{\Vect c}$,
  $P_p\in\Field$ is the coefficient of the proof-structure $p$ in this linear
  combination $P$. The sum which appears in the conclusion ranges over all
  possible proof-structures $p$, but there are only finitely many $p$'s such
  that $P_p\not=0$ so that this sum is actually finite. A particular case of
  this rule is $c\Rel\Redcut 0\,\Implies\,\Net{\Vect s}{c,\Vect b}\Rel\Redcut
  0$.
\end{remark}

\subsection{Promotion}
\label{sec:promotion-syntax}
Let $p=\Net{\Vect s}{\Vect c}$ be a simple proof-structure. The \emph{width} of
$p$ is the number of elements of the sequence $\Vect s$.

By definition, a proof-structure of width $n$ is a finite linear combination of
simple proof-structures of width $n$.

Observe that $0$ is a proof-structure of width $n$ for all $n$. 

Let $P$ be a proof-structure\footnote{To be completely precise, we should also
  provide a typing environment for the free variables of $P$; this can be
  implemented by equipping each variable with a type.} of width $n+1$.  We
introduce a new constructor\footnote{The definitions of the syntax of proof
  trees and of the signature $\Conset$ are mutually recursive when promotion
  is taken into account.} called \emph{promotion box}, of arity $n$:
\[
\PROM nP\in\Conset_n\,.
\] 
The presence of $n$ in the notation is useful only in the case where $P=0$ so
it can most often be omitted. The use of a non necessarily simple proof
structure $P$ in this construction is crucial: promotion is not a linear
construction and is actually the only non linear construction of (differential)
\LL{}.

\begin{figure}[t]
  \centering
  \begin{tikzpicture}
    \Prombox{2}{1.4}{prom}{pnode}{}
    \Somenet[above=0.2 of pnode]{1.8}{0.8}{pcont}{$P$}
    \node[left=0.6 of pcont.north](a1){};
    \node[right=0.6 of pcont.north](an){};
    \node[below=-0.05 of a1.center]{$\scriptstyle 1$};
    \node[below=-0.05 of an.center]{$\scriptstyle{n}$};
    \node[above=-0.05 of pcont.south]{$\Mainport$};
    \draw[inetwire](pcont.south)--(pnode.pax);

    \Sometree[above=0.4 of a1.center]{0.9}{t1}{$t_1$}
    \Sometree[above=0.4 of an.center]{0.9}{tn}{$t_n$}
    \draw[inetwire](t1.pal)--(a1.center);
    \draw[inetwire](tn.pal)--(an.center);
    \node[above=-0.07 of pcont.north]{$\cdots$};
  \end{tikzpicture}
  \caption{Graphical representation of a tree whose outermost constructor is a
    promotion box}
  \label{fig:prom-example}
\end{figure}
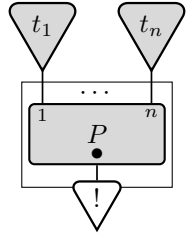

So if $\List t1n$ are trees, $\PROM nP(\List t1n)$ is a tree. Pictorially, this
tree will typically be represented as in Figure~\ref{fig:prom-example}.  A
simple net $p$ appearing in $P$ is typically of the form $\Net{\Vect s}{\Vect
  c}$ and its width is $n+1$, so that $\Vect s=(s_1,\dots,s_n,s)$.  The
indices $1,\dots,n$ and $\Mainport$ which appear on the gray rectangle
representing $P$ stand for the roots of these trees $\List s1n$ and $s$.

\paragraph{Typing rule.}
The typing rule for this construction is
\begin{center}
  \AxiomC{$\Typing{\Phi}{P}
    {\Int{\Orth{A_1}},\dots,\Int{\Orth{A_n},B}}$}
  \AxiomC{$\Typing{\Phi}{t_i}
    {\Excl{A_i}}\quad (i=1,\dots,n)$}
  \BinaryInfC{$\Typing{\Phi}
    {\PROM n{P}(\List t1n)}
    {\Excl B}$}
  \DisplayProof
\end{center}

\paragraph{Logical rule.}
The logical rule associated with this construction is the following.
\begin{center}
  \AxiomC{$\Logic{\Phi}{P}
    {\Int{\Orth{A_1}},\dots,\Int{\Orth{A_n},B}}$}
  \AxiomC{$\Logic{\Phi}{\Net{\Vect{t_i},t_i}{\Vect{c_i}}}
    {\Gamma_i,\Excl{A_i}}\quad (i=1,\dots,n)$}
  \BinaryInfC{$\Logic{\Phi}
    {\Net{\Vect{t_1},\dots,\Vect{t_n},\PROM n{P}(\List t1n)}
    {\Vect{c_1},\dots,\Vect{c_n}}}{\Gamma_1,\dots,\Gamma_n,\Excl B}$}
  \DisplayProof
\end{center}
\begin{remark}
  This promotion rule is of course highly debatable. We choose this
  presentation because it is compatible with our tree-based presentation of
  proof-structures.
\end{remark}

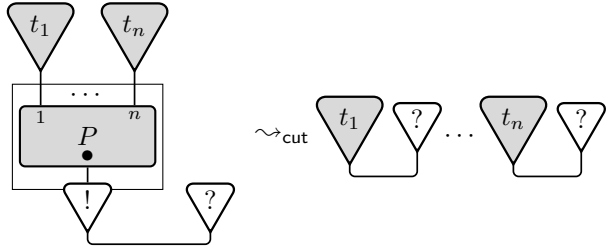
\begin{figure}[t]
  \centering
\begin{tikzpicture}
  \Prombox{2}{1.4}{prom}{pnode}{}
  \Somenet[above=0.2 of pnode]{1.8}{0.8}{pcont}{$P$}
  \draw[inetwire](pcont.south)--(pnode.pax);
  \node[above=-0.05 of pcont.south]{$\Mainport$};
  \node[above=-0.07 of pcont.north]{$\cdots$};
  \node[left=0.5 of pcont.north](anchl){};
  \node[right=0.5 of pcont.north](anchr){};
  \Sometree[above=0.3 of anchl]{0.9}{s1}{$t_1$}
  \Sometree[above=0.3 of anchr]{0.9}{sn}{$t_n$}
  \node[below=-0.05 of anchl.center]{$\scriptstyle 1$};
  \node[below=-0.05 of anchr.center]{$\scriptstyle n$};
  \draw[inetwire](s1.pal)--(anchl.center);
  \draw[inetwire](sn.pal)--(anchr.center);
  \inetcell[right=1.1 of pnode](w){$\wn$}
  \Inetcut{pnode.pal}{w.pal}

  \node[right=1.2 of pcont]{$\Redcut$};

  \node[right=3.7 of pcont](P){$\cdots$};
  \node[above=-0.1 of P](Pu){};
  \Sometree[left=1 of Pu]{0.9}{sr1}{$t_1$}
  \inetcell[left=0.2 of Pu](w1){$\wn$}
  \Inetcut{sr1.pal}{w1.pal}
  \Sometree[right=0.2 of Pu]{0.9}{srn}{$t_n$}
  \inetcell[right=1.2 of Pu](wn){$\wn$}
  \Inetcut{srn.pal}{wn.pal}
\end{tikzpicture}  
  \caption{Promotion/weakening reduction}
  \label{fig:prom-weak-red}
\end{figure}

\begin{figure}[t]
  \centering
\begin{tikzpicture}
  \Prombox{2}{1.4}{prom}{pnode}{}
  \Somenet[above=0.2 of pnode]{1.8}{0.8}{pcont}{$P$}
  \draw[inetwire](pcont.south)--(pnode.pax);
  \node[above=-0.05 of pcont.south]{$\Mainport$};
  \node[above=-0.07 of pcont.north]{$\cdots$};
  \node[left=0.5 of pcont.north](anchl){};
  \node[right=0.5 of pcont.north](anchr){};
  \Sometree[above=0.3 of anchl]{0.9}{s1}{$t_1$}
  \Sometree[above=0.3 of anchr]{0.9}{sn}{$t_n$}
  \node[below=-0.05 of anchl.center]{$\scriptstyle 1$};
  \node[below=-0.05 of anchr.center]{$\scriptstyle n$};
  \draw[inetwire](s1.pal)--(anchl.center);
  \draw[inetwire](sn.pal)--(anchr.center);
  \inetcell[right=1.5 of pnode](ctr){$\wn$}
  \node[above=0.7 of ctr.pax](reft){};
  \Sometree[left=0 of reft]{0.8}{t1}{$s_1$}
  \Sometree[right=0 of reft]{0.8}{t2}{$s_2$}
  \draw[inetwire](t1.pal)|-++(0.1,-0.2)-|(ctr.right pax);
  \draw[inetwire](t2.pal)|-++(-0.1,-0.2)-|(ctr.left pax);
  \Inetcut{pnode.pal}{ctr.pal}

  \node[below=2 of pcont](Pd){};
  \node[right=3 of Pd](P){$\cdots$};

  \node[above=-0.1 of P](Pu){};
  \Sometree[left=1 of Pu]{0.9}{sr1}{$t_1$}
  \inetcell[left=0.2 of Pu](ctr1){$\wn$}
  \Inetcut{sr1.pal}{ctr1.pal}
  \Sometree[right=0.2 of Pu]{0.9}{srn}{$t_n$}
  \inetcell[right=1.2 of Pu](ctrn){$\wn$}
  \Inetcut{srn.pal}{ctrn.pal}

  \Prombox[left=1.7 of P]{2}{1.4}{prom1}{pnode1}{}
  \Somenet[above=0.2 of pnode1]{1.8}{0.8}{pcont1}{$P$}
  \draw[inetwire](pcont1.south)--(pnode1.pax);
  \node[above=-0.05 of pcont1.south]{$\Mainport$};
  \node[above=-0.09 of pcont1.north]{$\cdots$};
  \node[left=0.5 of pcont1.north](anchl1){};
  \node[right=0.5 of pcont1.north](anchr1){};
  \node[below=-0.05 of anchl1.center]{$\scriptstyle 1$};
  \node[below=-0.05 of anchr1.center]{$\scriptstyle n$};

  \Sometree[left=0.15 of prom1]{0.8}{t1}{$s_1$}
  \Inetcutr{pnode1.pal}{t1.pal}

  \Prombox[right=1.7 of P]{2}{1.4}{prom2}{pnode2}{}
  \Somenet[above=0.2 of pnode2]{1.8}{0.8}{pcont2}{$P$}
  \draw[inetwire](pcont2.south)--(pnode2.pax);
  \node[above=-0.05 of pcont2.south]{$\Mainport$};
  \node[above=-0.09 of pcont2.north]{$\cdots$};
  \node[left=0.5 of pcont2.north](anchl2){};
  \node[right=0.5 of pcont2.north](anchr2){};
  \node[below=-0.05 of anchl2.center]{$\scriptstyle 1$};
  \node[below=-0.05 of anchr2.center]{$\scriptstyle n$};

  \Sometree[right=0.15 of prom2]{0.8}{t2}{$s_2$}
  \Inetcut{pnode2.pal}{t2.pal}

  \draw[inetwire](anchl1.center)|-++(0.5,0.4)-|(ctr1.right pax);
  \draw[inetwire](anchr1.center)|-++(0.5,0.5)-|(ctrn.right pax);
  \draw[inetwire](anchl2.center)|-++(-0.5,0.4)-|(ctr1.left pax);
  \draw[inetwire](anchr2.center)|-++(-0.5,0.5)-|(ctrn.left pax);

  \node[right=2 of pcont]{$\Redcut$};
\end{tikzpicture}  
  \caption{Promotion/contraction reduction}
  \label{fig:prom-contr-red}
\end{figure}
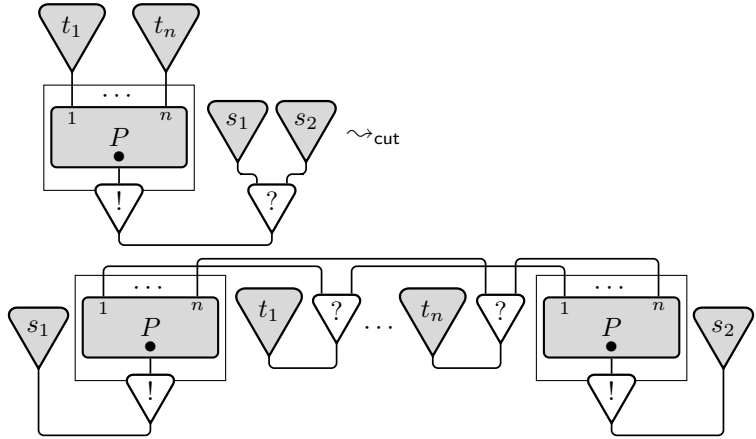

\paragraph{Cut elimination rules.}
The basic reductions associated with promotion are as follows.
\begin{align*}
  \Cut{\PROM n{P}(\List t1n)}{\WEAK}
  & \Rel\Redcut \Net{}{\Cut{t_1}{\WEAK},\dots,\Cut{t_n}{\WEAK}}
  \text{\quad see Figure~\ref{fig:prom-weak-red}.}\\
  \Cut{\PROM n{P}(\List t1n)}{\DER(s)}
  & \Rel\Redcut\\
  & \sum_{p=\Net{\Vect s,s'}{\Vect c}}
     P_{p}\Scalmult\Net{}{\Vect c,
       \Cut{s_1}{t_1},\dots,\Cut{s_n}{t_n},\Cut{s'}{s}}\\
  &\hspace{0.4\textwidth}\text{see Figure~\ref{fig:prom-der-red}.}\\
  \Cut{\PROM n{P}(\List t1n)}{\CONTR(s_1,s_2)} & \Rel\Redcut
  \Net{}{\Cut{\PROM nP(\List x1n)}{s_1},\Cut{\PROM nP(\List y1n)}{s_2},\\
  & \quad\quad\quad\Cut{t_1}{\CONTR(\Covar{x_1},\Covar{y_1})},\dots,
  \Cut{t_n}{\CONTR(\Covar{x_n},\Covar{y_n})}}\\
  &\hspace{0.4\textwidth}\text{see Figure~\ref{fig:prom-contr-red}.}
\end{align*}
In the second reduction rule, one has to avoid clashes of variables.

In the last reduction rules, the variables $\List x1n$ and $\List y1n$ that we
introduce together with their covariables are assumed to be pairwise distinct
and fresh. Up to $\alpha$-conversion, the choice of these variables is not
relevant.

\begin{figure}[t]
  \centering
\begin{tikzpicture}
  \Prombox{2}{1.4}{prom}{pnode}{}
  \Somenet[above=0.2 of pnode]{1.8}{0.8}{pcont}{$P$}
  \draw[inetwire](pcont.south)--(pnode.pax);
  \node[above=-0.05 of pcont.south]{$\Mainport$};
  \node[above=-0.07 of pcont.north]{$\cdots$};
  \node[left=0.5 of pcont.north](anchl){};
  \node[right=0.5 of pcont.north](anchr){};
  \Sometree[above=0.3 of anchl]{0.9}{s1}{$t_1$}
  \Sometree[above=0.3 of anchr]{0.9}{sn}{$t_n$}
  \node[below=-0.05 of anchl.center]{$\scriptstyle 1$};
  \node[below=-0.05 of anchr.center]{$\scriptstyle n$};
  \draw[inetwire](s1.pal)--(anchl.center);
  \draw[inetwire](sn.pal)--(anchr.center);
  \inetcell[right=1.2 of pnode](der){$\wn$}
  \Sometree[above=0.2 of der]{0.8}{t}{$s$}
  \draw[inetwire](t.pal)--(der.pax);
  \Inetcut{pnode.pal}{der.pal}
  \Somenet[right=4.7 of pcont]{1.8}{0.8}{P}{$P$}
  \node[left=1.3 of P](refp){};
  \node[above=-0.2 of refp](ref){};
  \node[below=-0.1 of ref](cd){$\cdots$};

  \node[left=0.7 of cd]{$\Redcut$};

  \Sometree[left=0.3 of ref]{0.9}{sr1}{$t_1$}
  \Sometree[right=0.3 of ref]{0.9}{srn}{$t_n$}
  \Sometree[right=3.4 of ref]{0.8}{tr}{$s$}
  \node[right=0.5of P.south](a0){};
  \node[left=0.5of P.south](an){};
  \node[left=-0.4 of P.south](a1){};
  \node[left=0.05 of P.south](ad){};
  \node[above=-0.05 of a1.center]{$\scriptstyle 1$};
  \node[above=-0.05 of a0.center]{$\Mainport$};
  \node[above=-0.05 of an.center]{$\scriptstyle n$};
  \node[below=-0.05 of ad.center]{$\cdots$};
  \Inetcutr{tr.pal}{a0.center}
  \Inetcut[-0.3]{sr1.pal}{a1.center}
  \Inetcut{srn.pal}{an.center}
\end{tikzpicture}  
  \caption{Promotion/dereliction reduction}
  \label{fig:prom-der-red}
\end{figure}
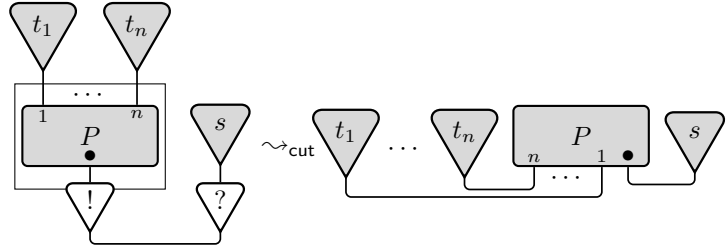

\paragraph{Commutative reductions.}
There are also auxiliary reduction rules sometimes called \emph{commutative
  reductions} which do not deal with cuts ---~at least in the formalization of
nets we present here.

The format of these reductions is 
\[t\Rel\Redcom P\] where $t$ is a simple tree and $P$ is a (not necessarily
simple) proof-structure whose width is exactly $1$.

\begin{figure}[t]
  \centering
\begin{tikzpicture}
  \Prombox{2.2}{1.4}{prom1}{pnode1}{}
  \Somenet[above=0.2 of pnode1]{2}{0.8}{pcont1}{$P$}  
  \Prombox[above=0.7 of prom1]{2}{1.4}{prom2}{pnode2}{}
  \Somenet[above=0.2 of pnode2]{1.8}{0.8}{pcont2}{$Q$}  
  \node[above=0.4 of pnode2](reft){};

  \Sometree[left=1.1 of reft]{0.9}{t1}{$v_1$}
  \Sometree[right=1.1 of reft]{0.9}{tn}{$v_{n}$}

  \node[left=0.6 of pcont1.north](a1){};
  \node[right=0.6 of pcont1.north](an){};
  \node[left=0.2 of pcont1.north](adl){};
  \node[right=0.24 of pcont1.north](adr){};
  \node[above=-0.07 of adl.center]{$\cdots$};
  \node[above=-0.07 of adr.center]{$\cdots$};

  \draw[inetwire](t1.pal)|-++(0.3,-0.3)-|(a1.center);
  \draw[inetwire](pnode2.pal)--(pcont1.north);  
  \draw[inetwire](tn.pal)|-++(-0.3,-0.3)-|(an.center);

  \node[below=-0.05 of a1.center]{$\scriptstyle 1$};
  \node[below=-0.05 of an.center]{$\scriptstyle{n}$};
  \node[below=-0.05 of pcont1.north]{$\scriptstyle{i}$};
  \node[above=-0.05 of pcont1.south]{$\Mainport$};
  \draw[inetwire](pcont1.south)--(pnode1.pax);

  \node[left=0.5 of pcont2.north](b1){};
  \node[right=0.5 of pcont2.north](bk){};

  \node[below=-0.05 of b1.center]{$\scriptstyle 1$};
  \node[below=-0.05 of bk.center]{$\scriptstyle{k}$};
  \node[above=-0.05 of pcont2.south]{$\Mainport$};
  \draw[inetwire](pcont2.south)--(pnode2.pax);

  \Sometree[above=0.3 of b1]{0.9}{s1}{$u_1$}
  \Sometree[above=0.3 of bk]{0.9}{sk}{$u_k$}
  \draw[inetwire](s1.pal)--(b1.center);
  \draw[inetwire](sk.pal)--(bk.center);
  \node[above=-0.07 of pcont2.north]{$\cdots$};

  \node[above=0.3 of prom1](iref){};
  \node[right=1.8 of iref]{$\Redcom$};

  \Prombox[right=3.7 of iref]{2.6}{3.7}{r-prom1}{r-pnode1}{}
  \Somenet[above=0.2 of r-pnode1]{2}{0.8}{r-pcont1}{$P$}  
  \draw[inetwire](r-pcont1.south)--(r-pnode1.pax);

  \node[left=0.6 of r-pcont1.north](r-a1){};
  \node[right=0.6 of r-pcont1.north](r-an){};
  \node[left=0.2 of r-pcont1.north](r-adl){};
  \node[right=0.24 of r-pcont1.north](r-adr){};
  \node[right=0.6 of r-pcont1.south](r-ai){};
  \node[above=-0.07 of r-pcont1.north]{$\cdots$};
  \node[below=-0.05 of r-a1.center]{$\scriptstyle 1$};
  \node[below=-0.05 of r-an.center]{$\scriptstyle{n}$};
  \node[above=-0.05 of r-ai.center]{$\scriptstyle{i}$};
  \node[above=-0.05 of r-pcont1.south]{$\Mainport$};

  \Prombox[above=1 of r-pcont1]{2}{1.4}{r-prom2}{r-pnode2}{}
  \Somenet[above=0.2 of r-pnode2.pax]{1.8}{0.8}{r-pcont2}{$Q$}  

  \node[left=0.5 of r-pcont2.north](r-b1){};
  \node[right=0.5 of r-pcont2.north](r-bk){};

  \node[below=-0.05 of r-b1.center]{$\scriptstyle 1$};
  \node[below=-0.05 of r-bk.center]{$\scriptstyle{k}$};
  \node[above=-0.05 of r-pcont2.south]{$\Mainport$};
  \draw[inetwire](r-pcont2.south)--(r-pnode2.pax);
  \draw[inetwire](r-pnode2.pal)|-++(0.3,-0.2)-|++(0.85,-0.4)|-++(-0.2,-0.85)-|(r-ai.center);
  
  \Sometree[above=0.5 of r-b1]{0.9}{r-s1}{$u_1$}
  \Sometree[above=0.5 of r-bk]{0.9}{r-sk}{$u_k$}
  \draw[inetwire](r-s1.pal)--(r-b1.center);
  \draw[inetwire](r-sk.pal)--(r-bk.center);
  \node[above=-0.07 of r-pcont2.north]{$\cdots$};

  \node[above=1.2 of r-pcont1.north](r-ref){};
  \Sometree[left=1.4 of r-ref]{0.9}{r-t1}{$v_1$}
  \Sometree[right=1.4 of r-ref]{0.9}{r-tn}{$v_n$}
  \draw[inetwire](r-t1.pal)|-++(0.5,-0.2)-|(r-a1.center);
  \draw[inetwire](r-tn.pal)|-++(-0.5,-0.2)-|(r-an.center);
\end{tikzpicture}  
  \caption{Promotion/promotion commutative reduction}
  \label{fig:prom-prom-com-red}
\end{figure}
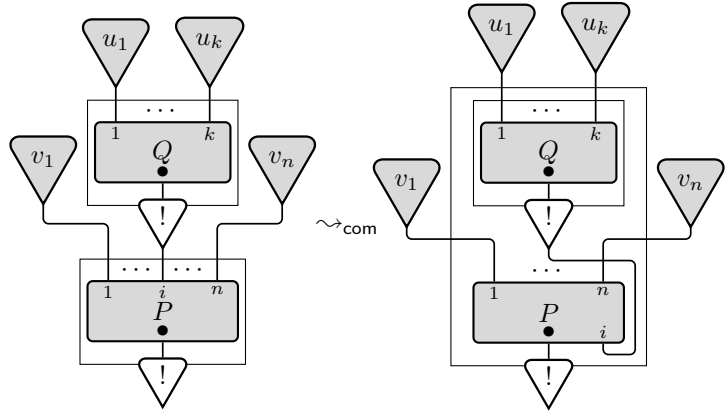

The first of these reductions is illustrated in
Figure~\ref{fig:prom-prom-com-red} and deals with the interaction between two
promotions.
\begin{multline}
  \label{eq:box-box-reduction}
  \PROM{n+1}P(\List t1{i-1},\PROM kQ(\List ti{k+i-1}),\List t{k+i}{k+n})\\
  \Rel\Redcom\Net{\PROM{k+n}R(\List t1{k+n})}{}
\end{multline}
where
\begin{align*}
  &R=\sum_{p=\Net{\List s1{n+1},s}{\Vect c}}
  P_p\Scalmult(\Vect c,\Cut{s_i}{\PROM kQ(\List x1k)}\,;\\
  &\hspace{12em}  \List s1{i-1},\Covar{x_1},
               \dots,\Covar{x_k},\List s{i+1}{n+1},s)\,.
\end{align*}

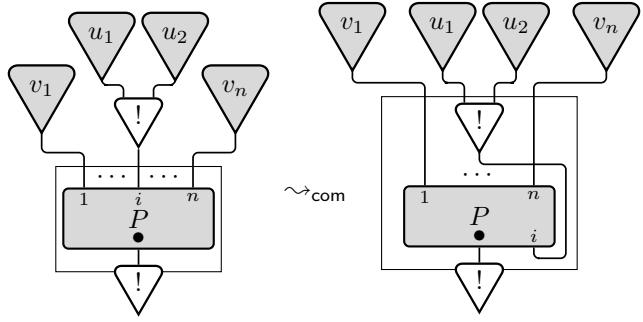
\begin{figure}[t]
  \centering
\begin{tikzpicture}
  \Prombox{2.2}{1.4}{prom1}{pnode1}{}
  \Somenet[above=0.2 of pnode1]{2}{0.8}{pcont1}{$P$}  
  \node[above=1.2 of pcont1.north](reft){};

  \Sometree[left=0.8 of reft]{0.9}{t1}{$v_1$}
  \Sometree[right=0.8 of reft]{0.9}{tn}{$v_n$}

  \node[left=0.6 of pcont1.north](a1){};
  \node[right=0.6 of pcont1.north](an){};
  \node[left=0.2 of pcont1.north](adl){};
  \node[right=0.24 of pcont1.north](adr){};
  \node[above=-0.07 of adl.center]{$\cdots$};
  \node[above=-0.07 of adr.center]{$\cdots$};

  \inetcell[below=0 of reft](coctr){$\oc$}
  \node[above=0.7 of coctr.pax](refs){};

  \Sometree[left=0 of refs]{0.9}{s1}{$u_1$}
  \Sometree[right=0 of refs]{0.9}{s2}{$u_2$}
  \draw[inetwire](s1.pal)|-++(0.1,-0.1)-|(coctr.right pax);
  \draw[inetwire](s2.pal)|-++(-0.1,-0.1)-|(coctr.left pax);

  \draw[inetwire](t1.pal)|-++(0.3,-0.3)-|(a1.center);
  \draw[inetwire](tn.pal)|-++(-0.3,-0.3)-|(an.center);
  \draw[inetwire](coctr)--(pcont1.north);

  \node[below=-0.05 of a1.center]{$\scriptstyle 1$};
  \node[below=-0.05 of an.center]{$\scriptstyle{n}$};
  \node[below=-0.05 of pcont1.north]{$\scriptstyle{i}$};
  \node[above=-0.05 of pcont1.south]{$\Mainport$};
  \draw[inetwire](pcont1.south)--(pnode1.pax);

  \node[above=-0.5 of prom1](iref){};
  \node[above=-0.1 of iref](r-ref){};
  \node[right=1.7 of iref]{$\Redcom$};

  \Prombox[right=3.1 of r-ref]{2.6}{2.3}{r-prom1}{r-pnode1}{}
  \Somenet[above=0.2 of r-pnode1]{2}{0.8}{r-pcont1}{$P$}  
  \node[above=2 of r-pcont1.north](r-reft){};

  \Sometree[left=1.2 of r-reft]{0.9}{r-t1}{$v_1$}
  \Sometree[right=1.2 of r-reft]{0.9}{r-tn}{$v_n$}

  \node[left=0.6 of r-pcont1.north](r-a1){};
  \node[right=0.6 of r-pcont1.north](r-an){};
  \node[right=0.6 of r-pcont1.south](r-ai){};
  \node[left=0.2 of r-pcont1.north](r-adl){};
  \node[right=0.24 of r-pcont1.north](r-adr){};
  \node[above=-0.07 of r-pcont1.north]{$\cdots$};

  \inetcell[below=0.9 of r-reft](r-coctr){$\oc$}

  \Sometree[left=0 of r-reft]{0.9}{r-s1}{$u_1$}
  \Sometree[right=0 of r-reft]{0.9}{r-s2}{$u_2$}
  \draw[inetwire](r-s1.pal)|-++(0.1,-0.2)-|(r-coctr.right pax);
  \draw[inetwire](r-s2.pal)|-++(-0.1,-0.2)-|(r-coctr.left pax);

  \draw[inetwire](r-t1.pal)|-++(0.3,-0.2)-|(r-a1.center);
  \draw[inetwire](r-tn.pal)|-++(-0.3,-0.2)-|(r-an.center);
  \draw[inetwire](r-coctr.pal)|-++(0.3,-0.2)-|++(0.85,-0.4)|-++(-0.3,-0.85)-|(r-ai.center);

  \node[below=-0.05 of r-a1.center]{$\scriptstyle 1$};
  \node[below=-0.05 of r-an.center]{$\scriptstyle{n}$};
  \node[above=-0.05 of r-ai.center]{$\scriptstyle{i}$};
  \node[above=-0.05 of r-pcont1.south]{$\Mainport$};
  \draw[inetwire](r-pcont1.south)--(r-pnode1.pax);
\end{tikzpicture}  
  \caption{Promotion/cocontraction commutative reduction rules}
  \label{fig:prom-cocontr-com-red}
\end{figure}

\begin{remark}
  In Figure~\ref{fig:prom-prom-com-red} and~\ref{fig:prom-cocontr-com-red}, for
  graphical reasons, we don't follow exactly the notations used in the
  text. For instance in Figure~\ref{fig:prom-prom-com-red}, the correspondence
  with the notations of~\Eqref{eq:box-box-reduction} is given by
  $v_1=t_1$,\dots,$v_{i-1}=t_{i-1}$, $u_1=t_i$,\dots, $u_k=t_{k+i-1}$,
  $v_i=t_{k+i}$,\dots,$v_n=t_{k+n}$.
\end{remark}

\begin{remark}
  Figure~\ref{fig:prom-prom-com-red} is actually slightly incorrect as the
  connections between the ``auxiliary ports'' of the cocontraction rule within
  the promotion box of the right hand proof-structure and the main ports of the
  trees $u_1,\dots,u_k$ are represented as vertical lines whereas they involve
  axioms (corresponding to the pairs $(x_i,\Covar{x_i})$ for $i=1,\dots,k$ in
  the formula above). The same kind of slight incorrectness occurs in
  figure~\ref{fig:prom-cocontr-com-red}.
\end{remark}

The three last commutative reductions deal with the interaction between a
promotion and the costructural rules.

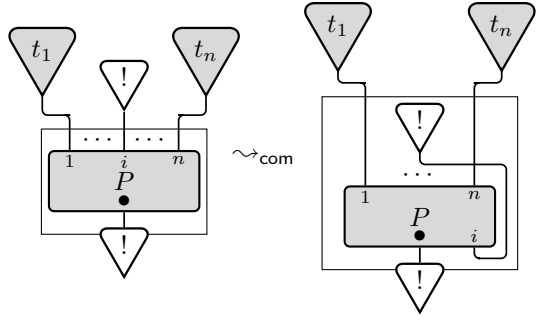
\begin{figure}[t]
  \centering
\begin{tikzpicture}
  \Prombox{2.2}{1.4}{prom1}{pnode1}{}
  \Somenet[above=0.2 of pnode1]{2}{0.8}{pcont1}{$P$}  
  \node[above=1.2 of pcont1.north](reft){};

  \Sometree[left=0.6 of reft]{0.9}{t1}{$t_1$}
  \Sometree[right=0.6 of reft]{0.9}{tn}{$t_n$}

  \node[left=0.6 of pcont1.north](a1){};
  \node[right=0.6 of pcont1.north](an){};
  \node[left=0.2 of pcont1.north](adl){};
  \node[right=0.24 of pcont1.north](adr){};
  \node[above=-0.07 of adl.center]{$\cdots$};
  \node[above=-0.07 of adr.center]{$\cdots$};

  \inetcell[below=0 of reft](cow){$\oc$}

  \draw[inetwire](t1.pal)|-++(0.3,-0.3)-|(a1.center);
  \draw[inetwire](tn.pal)|-++(-0.3,-0.3)-|(an.center);
  \draw[inetwire](cow)--(pcont1.north);

  \node[below=-0.05 of a1.center]{$\scriptstyle 1$};
  \node[below=-0.05 of an.center]{$\scriptstyle{n}$};
  \node[below=-0.05 of pcont1.north]{$\scriptstyle{i}$};
  \node[above=-0.05 of pcont1.south]{$\Mainport$};
  \draw[inetwire](pcont1.south)--(pnode1.pax);

  \node[above=-0.5 of prom1](iref){};
  \node[above=-0.6 of iref](r-ref){};
  \node[right=1.2 of iref]{$\Redcom$};

  \Prombox[right=2.5 of r-ref]{2.6}{2.3}{r-prom1}{r-pnode1}{}
  \Somenet[above=0.2 of r-pnode1]{2}{0.8}{r-pcont1}{$P$}  
  \node[above=2 of r-pcont1.north](r-reft){};

  \Sometree[left=0.6 of r-reft]{0.9}{r-t1}{$t_1$}
  \Sometree[right=0.6 of r-reft]{0.9}{r-tn}{$t_n$}

  \node[left=0.6 of r-pcont1.north](r-a1){};
  \node[right=0.6 of r-pcont1.north](r-an){};
  \node[right=0.6 of r-pcont1.south](r-ai){};
  \node[left=0.2 of r-pcont1.north](r-adl){};
  \node[right=0.24 of r-pcont1.north](r-adr){};
  \node[above=-0.07 of r-pcont1.north]{$\cdots$};

  \inetcell[below=0.9 of r-reft](r-cow){$\oc$}

  \draw[inetwire](r-t1.pal)|-++(0.3,-0.2)-|(r-a1.center);
  \draw[inetwire](r-tn.pal)|-++(-0.3,-0.2)-|(r-an.center);
  \draw[inetwire](r-cow.pal)|-++(0.3,-0.2)-|++(0.85,-0.4)|-++(-0.3,-0.85)-|(r-ai.center);

  \node[below=-0.05 of r-a1.center]{$\scriptstyle 1$};
  \node[below=-0.05 of r-an.center]{$\scriptstyle{n}$};
  \node[above=-0.05 of r-ai.center]{$\scriptstyle{i}$};
  \node[above=-0.05 of r-pcont1.south]{$\Mainport$};
  \draw[inetwire](r-pcont1.south)--(r-pnode1.pax);
\end{tikzpicture}  
  \caption{Promotion/coweakening commutative reduction}
  \label{fig:prom-coweak-red}
\end{figure}

Interaction between a promotion and a coweakening, see
Figure~\ref{fig:prom-coweak-red}:
\begin{equation*}
  \PROM{n+1}P(\List t1{i-1},\COWEAK,\List t{i}{n})\\
  \Rel\Redcom\Net{\PROM{n}R(\List t1n)}{}
\end{equation*}
where
\begin{equation*}
  R=\sum_{p=\Net{\Vect s,s}{\Vect c}}P_p
  \Scalmult\Net{\List s1{i-1},\List s{i+1}{n+1},s}
               {\Vect c,\Cut{s_i}{\COWEAK}}\,.
\end{equation*}

Interaction between a promotion and a cocontraction, see
Figure~\ref{fig:prom-cocontr-com-red}:
\begin{equation}
  \PROM{n+1}P(\List t1{i-1},\COCONTR(t_i,t_{i+1}),\List t{i+2}{n+2})\\
  \Rel\Redcom\Net{\PROM{n+2}R(\List t1{n+2})}{}
\end{equation}
where
\begin{equation*}
  R=\sum_{p=\Net{\Vect s,s}{\Vect c}}P_p\Net{\List s1{i-1},\Covar x,\Covar
    y,\List sin,s}{\Vect c,\Cut{s_i}{\COCONTR(x,y)}}\,.
\end{equation*}

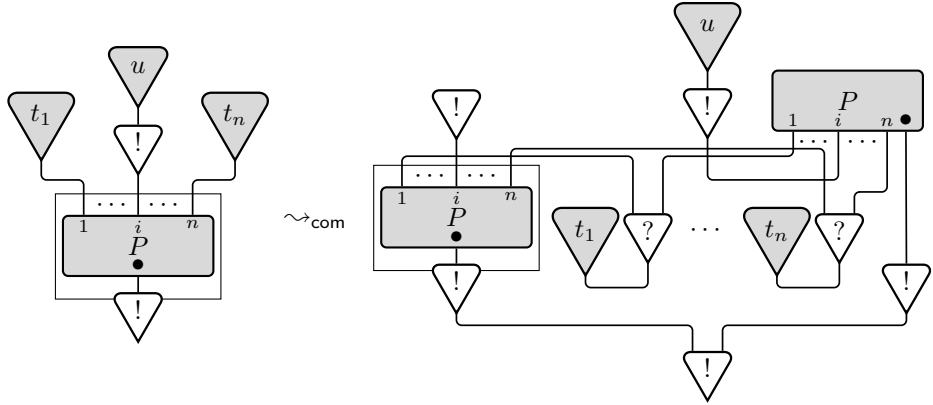
\begin{figure}[t]
  \centering
\begin{tikzpicture}
  \Prombox{2.2}{1.4}{prom1}{pnode1}{}
  \Somenet[above=0.2 of pnode1]{2}{0.8}{pcont1}{$P$}  
  \node[above=1.2 of pcont1.north](reft){};

  \Sometree[left=0.8 of reft]{0.9}{t1}{$t_1$}
  \Sometree[right=0.8 of reft]{0.9}{tn}{$t_n$}

  \node[left=0.6 of pcont1.north](a1){};
  \node[right=0.6 of pcont1.north](an){};
  \node[left=0.2 of pcont1.north](adl){};
  \node[right=0.24 of pcont1.north](adr){};
  \node[above=-0.07 of adl.center]{$\cdots$};
  \node[above=-0.07 of adr.center]{$\cdots$};

  \inetcell[below=0 of reft](coder){$\oc$}

  \Sometree[above=0.2 of coder]{0.8}{s}{$u$}
  \draw[inetwire](s.pal)--(coder.pax);

  \draw[inetwire](t1.pal)|-++(0.3,-0.3)-|(a1.center);
  \draw[inetwire](tn.pal)|-++(-0.3,-0.3)-|(an.center);
  \draw[inetwire](coder.pal)--(pcont1.north);

  \node[below=-0.05 of a1.center]{$\scriptstyle 1$};
  \node[below=-0.05 of an.center]{$\scriptstyle{n}$};
  \node[below=-0.05 of pcont1.north]{$\scriptstyle{i}$};
  \node[above=-0.05 of pcont1.south]{$\Mainport$};
  \draw[inetwire](pcont1.south)--(pnode1.pax);

  \node[above=-0.5 of prom1](iref){};
  \node[above=-0.2 of iref](r-ref){};
  \node[right=1.7 of iref]{$\Redcom$};

  \Prombox[right=3 of r-ref]{2.2}{1.4}{r-prom1}{r-pnode1}{}
  \Somenet[above=0.2 of r-pnode1]{2}{0.8}{r-pcont1}{$P$}  

  \node[left=0.6 of r-pcont1.north](r-a1){};
  \node[right=0.6 of r-pcont1.north](r-an){};
  \node[left=0.2 of r-pcont1.north](r-adl){};
  \node[right=0.24 of r-pcont1.north](r-adr){};
  \node[above=-0.07 of r-adl.center]{$\cdots$};
  \node[above=-0.07 of r-adr.center]{$\cdots$};

  \node[below=-0.05 of r-a1.center]{$\scriptstyle 1$};
  \node[below=-0.05 of r-an.center]{$\scriptstyle{n}$};
  \node[below=-0.05 of r-pcont1.north]{$\scriptstyle{i}$};
  \node[above=-0.05 of r-pcont1.south]{$\Mainport$};
  \draw[inetwire](r-pcont1.south)--(r-pnode1.pax);

  \inetcell[above=0.6 of r-pcont1.north](r-cow){$\oc$}

  \node[right=2.2 of r-pcont1](r-preft){};
  \node[above=-0.5 of r-preft](r-reft){$\cdots$};
  \draw[inetwire](r-cow.pal)--(r-pcont1.north);

  \Sometree[left=0.9 of r-reft]{0.9}{r-t1}{$t_1$}
  \inetcell[right=0.2 of r-t1](r-ctr1){$\wn$}
  \Sometree[right=0.2 of r-reft]{0.9}{r-tn}{$t_n$}
  \inetcell[right=0.2 of r-tn](r-ctrn){$\wn$}
  \Inetcut{r-t1.pal}{r-ctr1.pal}
  \Inetcut{r-tn.pal}{r-ctrn.pal}

  \draw[inetwire](r-a1.center)|-++(0.5,0.4)-|(r-ctr1.right pax);
  \draw[inetwire](r-an.center)|-++(0.5,0.5)-|(r-ctrn.right pax);

  \node[above=1.4 of r-reft](r-pcp2){};
  \Somenet[right=0.725 of r-pcp2]{2}{0.8}{r-pcont2}{$P$}  

  \node[left=0.6 of r-pcont2.south](rr-a1){};
  \node[left=0 of r-pcont2.south](rr-ai){};
  \node[right=0.4 of r-pcont2.south](rr-an){};
  \node[right=0.65 of r-pcont2.south](rr-a0){};
  \node[left=0.3 of r-pcont2.south](rr-adl){};
  \node[right=0.1 of r-pcont2.south](rr-adr){};
  \node[below=-0.07 of rr-adl.center]{$\cdots$};
  \node[below=-0.07 of rr-adr.center]{$\cdots$};

  \node[above=-0.05 of rr-a1.center]{$\scriptstyle 1$};
  \node[above=-0.05 of rr-an.center]{$\scriptstyle{n}$};
  \node[above=-0.05 of rr-ai.center]{$\scriptstyle{i}$};
  \node[above=-0.05 of rr-a0.center]{$\Mainport$};

  \draw[inetwire](rr-a1.center)|-++(-0.5,-0.325)-|(r-ctr1.left pax);
  \draw[inetwire](rr-an.center)|-++(-0.3,-0.8)-|(r-ctrn.left pax);

  \inetcell[above=1 of r-reft](r-coder1){$\oc$};
  \draw[inetwire](r-coder1.pal)|-++(0.4,-0.6)-|(rr-ai.center);
  \Sometree[above=0.2 of r-coder1.pax]{0.9}{r-s}{$u$}
  \draw[inetwire](r-s.pal)--(r-coder1.pax);

  \inetcell[right=5.45 of r-pnode1](r-coder2){$\oc$}
  \draw[inetwire](rr-a0.center)--(r-coder2.pax);

  \inetcell[below=1.4 of r-reft](coctr){$\oc$}
  \draw[inetwire](r-pnode1.pal)|-++(0.5,-0.2)-|(coctr.right pax);
  \draw[inetwire](r-coder2.pal)|-++(-0.5,-0.2)-|(coctr.left pax);
\end{tikzpicture}
  \caption{Promotion/codereliction commutative reduction (Chain Rule)}
  \label{fig:prom-chain}
\end{figure}

The interaction between a promotion and a codereliction is a syntactic version
of the \emph{chain rule} of calculus, see Figure~\ref{fig:prom-chain}.
\begin{align*}
  &\PROM{n+1}P(\List t1{i-1},\CODER(u),\List t{i+1}{n+1})\\
  &\hspace{1em}\Rel\Redcom
  \sum_{p=\Net{\List s1{n+1},s}{\Vect c}}
  P_p\Scalmult(
  \Cut{s_i}{\CODER(u)},
   \\
&\hspace{5em}\Cut{\CONTR(\Covar{x_1},s_1)}{t_1},\dots,
    \widehat{\Cut{\CONTR(\Covar{x_{i}},s_i)}{t_i}},\dots,
    \Cut{\CONTR(\Covar{x_{n+1}},s_{n+1})}{t_{n+1}}\,;\\
&\hspace{5em}\COCONTR(\PROM{n+1}P(\List x1{i-1},\COWEAK,\List x{i+1}{n+1}),
                        \CODER(s)))
\end{align*}
where we use the standard notation $a_1,\dots,\widehat{a_i},\dots,a_n$ for the
sequence \[a_1,\dots,a_{i-1},a_{i+1},\dots,a_n\,.\]

We also have to explain how these commutative reductions can be used in
arbitrary contexts. We deal first with the case where such a reduction occurs
under a constructor symbol $\phi\in\Conset_{n+1}$.
\begin{center}
  \AxiomC{$t\Rel\Redcom P$}
  \UnaryInfC{$\phi(\Vect u,t,\Vect v)\Rel\Redcom
    \sum_{p=\Net{w}{\Vect c}}
    P_p\Scalmult\Net{\phi(\Vect u,w,\Vect v)}{\Vect c}$}
  \DisplayProof
\end{center}
Next we deal with the case where $t$ occurs in outermost position in a
proof-structure. There are actually two possibilities.
\begin{center}
  \AxiomC{$t\Rel\Redcom P$}
  \UnaryInfC{$\Net{\Vect u,t,\Vect v}{\Vect c}\Rel\Redcom
    \sum_{p=\Net{w}{\Vect d}}P_p\Scalmult\Net{\Vect u,w,\Vect v}
      {\Vect c,\Vect d}$}
  \DisplayProof
\end{center}

\begin{center}
  \AxiomC{$t\Rel\Redcom P$}
  \UnaryInfC{$\Net{\Vect t}{\Cut t{t'},\Vect c}\Rel\Redcom
    \sum_{p=\Net{w}{\Vect d}}
    P_p\Scalmult\Net{\Vect t}{\Vect c,\Vect d,\Cut w{t'}}$}
  \DisplayProof
\end{center}

We use $\Red$ for the union of the reduction relations $\Redcut$ and $\Redcom$.

This formalization of nets enjoys a subject reduction property.
\begin{theorem}
  If $\Logic\Phi p\Gamma$ and $p\Rel\Red P$ then $\Logic{\Phi'}{P}\Gamma$ for
  some $\Phi'$ which extends $\Phi$.
\end{theorem}
The proof is a rather long case analysis. We need to consider possible
extensions of $\Phi$ because of the fresh variables which are introduced by
several reduction rules.

\subsection{Correctness criterion and properties of the 
reduction}
Let $P$ be proof-structure, $\Phi$ be a closed typing context and $\Gamma$ be a
sequence of formulas such that $\Typing\Phi P\Gamma$. One says that $P$ is a
\emph{proof-net} if it satisfies $\Logic\Phi P\Gamma$. A \emph{correctness
  criterion} is a criterion on $P$ which guarantees that $P$ is a proof-net; of
course, saying that $\Logic\Phi P\Gamma$ is a correctness criterion, but is not
a satisfactory one because it is not easy to prove that it is preserved by
reduction.

Various such criteria can be found in the literature, but most of them apply to
proof-structures considered as graphical objects and are not very suitable to
our term-based approach. We rediscovered recently a correctness criterion
initially due to R\'etor\'e~\cite{Retore03} which seems more convenient for the
kind of presentation of proof-structures that we use here,
see~\cite{Ehrhard14}. This criterion, which is presented for $\MLL$, can easily
be extended to the whole of $\DILL$.

So far, the reduction relation $\Red$ is defined as a relation between simple
proof-structures and proof-structures. It must be extended to a relation
between arbitrary proof-structures. This is done by means of the following
rules
\begin{center}
  \AxiomC{$p\Rel\Red P$}
  \UnaryInfC{$p+Q\Rel\Red P+Q$}
  \DisplayProof
  \quad
  \AxiomC{$p\Rel\Red P$}
  \AxiomC{$\mu\in\Field\setminus\{0\}$}
  \BinaryInfC{$\mu\Scalmult p\Rel\Red\mu\Scalmult P$}
  \DisplayProof
\end{center}
As it is defined, our reduction relation does not allow us to perform the
reduction within boxes. To this end, one should add the following rule.
\begin{center}
  \AxiomC{$P\Rel\Red Q$}
  \UnaryInfC{$\PROM nP(\List t1n)\Rel\Red\PROM n{Q}(\List t1n)$}
  \DisplayProof
\end{center}

It is then possible to prove basic properties such as confluence and
normalization\footnote{For confluence, one needs to introduce an equivalence
  relation on proof-structures which expresses typically that contraction is
  associative, see~\cite{Tranquilli09}. For normalization, some conditions have
  to be satisfied by $\Field$; typically, it holds if one assumes that
  $\Field=\Nat$ but difficulties arise if $\Field$ has additive inverses.}. For
these topics, we refer mainly to the work of Pagani~\cite{Pagani09},
Tranquilli~\cite{PaganiTranquilli09,Tranquilli09,PaganiTranquilli09a},
Gimenez~\cite{Gimenez11}. We also refer to Vaux~\cite{Vaux08} for the link
between the algebraic properties of $\Field$ and the properties of $\Red$, in a
simpler $\lambda$-calculus setting.

Of course these proofs should be adapted to our presentation of proof
structures. This has not been done yet but we are confident that it should
not lead to difficulties.


\section{Categorical denotational semantics}
We describe now the denotational semantics of $\DILL$ in a general categorical
setting. This will give us an opportunity to provide more intuitions about the
rules of this system. More intuition about the meaning of the differential
constructs of $\DILL$ is given in Section~\ref{sec:more-exp-struct}.

\subsection{Notations and conventions}
Let $\cC$ be a category. Given objects $X$ and $Y$ of $\cC$, we use $\cC(X,Y)$
for the set of morphisms from $X$ to $Y$. Given $f\in\cC(X,Y)$ and
$g\in\cC(Y,Z)$, we use $g\Complin f$ for the composition of $f$ and $g$, which
belongs to $\cC(X,Z)$. In specific situations, we use also the notation $g\Comp
f$. When there are no ambiguities, we use $X$ instead of $\Id_X$ to denote the
identity from $X$ to $X$.

Given $n\in\Nat$ and a functor $F:\cC\to\cD$, we use the same notation $F$ for
the functor $\cC^n\to\cD^n$ defined in the obvious manner: $F(\List
X1n)=(F(X_1),\dots,F(X_n))$ and similarly for morphisms. If $F,G:\cC\to\cD$ are
functors and if $T$ is a natural transformation, we use again the same notation
$T$ for the corresponding natural transformation between the functors
$F,G:\cC^n\to\cD^n$, so that $T_{\List X1n}=(T_{X_1},\dots,T_{X_n})$.

\subsection{Monoidal structure}\label{sec:monoidal-struct}
A symmetric monoidal category is a structure
$(\cL,I,\square,\lambda,\rho,\alpha,\sigma)$ where $\cL$ is a category, $I$ is
an object of $\cL$, $\square:\cL^2\to\cL$ is a functor and
$\lambda_X\in\cL(I\Rel\square X,X)$, $\rho_X\in\cL(X\Rel\square I,X)$,
$\alpha_{X,Y,Z}\in\cL((X\Rel\square Y)\Rel\square Z,X\Rel\square(Y\Rel\square
Z))$ and $\sigma_{X,Y}\in\cL(X\Rel\square Y,Y\Rel\square X)$ are natural
isomorphisms satisfying coherence conditions which can be expressed as
commutative diagrams, and that we do not recall here. Following
McLane~\cite{MacLane71}, we present these coherence conditions using a notion
of \emph{monoidal trees} (called \emph{binary words} in~\cite{MacLane71}).

\emph{Monoidal trees} (or simply \emph{trees} when there are no ambiguities)
are defined by the following syntax.
\begin{itemize}
\item $\TreeZ$ is the empty tree
\item $\TreeO$ is the tree consisting of just one leaf
\item and, given trees $\tau_1$ and $\tau_2$, $\TreeB{\tau_1}{\tau_2}$ is a
  tree. 
\end{itemize}

Let $\Nleaves(\tau)$ be the number of leaves of $\tau$, defined by
\begin{align*}
  \Nleaves(\TreeZ)&=0\\
  \Nleaves(\TreeO)&=1\\
  \Nleaves(\TreeB{\tau_1}{\tau_2})&=\Nleaves(\tau_1)+\Nleaves(\tau_2)\,.
\end{align*}
Let $\Trees n$ be the set of trees $\tau$ such that $\Nleaves(\tau)=n$. This
set is infinite for all $n$. 

Let $\tau\in\Trees n$. Then we define in an obvious way a functor
$\TreeExt\square\tau:\cL^n\to\cL$. On object, it is defined as follows:
\begin{align*}
\TreeExt\square{\TreeZ}&=I\\
\TreeExt\square\TreeO X&=X\\
\TreeExt\square{\TreeB{\tau_1}{\tau_2}}
  (\List X1{\Nleaves(\tau_1)},\List Y1{\Nleaves(\tau_2)})&=
  (\TreeExt\square{\tau_1}(\Vect X))
    \mathrel\square(\TreeExt\square{\tau_2}{(\Vect Y)})\,. 
\end{align*}
The definition on morphisms is similar.

\paragraph{Generalized associativity.}
Given $\tau_1,\tau_2\in\Trees n$, the isomorphisms $\lambda$, $\rho$ and
$\alpha$ of the monoidal structure of $\cL$ allow us to build an unique natural
isomorphism $\Treeiso\square{\tau_1}{\tau_2}$ from $\TreeExt\square{\tau_1}$ to
$\TreeExt\square{\tau_2}$. We have in particular
\begin{align*}
  \lambda_X &= \Treeiso\square{\TreeB\TreeZ\TreeO}{\TreeO}_X\\
  \rho_X &= \Treeiso\square{\TreeB\TreeO\TreeZ}{\TreeO}_X\\
  \alpha_{X,Y,Z} &= \Treeiso\square{\TreeB{\TreeB\TreeO\TreeO}\TreeO}
  {\TreeB\TreeO{\TreeB\TreeO\TreeO}}_{X,Y,Z}
\end{align*}

The coherence commutation diagrams (which include the McLane Pentagon) allow
one indeed to prove that all the possible definitions of an isomorphism
$\TreeExt\square{\tau_1}(\Vect X)\to\TreeExt\square{\tau_2}(\Vect X)$ using
these basic ingredients give rise to the same result. This is McLane
coherence Theorem for monoidal categories. In particular the following
properties will be quite useful:
\begin{equation}\label{eq:treeiso-comp}
  \Treeiso\square\tau\tau_{\Vect X}=\Id_{\TreeExt\square\tau(\Vect X)}
  \quad\text{and}\quad
  \Treeiso\square{\tau_2}{\tau_3}_{\Vect X}
    \Complin\Treeiso\square{\tau_1}{\tau_2}_{\Vect X}
  =\Treeiso\square{\tau_1}{\tau_3}_{\Vect X}\,. 
\end{equation}
We shall often omit the indexing sequence $\Vect X$ when using these natural
isomorphisms, writing $\Treeiso\square\sigma\tau$ instead of
$\Treeiso\square\sigma\tau_{\Vect X}$. 

\paragraph{Generalized symmetry.}\label{sec:gen-monoidal-sym}
Let $n\in\Nat$.
Let $\phi\in\Symgrp n$, we define a functor $\Symfunc\phi:\cL^n\to\cL^n$ by
$\Symfunc\phi(X_1,\dots,X_n)=(X_{\phi(1)},\dots,X_{\phi(n)})$

Assume that the monoidal category $\cL$ is also symmetric. The corresponding
additional structure allows one to define a natural isomorphism
$\Symiso\square\phi\tau$ from the functor $\TreeExt\square\tau$ to the functor
$\TreeExt\square\tau\Comp\Symfunc\phi$. The correspondence
$\phi\mapsto\Symiso\square\phi\tau$ is of course functorial. Moreover, given
$\sigma,\tau\in\Trees n$ and $\phi\in\Symgrp n$, the following diagram is
commutative
\begin{equation}\label{eq:gen-sym-coherence}
  \begin{tikzpicture}[->, >=stealth]
    \node (1) {$\TreeExt\square\sigma\Vect X $};
    \node (2) [right of=1, node distance=24mm] 
      {$\TreeExt\square\sigma\Symfunc\phi(\Vect X) $};
    \node (3) [below of=1, node distance=12mm]
      {$\TreeExt\square\tau{\Vect X} $}; 
    \node (4) [below of=2, node distance=12mm]
      {$\TreeExt\square\tau\Symfunc\phi{(\Vect X)} $}; 
    \tikzstyle{every node}=[midway,auto,font=\scriptsize]
    \draw (1) -- node {$\Symiso\square\phi\sigma $} (2);
    \draw (1) -- node [swap] {$\Treeiso\square\sigma\tau $} (3);
    \draw (2) -- node {$\Treeiso\square\sigma\tau $} (4);
    \draw (3) -- node {$\Symiso\square\phi\tau $} (4);
  \end{tikzpicture}  
\end{equation}

This is a consequence of McLane coherence Theorem for symmetric monoidal
categories.

\subsection{*-autonomous categories}
\label{sec:star-autonomous}

A \emph{*-autonomous category} is a symmetric monoidal category
$(\cL,\ITens,\lambda,\rho,\alpha,\sigma)$ equipped with
the following structure:
\begin{itemize}
\item an endomap on the objects of $\cL$ that we denote as $X\mapsto\Orth X$;
\item for each object $X$, an evaluation morphism $\Evdual\in\cL(\Tens{\Orth
    X}{X},\Bot)$, where $\Bot=\Orth\One$;
\item a curryfication function $\Curdual:\cL(\Tens UX,\Bot)\to\cL(U,\Orth X)$
\end{itemize}
subject to the following equations (with $f\in\cL(\Tens UX,\Bot)$ and
$g\in\cL(V,U)$, so that $\Tens gX\in\cL(\Tens VX,\Tens UX)$):
\begin{align*}
  \Evdual\Complin\Tensp{\Curdual(f)}{X} &= f\\
  \Curdual(f)\Complin g &= \Curdual(f\Complin\Tensp{g}{X})\\
  \Curdual(\Evdual) &= \Id\,.
\end{align*}
Then $\Curdual$ is a bijection. Indeed, let $g\in\cL(U,\Orth X)$. Then $\Tens
gX\in\cL(\Tens UX,\Tens{\Orth X}X)$ and hence
$\Evdual\Complin\Tensp{g}{X}\in\cL(\Tens UX,\Bot)$. The equations allow one to
prove that the function $g\mapsto\Evdual\Complin\Tensp{g}{X}$ is the inverse of
the function $\Curdual$.

For any object $X$ of $\cL$, let
$\Biorthiso_X=\Curdual(\Evdual\Complin\Tsym_{X,\Orth X})\in\cL(X,\Biorth X)$.

The operation $X\mapsto\Orth X$ can be extended into a functor
$\Op\cL\to\cL$ as follows. Let $f\in\cL(X,Y)$, then $\Biorthiso_Y\Complin
f\in\cL(X,\Biorth Y)$, so $\Evdual\Complin\Tensp{(\Biorthiso_Y\Complin
  f)}{\Orth Y}\in\cL(\Tens{X}{\Orth Y},\Bot)$ and we set $\Orth
f=\Curdual(\Evdual\Complin\Tensp{(\Biorthiso_Y\Complin f)}{\Orth
  Y}\Complin\Tsym_{\Orth Y,X})\in\cL(\Orth Y,\Orth X)$. It can be checked that
this operation is functorial.

We assume last that $\Biorthiso_X$ is an iso for each object $X$.

One sets $\Limpl XY=\Orth{(\Tens X{\Orth Y})}$ and one defines an evaluation
morphism $\Evlin\in\cL(\Tens{(\Limpl XY)}{X},Y)$ as follows. We have
\[
\Evdual\in\cL(\Tens{\Orth{(\Tens X{\Orth Y})}}{(\Tens X{\Orth Y})},\Bot)
\]
hence 
\[\Evdual\Complin\Treeiso\ITens{\TreeB{\TreeB\TreeO\TreeO}\TreeO}
{\TreeB\TreeO{\TreeB\TreeO\TreeO}}\in\cL(\Tens{\Tensp{\Orth{\Tensp{X}{\Orth
      Y}}}{X}}{\Orth Y},\Bot)\,,
\]
therefore
\[
\Curdual(\Evdual\Complin\Treeiso\ITens{\TreeB{\TreeB\TreeO\TreeO}\TreeO}
{\TreeB\TreeO{\TreeB\TreeO\TreeO}})
\in\cL(\Tens{\Orth{\Tensp{X}{\Orth Y}}}{X},\Biorth Y)
\]
and we set
\[
\Evlin=\Funinv\Biorthiso\Complin
\Curdual(\Evdual\Complin\Treeiso\ITens{\TreeB{\TreeB\TreeO\TreeO}\TreeO}
{\TreeB\TreeO{\TreeB\TreeO\TreeO}}) \in\cL(\Tens{\Orth{\Tensp{X}{\Orth
      Y}}}{X},Y).
\]

Let $f\in\cL(\Tens UX,Y)$.  We have $\Biorthiso\Complin f\in\cL(\Tens
UX,\Biorth Y)$, hence $\Funinv\Curdual(\Biorthiso\Complin f)\in\cL(\Tens{\Tensp
  UX}{\Orth Y},\Bot)$, so $\Funinv\Curdual(\Biorthiso\Complin
f)\Complin\Treeiso\ITens{\TreeB{\TreeO}{\TreeB\TreeO\TreeO}}
{\TreeB{\TreeB\TreeO\TreeO}\TreeO}\in\cL(\Tens U{\Tensp X{\Orth Y}},\Bot)$ and
we can define a linear curryfication of $f$ as
\[
\Curlin(f)=\Curdual(\Funinv\Curdual(\Biorthiso\Complin
f)\Complin\Treeiso\ITens{\TreeB{\TreeO}{\TreeB\TreeO\TreeO}}
{\TreeB{\TreeB\TreeO\TreeO}\TreeO})\in
\cL(U,\Limpl XY)\,.
\]

One can prove then that the following equations hold, showing that the
symmetric monoidal category $\cL$ is closed.
\begin{align*}
  \Evlin\Complin\Tensp{\Curlin(f)}{X} & =f \\
  \Curlin(f)\Complin g &= \Curlin(f\Complin\Tensp gX) \\
  \Curlin(\Evlin) &= \Id
\end{align*}
where $g\in\cL(V,U)$.

It follows as usual that $\Curlin$ is a bijection from $\cL(\Tens UX,Y)$ to
$\cL(U,\Limpl XY)$. 

We set $\Par XY=\Orth{\Tensp{\Orth X}{\Orth Y}}=\Limpl{\Orth X}Y$; this
operation is the \emph{cotensor product}, also called \emph{par} in linear
logic. Using the above properties one shows that this operation is a functor
$\cL^2\to\cL$ which defines another symmetric monoidal structure on $\cL$. The
operation $X\mapsto\Orth X$ is an equivalence of symmetric monoidal categories
from $(\Op\cL,\ITens)$ to $(\cL,\IPar)$.

\paragraph{MIX.}\label{par:mix-star-auto}
A \emph{mix *-autonomous category} is a *-autonomous category $\cL$ where
$\Bot$ is endowed with a structure of commutative $\ITens$-monoid\footnote{If
  we see $\Bot$ as the object of scalars, which is compatible with the
  intuition that $\Limpl X\Bot$ is the dual of $X$, that is, the ``space of
  linear forms on $X$'', then this monoid structure is an internal
  multiplication law on scalars.}. So we have two morphisms
$\MIXZ\in\cL(\One,\Bot)$ and $\MIXB\in\cL(\Tens\Bot\Bot,\Bot)$ and some
standard diagrams must commute, which express that $\MIXZ$ is left and right
neutral for the binary operation $\MIXB$, and that this binary operation is
associative and commutative. Observe that
$\Orthp{\MIXB}\in\cL(\Orth\Bot,\Orth{\Tensp\Bot\Bot})$ so that
\begin{equation*}
  \MIXB'=\Orthp{\MIXB}\Complin\Biorthiso_\One\in\cL(\One,\Par\One\One)
\end{equation*}
and $(\One,\MIXZ,\MIXB')$ is a commutative $\IPar$-comonoid.

\paragraph{Vectors.}
Let $\cL$ be a *-autonomous category and let $\List X1n$ be objects of $\cL$. 

A \emph{$(\List X1n)$-vector} is a family $(u_\tau)_{\tau\in\Trees n}$ where
$u_\tau\in\cL(\One,\TreeExt\IPar\tau(\List X1n))$ satisfies
$u_{\tau'}=\Treeiso\IPar{\tau}{\tau'}\Complin u_\tau$ for all
$\tau,\tau'\in\Trees n$. Of course such a vector $u$ is determined as soon as
one of the $u_\tau$'s is given. The point of this definition is that none of
these $u_\tau$'s is more canonical than the others, that is why we find more
convenient to deal with the whole family $u$. Let $\Vectors\cL{\List X1n}$ be
the set of these vectors. Notice that, since $\Trees n$ is infinite for all
$n$, all vectors are infinite families.

\paragraph{\MLL{} vector constructions.}
Let $X\in\cL$. 
We define $\Idpar\in\Vectors\cL{\Orth X,X}$ by setting
$\Idpar_\tau=\Treeiso\IPar{\TreeB\TreeO\TreeO}\tau\Complin
\Curlin(\Funinv{\Biorthiso_X}\Complin
\Treeiso\ITens{\TreeB\TreeZ\TreeO}{\TreeO})\in\cL(\One,\Limpl{\Biorth
  X}{X})=\cL(\One,\Par{\Orth X}{X})$ for all $\tau\in\Trees 2$.

Let $u\in\Vectors\cL{\List X1n,X,Y}$. We define $\Parvect u\in\Vectors\cL{\List
  X1n,\Par
  XY}$ as follows. Let $\tau\in\Trees n$, we know that
\[
u_{\TreeB{\tau}{\TreeB{\TreeO}{\TreeO}}}\in\cL(\One,\Par{(\TreeExt\IPar
  \tau(\List X1n)}{\Parp XY})=\cL(\One,{(\TreeExt\IPar
  {\TreeB\tau\TreeO}(\List X1n,\Par XY)})\,.
\] 
For any $\theta\in\Trees{n+1}$, we set
\[
\Parvect
u_{\theta}=\Treeiso\IPar{\TreeB\tau\TreeO}{\theta}
\Complin u_{\TreeB{\tau}{\TreeB{\TreeO}{\TreeO}}}
\in\cL(\One,\TreeExt\IPar\theta(\List X1n,\Par XY))\,.
\]
One sees easily that this definition does not depend on
the choice of $\tau$: let $\tau'\in\Trees n$, we have
\begin{align*}
  \Treeiso\IPar{\TreeB\tau\TreeO}{\theta}
    \Complin\Parvect u_{\TreeB\tau{\TreeB\TreeO\TreeO}}
  &= \Treeiso\IPar{\TreeB\tau\TreeO}{\theta}\Complin
  \Treeiso\IPar{\TreeB{\tau'}\TreeO}{\TreeB\tau\TreeO}
  \Complin\Parvect u_{\TreeB{\tau'}{\TreeB\TreeO\TreeO}}\\
  &= \Treeiso\IPar{\TreeB{\tau'}\TreeO}{\theta}
  \Complin\Parvect u_{\TreeB{\tau'}{\TreeB\TreeO\TreeO}}\,.
\end{align*}
thanks to the definition of vectors and to Equation (\ref{eq:treeiso-comp}). 

Let $U_i,X_i$ be objects of $\cL$ for $i=1,2$. Given
\[
u_i\in\cL(\One,\Par{U_i}{X_i})=\cL(\One,\Limpl{\Orth{U_i}}{X_i})
\]
for $i=1,2$, we define
\[
\Tenscont{U_1}{X_1}{U_2}{X_2}{u_1}{u_2}
\in\cL(\One,\Par{\Parp{U_1}{U_2}}{\Tensp{X_1}{X_2}})
\]
as follows. We have
$\Funinv\Curlin(u_i)\Complin\Treeiso\ITens\TreeO{\TreeB\TreeZ\TreeO}
\in\cL(\Orth{U_i},X_i)$ and hence
\[
v=\Tens{(\Funinv\Curlin(u_1)\Complin\Treeiso\ITens\TreeO{\TreeB\TreeZ\TreeO})}
{(\Funinv\Curlin(u_2)\Complin\Treeiso\ITens\TreeO{\TreeB\TreeZ\TreeO})}
\in\cL(\Tens{\Orth{U_1}}{\Orth{U_2}},\Tens{X_1}{X_2})\,.
\]
We have 
\[
\Curlin(v\Complin\Treeiso\ITens{\TreeB\TreeZ\TreeO}{\TreeO})
\in\cL(\One,\Orth{\Tensp{\Tensp{\Orth{U_1}}{\Orth{U_2}}}
  {\Orth{\Tensp{X_1}{X_2}}}})
\]
So we set
\begin{align*}
\Tenscont{U_1}{X_1}{U_2}{X_2}{u_1}{u_2} &=
\Orth{\Tensp{\Funinv{\Biorthiso_{\Tensp{\Orth{U_1}}{\Orth{U_2}}}}}
  {\Orth{\Tensp{X_1}{X_2}}}}\Complin
\Curlin(v\Complin\Treeiso\ITens{\TreeB\TreeZ\TreeO}{\TreeO})\\
&\in\cL(\One,\Par{\Parp{U_1}{U_2}}{\Tensp{X_1}{X_2}})
\end{align*}
where the natural iso $\Biorthiso$ is defined in
Section~\ref{sec:star-autonomous}. 

This construction is natural in the sense that, given $f_i\in\cL(X_i,X'_i)$,
$g_i\in\cL(U_i,U'_i)$, one has
\begin{equation}\label{eq:tenscont-natural}
\begin{aligned}
&\Parp{\Parp{g_1}{g_2}}{\Tensp{f_1}{f_2}}
\Complin\Tenscont{U_1}{X_1}{U_2}{X_2}{u_1}{u_2}\\
&\quad\quad\quad\quad=
\Tenscont{U'_1}{X'_1}{U'_2}{X'_2}{\Parp{f_1}{g_1}
  \Complin u_1}{\Parp{f_2}{g_2}\Complin u_2}
\end{aligned}  
\end{equation}

Let $u\in\Vectors\cL{\List X1n,X}$ and $v\in\Vectors\cL{\List Y1p,Y}$. Let
$\sigma\in\Trees n$ and $\tau\in\Trees p$. Then we have
\begin{align*}
u_{\TreeB{\sigma}{\TreeO}}&\in\cL(\One,\Par {(\TreeExt{\IPar}{\sigma}(\List
  X1n))}X)\quad\text{and}\\
v_{\TreeB\tau\TreeO}&\in\cL(\One,\Par{(\TreeExt\IPar\tau{(\List Y1p)})}Y)
\end{align*}
and we set
\begin{align*}
  \Tensvect uv_{\TreeB{\TreeB\sigma\tau}{\TreeO}}
  &=\Tenscont {\TreeExt{\IPar}{\sigma}(\List
    X1n)}X{\TreeExt\IPar\tau{(\List Y1p)}}Y
  {u_{\TreeB\sigma\TreeO}}{v_{\TreeB\tau\TreeO}}\\
&\in\cL(\One,\TreeExt\IPar{\TreeB{\TreeB\sigma\tau}{\TreeO}}(\List
X1n,\List Y1p,\Tens XY)
\end{align*}
since
\begin{align*}
  &\Par{\Parp{\TreeExt{\IPar}{\sigma}(\List
      X1n)}{(\TreeExt\IPar\tau{(\List
        Y1p)})}}{\Tensp{X}{Y}}\\
  &\quad\quad\quad\quad\quad\quad
  =\TreeExt\IPar{\TreeB{\TreeB\sigma\tau}{\TreeO}}(\List X1n,\List
  Y1p,\Tens XY)\,.
\end{align*}
Then, given $\theta\in\Trees{n+p+1}$, one sets of course
\[
\Tensvect uv_\theta=\Treeiso\IPar{\TreeB{\TreeB{\sigma}{\tau}}{\TreeO}}\theta
\Complin\Tensvect uv_{\TreeB{\TreeB\sigma\tau}{\TreeO}}\,.
\]
One checks easily that this definition does not depend on the choice of
$\sigma$ and $\tau$, using Equations (\ref{eq:treeiso-comp}) and
(\ref{eq:tenscont-natural}), and one can check that
\[
\Tensvect uv\in\Vectors\cL{\List X1n,\List Y1p,\Tens XY}\,.
\]

Let $u\in\Vectors\cL{\List X1n}$ and $\phi\in\Symgrp n$. Given $\sigma\in\Trees
n$, we have $u_\sigma\in\cL(\One,\TreeExt\IPar\sigma{(\List X1n)})$ and hence
$\Symiso\IPar\phi\sigma\Complin
u_\sigma\in\cL(\One,\TreeExt\IPar\sigma{(X_{\phi(1)},\dots,X_{\phi(n)})})$.
Given $\theta\in\Trees n$, we set therefore
\begin{equation*}
  \Permvect\phi u_\theta
  =\Treeiso\IPar\sigma\theta\Complin\Symiso\IPar\phi\sigma\Complin
   u_\sigma\in\cL(\One,\TreeExt\IPar\theta{(X_{\phi(1)},\dots,X_{\phi(n)})})
\end{equation*}
defining an element $\Permvect\phi u$ of
$\Vectors\cL{X_{\phi(1)},\dots,X_{\phi(n)}}$ which does not depend on the
choice of $\sigma$. 
Indeed, let $\tau\in\Trees n$, we know that $u_\sigma=\Treeiso\IPar\tau\sigma
u_\tau$ and hence $\Permvect\phi u_\theta=
\Treeiso\IPar\sigma\theta\Complin\Symiso\IPar\phi\sigma
\Complin\Treeiso\IPar\tau\sigma u_\tau=\Treeiso\IPar\sigma\theta\Complin
\Treeiso\IPar\tau\sigma\Complin\Symiso\IPar\phi\tau
u_\tau=\Treeiso\IPar\tau\theta\Complin\Symiso\IPar\phi\tau u_\tau$, 
using Diagram~\Eqref{eq:gen-sym-coherence}.

Let $u\in\Vectors\cL{\List X1n,\Orth X}$ and $v\in\Vectors\cL{\List Y1p,X}$. We
have 
\[
\Tensvect uv\in\Vectors\cL{\List X1n,\List Y1p,\Tens{\Orth
    X}{X}}
\]
Given $\sigma\in\Trees n$ and $\tau\in\Trees p$ we have
\[
\Tensvect uv_{\TreeB{\TreeB\sigma\tau}{\TreeO}}\in\cL(\One,\Par{\Parp{(\TreeExt\IPar{\sigma}{(\List X1n)})}{(\TreeExt\IPar\tau{\List
      Y1p})}}{\Tensp{\Orth
    X}{X}})
\]
so that 
\[\Parp{\Id}{\Evdual}\Complin\Tensvect
uv_{\TreeB{\TreeB\sigma\tau}{\TreeO}}
\in\cL(\One,\TreeExt\IPar{\TreeB{\TreeB{\sigma}{\tau}}{\TreeZ}}{(\List
  X1n,\List Y1p)})\,.
\] 
Given $\theta\in\Trees{n+p}$, we set
\begin{equation*}
  \Cutvect uv_\theta
  =\Treeiso\IPar{\TreeB{\TreeB{\sigma}{\tau}}\TreeZ}{\theta}\Complin
  \Parp{\Id}{\Evdual}\Complin\Tensvect
uv_{\TreeB{\TreeB\sigma\tau}{\TreeO}}
\end{equation*}
and we define in that way an element $\Cutvect uv$ of $\Vectors\cL{\List
  X1n,\List Y1p}$.

Assume now that $\cL$ is a mix *-autonomous category (in the sense of
Paragraph~\ref{par:mix-star-auto}).  

Let $\List X1n$ and $\List Y1p$ be objects of $\cL$. Let $u\in\Vectors\cL{\List
  X1n}$ and $v\in\Vectors\cL{\List Y1p}$. Let $\sigma\in\Trees n$ and
$\tau\in\Trees p$. We have $u_\sigma\in\cL(\One,\TreeExt\IPar\sigma(\List
X1n))$ and $v_\tau\in\cL(\One,\TreeExt\IPar\tau(\List Y1p))$. Hence
\begin{equation*}
  \Parp{u_\sigma}{v_\tau}
  \Complin\MIXB'\in\cL(\One,\Par{(\TreeExt\IPar\sigma{(\List
      X1n)})}{(\TreeExt\IPar\tau{(\List Y1p)})})
\end{equation*}
and we define therefore $\Mixvect uv\in\Vectors\cL{\List X1n,\List Y1p}$ by
setting
\begin{equation*}
  \Mixvect uv_\theta=\Treeiso\IPar{\TreeB\sigma\tau}{\theta}
\Complin\Parp{u_\sigma}{v_\tau}\Complin
\MIXB'\in\cL(\One,\TreeExt\IPar\theta{(\List X1n,\List Y1p)})  
\end{equation*}
for each $\theta\in\Trees{n+p}$. As usual, this definition does not depend on
the choice of $\sigma$ and $\tau$.

\paragraph{Interpreting \MLL{} derivations.}\label{par:MLL-der-interp}
We start with a valuation which, with each $\alpha\in\Atoms$, associates
$\Typesem\alpha\in\cL$ in such a way that
$\Typesem{\Coatom\alpha}=\Orth{\Typesem\alpha}$. We extend this valuation to an
interpretation of all \MLL{} types as objects of $\cL$ in the obvious manner,
so that we have a \emph{De Morgan} iso $\Demorgan_A\in\cL(\Typesem{\Orth
  A},\Orth{\Typesem A})$ defined inductively as follows.

We set first $\Demorgan_\alpha = \Id_{\Orth{\Typesem\alpha}}$. We have
$\Demorgan_A\in\cL(\Typesem{\Orth A},\Orth{\Typesem A})$ and
$\Demorgan_B\in\cL(\Typesem{\Orth B},\Orth{\Typesem B})$, therefore
$\Par{\Demorgan_A}{\Demorgan_B}\in\cL(\Typesem{\Orth{\Tensp
    AB}},\Par{\Orth{\Typesem A}}{\Orth{\Typesem B}})$. We have
\[
\Par{\Orth{\Typesem
    A}}{\Orth{\Typesem B}}=\Orth{\Tensp{\Biorth{\Typesem A}}{\Biorth{\Typesem
      B}}}
\]
by definition of $\IPar$ and remember that ${\Biorthiso_{\Typesem
    A}}\in\cL(\Typesem A,\Biorth{\Typesem A})$ and so we set
\begin{equation*}
  \Demorgan_{\Tens AB}=
  \Orth{(\Tens{\Biorthiso_{\Typesem A}}{\Biorthiso_{\Typesem B}})}
  \Complin(\Par{\Demorgan_A}{\Demorgan_B})\,.
\end{equation*}
We have $\Typesem{\Orth{\Parp AB}}=\Tens{\Typesem{\Orth A}}{\Typesem{\Orth B}}$
so $\Tens{\Demorgan_A}{\Demorgan_B}\in\cL(\Typesem{\Orth{\Parp
    AB}},\Tens{\Orth{\Typesem A}}{\Orth{\Typesem A}})$. By definition we
have $\Orth{\Typesem{\Par AB}}=\Biorth{\Tensp{\Orth{\Typesem A}}{\Orth{\Typesem
      B}}}$. So we set
\begin{equation*}
  \Demorgan_{\Par AB}=\Biorthiso_{\Tens{\Orth{\Typesem A}}{\Orth{\Typesem
      B}}}\Complin\Tensp{\Demorgan_A}{\Demorgan_B}\,.
\end{equation*}

Given a sequence $\Gamma=(\List A1n)$ of types, we denote as $\Typesem\Gamma$
the sequence of objects $(\Typesem{A_1},\dots,\Typesem{A_n})$.

\smallskip

Given a derivation $\pi$ of a logical judgment $\Logic {\Phi}p\Gamma$
we define now $\Dersem\pi\in\Vectors\cL{\Typesem\Gamma}$, by induction on the
structure of $\pi$.

Assume first that $\Gamma=(\Orth A,A)$, $p=\Net{\Covar x, x}{}$ and that $\pi$
is the axiom
\begin{center}
  \AxiomC{}
  \RightLabel{\quad\AXIOM}
  \UnaryInfC{$\Logic{\Phi}p\Gamma$}
  \DisplayProof
\end{center}
We have $\Idpar_{\Typesem{A}}\in\cL(1,\Par{\Orth{\Typesem A}}{\Typesem{A}})$
and $\Demorgan_A\in\cL(\Typesem{\Orth A},\Orth{\Typesem{A}})$ so that we can
set 
\[
\Dersem\pi_\sigma= \Treeiso\IPar{\TreeB\TreeO\TreeO}{\sigma}
\Complin\Parp{\Funinv\Demorgan_A}{\Id_{\Typesem A}}\Complin\Idpar_{\Typesem A}
\]
and we have $\Dersem\pi\in\Vectors\cL{\Typesem\Gamma}$ as required.

Assume next that $\Gamma=(\Delta,\Par AB)$, that $p=\Net{\Vect s,\Par st}{\Vect
  c}$ and that $\pi$ is the following derivation, where $\lambda$ is the
derivation of the premise:
\begin{center}
  \AxiomC{$\Logic{\Phi}{\Net{\Vect s,s,t}{\Vect c}}{\Delta,A,B}$}
  \RightLabel{\quad\PARRULE}
  \UnaryInfC{$\Logic{\Phi}{\Net{\Vect s,\Par st}{\Vect c}}{\Delta,\Par AB}$}
  \DisplayProof
\end{center}
then by inductive hypothesis we have
$\Dersem\lambda\in\Vectors\cL{\Typesem{\Delta,A,B}}$ and hence we set
\begin{equation*}
  \Dersem\pi
  =\Parvect{\Dersem\lambda}\in\Vectors\cL{\Typesem{\Delta,\Par AB}}\,.
\end{equation*}

Assume now that $\Gamma=(\Delta,\Lambda,\Tens AB)$, that $p=\Net{\Vect
  s,\Vect t,\Tens st}{\Vect c,\Vect d}$ and $\pi$ is the following derivation,
where $\lambda$ is the derivation of the left premise and $\rho$ is the
derivation of the right premise:
\begin{center}
  \AxiomC{$\Logic{\Phi}{\Net{\Vect s,s}{\Vect c}}{\Delta,A}$}
  \AxiomC{$\Logic{\Phi}{\Net{\Vect t,t}{\Vect d}}{\Lambda,B}$}
  \RightLabel{\quad\TENSRULE}
  \BinaryInfC{$\Logic{\Phi}
    {\Net{\Vect s,\Vect t,\Tens st}{\Vect c,\Vect d}}{\Delta,\Lambda,\Tens
      AB}$} \DisplayProof
\end{center}
then by inductive hypothesis we have
$\Dersem\lambda\in\Vectors\cL{\Typesem{\Delta,A}}$ and
$\Dersem\rho\in\Vectors\cL{\Typesem{\Lambda,B}}$ and hence we set
\begin{equation*}
  \Dersem\pi
  =\Tensvect{\Dersem\lambda}{\Dersem\rho}
  \in\Vectors\cL{\Typesem{\Delta,\Lambda,\Tens AB}}\,.
\end{equation*}

Assume that $\phi\in\Symgrp n$, $\Gamma=(A_{\phi(1)},\dots,A_{\phi(n)})$,
$p=\Net{s_{\phi(1)},\dots,s_{\phi(n)}}{\Vect c}$
and that $\pi$ is the following derivation, where $\lambda$ is the derivation
of the premise:
\begin{center}
  \AxiomC{$\Logic{\Phi}{\Net{\List s1n}{\Vect c}}{\List A1n}$}
  \RightLabel{\quad\PERMRULE}
  \UnaryInfC{$\Logic{\Phi}{p}{\Gamma}$}
  \DisplayProof
\end{center}
By inductive hypothesis we have
$\Dersem\lambda\in\Vectors\cL{\Typesem{A_1},\dots,\Typesem{A_n}}$ and we set
\begin{equation*}
  \Dersem\pi=\Permvect\phi{\Dersem\lambda}\in\Vectors\cL{\Typesem\Gamma}\,.
\end{equation*}

Assume that $\Gamma=(\Delta,\Lambda)$, that $p=\Net{\Vect s,\Vect t}{\Vect
  c,\Vect d,\Cut st}$ and that $\pi$ is the following derivation, where
$\lambda$ is the derivation of the left premise and $\rho$ is the derivation of
the right premise:
\begin{center}
  \AxiomC{$\Logic{\Phi}{\Net{\Vect s,s}{\Vect c}}{\Delta,\Orth A}$}
  \AxiomC{$\Logic{\Phi}{\Net{\Vect t,t}{\Vect d}}{\Lambda,A}$}
  \RightLabel{\quad\CUTRULE}
  \BinaryInfC{$\Logic{\Phi}{\Net{\Vect s,\Vect t}{\Vect c,\Vect
        d,\Cut st}}{\Delta,\Lambda}$} \DisplayProof
\end{center}
By inductive hypothesis we have
$\Dersem\lambda\in\Vectors\cL{\Typesem{\Delta,\Orth A}}$ and
$\Dersem\rho\in\Vectors\cL{\Typesem{\Lambda,A}}$. Let $n$ be the length of
$\Delta$. Let $\sigma\in\Trees n$. We have
$\Dersem\lambda_{\TreeB\sigma\TreeO}
\in\cL(\One,\Par{(\TreeExt\IPar\sigma(\Typesem\Delta))}{\Typesem{\Orth
    A}})$ and hence $\Parp{\Id}{\Demorgan_A}\Complin
\Dersem\lambda_{\TreeB\sigma\TreeO}\in\cL(\One,
\Par{(\TreeExt\IPar\sigma(\Typesem\Delta))}{\Orth{\Typesem{A}}})$. We define
therefore $l\in\Vectors\cL{\Typesem\Delta,\Orth{\Typesem A}}$ by
$l_{\TreeB{\sigma}{\TreeO}}=\Parp{\Id}{\Demorgan_A}\Complin
\Dersem\lambda_{\TreeB\sigma\TreeO}$ (this definition of $l$ does not depend on
the choice of $\sigma$). We set
\begin{equation*}
  \Dersem\pi
  =\Cutvect{l}{\Dersem\rho}
  \in\Vectors\cL{\Typesem{\Delta,\Lambda}}\,.
\end{equation*}

Assume last that $\Gamma=(\Delta,\Lambda)$, that $p=\Net{\Vect s,\Vect t}{\Vect
  c,\Vect d}$ and that $\pi$ is the following derivation, where
$\lambda$ is the derivation of the left premise and $\rho$ is the derivation of
the right premise:
\begin{center}
  \AxiomC{$\Logic{\Phi}{\Net{\Vect s}{\Vect c}}{\Delta}$}
  \AxiomC{$\Logic{\Phi}{\Net{\Vect t}{\Vect d}}{\Lambda}$}
  \RightLabel{\quad\MIXRULE}
  \BinaryInfC{$\Logic{\Phi}{\Net{\Vect s,\Vect t}{\Vect c,\Vect
        d}}{\Delta,\Lambda}$} \DisplayProof
\end{center}
so that by inductive hypothesis
$\Dersem\lambda\in\Vectors\cL{\Typesem{\Delta}}$ and
$\Dersem\rho\in\Vectors\cL{\Typesem{\Lambda}}$. We set
\begin{equation*}
  \Dersem\pi=
\Mixvect{\Dersem\lambda}{\Dersem\rho}\in\Vectors\cL{\Dersem{\Delta,\Lambda}}\,.
\end{equation*}

The first main property of this interpretation of derivations is that they only
depend on the underlying nets.
\begin{theorem}
  Let $\pi$ and $\pi'$ be derivations of $\Logic{\Phi}p\Gamma$. Then
  $\Dersem\pi=\Dersem{\pi'}$. 
\end{theorem}
The proof is a (tedious) induction on the structure of the derivations $\pi$
and $\pi'$.

We use therefore $\Netsem p$ to denote the value of $\Dersem\pi$ where
$\pi$ is an arbitrary derivation of $\Logic{}p\Gamma$.

\begin{remark}
  It would be much more satisfactory to be able to define $\Netsem p$ directly,
  without using the intermediate and non canonical choice of a derivation
  $\pi$. Such a definition would use directly the fact that $p$ fulfills a
  correctness criterion in order to build a morphism of $\cL$. It is not very
  clear yet how to do that in general, though such definitions are available in
  many concrete models of \LL{}, such as coherence spaces.
\end{remark}

The second essential property of this interpretation is that it is invariant
under reductions (subject reduction)
\begin{theorem}
  Assume that $\Logic{\Phi}p\Gamma$, $\Logic{\Phi}{p'}\Gamma$ and that
  $p\Rel\Red p'$. Then $\Netsem p=\Netsem{p'}$.
\end{theorem}

\subsection{Preadditive models}\label{sec:additive-lin-cat}
Let $\cL$ be a *-autonomous category. We say that $\cL$ is \emph{preadditive}
if each hom-set $\cL(X,Y)$ is equipped with a structure of $\Field$-module
(we use standard additive notations: $0$ for the neutral element and $+$ for
the operation), which is compatible with composition of morphisms and tensor
product:
\begin{align*}
  \left(\sum_{j\in J} \nu_jt_j\right)\Complin\left(\sum_{i\in I}
    \mu_is_i\right)&=
  \sum_{(i,j)\in I\times J}\nu_j\mu_i\,(t_j\Complin s_i)\\
  \left(\sum_{i\in I}
    \mu_is_i\right)\ITens\left(\sum_{j\in J} \nu_jt_j\right)&=
  \sum_{(i,j)\in I\times J}\mu_i\nu_j\,\Tensp{s_i}{t_j}\\
\end{align*}
where the $\mu_i$'s and the $\nu_j$'s are elements of $\Field$.  It follows
that, given a finite family $(s_i)_{i\in I}$ of morphisms $s_i\in\cL(\Tens
UX,\Bot)$, one has $\Curdual\left(\sum_{i\in I}\mu_is_i\right)=\sum_{i\in
  I}\mu_i\Curdual(s_i)$, and that the cotensor product is bilinear
\begin{equation*}
  \left(\sum_{i\in I} \mu_is_i\right)\IPar\left(\sum_{j\in J} \nu_jt_j\right)=
  \sum_{(i,j)\in I\times J}\mu_i\nu_j\,(\Par{s_i}{t_j})\,.
\end{equation*}

Let $(X_i)_{i=1}^n$ be a family of objects of $\cL$. The set $\Vectors\cL{\List
X1n}$ inherits canonically a $\Field$-module structure.

\subsection{Exponential structure}\label{sec:exp-struct}
If $\cC$ is a category, we use $\Groupoid\cC$ to denote the category whose
objects are those of $\cC$ and whose morphisms are the isos of $\cC$ (so
$\Groupoid\cC$ is a groupoid).

Let $\cL$ be a preadditive *-autonomous category. An \emph{exponential
  structure} on $\cL$ is a tuple
$(\Excl{\_},\WEAK,\CONTR,\COWEAK,\COCONTR,\DER,\CODER)$ where $\Excl{\_}$ is a
functor $\Groupoid\cL\to\Groupoid\cL$ and the other ingredients are natural
transformations: $\WEAK_X\in\cL(\Excl X,\One)$ (\emph{weakening}),
$\CONTR_X\in\cL(\Excl X,\Tens{\Excl X}{\Excl X})$ (\emph{contraction}),
$\COWEAK_X\in\cL(\One,\Excl X)$ (\emph{coweakening}),
$\COCONTR_X\in\cL(\Tens{\Excl X}{\Excl X},\Excl X)$ (\emph{cocontraction}),
$\DER_X\in\cL(\Excl X,X)$ (\emph{dereliction}) and $\CODER_X\in\cL(X,\Excl X)$
(\emph{codereliction}). 

These morphisms are assumed moreover to satisfy the following properties.


The structure $(\Excl X,\WEAK_X,\CONTR_X,\COWEAK_X,\COCONTR_X)$ is required to
be a commutative bialgebra. This means that $(\Excl X,\WEAK_X,\CONTR_X)$ is a
commutative comonoid, $(\Excl X,\COWEAK_X,\COCONTR_X)$ is a commutative monoid
and that the following diagrams commute (where $\phi=(1,3,2,4)\in\Symgrp 4$)

\begin{center}
     \begin{tikzpicture}[->, >=stealth]
    \node (1) {$\Tens{\Excl X}{\Excl X}$};
    \node (2) [right of=1, node distance=38mm]
      {$\Tens{(\Tens{\Excl X}{\Excl X})}{(\Tens{\Excl X}{\Excl X})}$};
    \node (3) [below of=1, node distance=12mm] {$\Excl{X}$};
    \node (4) [below of=3, node distance=12mm] {$\Tens{\Excl X}{\Excl X}$};
    \node (5) [right of=4, node distance=38mm]
      {$\Tens{(\Tens{\Excl X}{\Excl X})}{(\Tens{\Excl X}{\Excl X})}$};
    \node (11) [right of=3, node distance=88mm] {$\Excl X$};
    \node (11a) [above of=11, node distance=6mm] {};
    \node (11b) [above of=11, node distance=-6mm] {};
    \node (12) [left of=11a, node distance=16mm] {$\One$};
    \node (13) [left of=11b, node distance=16mm] {$\One$};
    \tikzstyle{every node}=[midway,auto,font=\scriptsize]
    \draw (1) -- node {$\Tens{\CONTR_X}{\CONTR_X}$} (2);
    \draw (2) -- node {$\Symiso\ITens\phi{\TreeB{\TreeB\TreeO\TreeO}
        {\TreeB\TreeO\TreeO}}$} (5);
    \draw (1) -- node [swap] {$\COCONTR_X$}(3);
    \draw (3) -- node [swap] {$\CONTR_X$}(4);
    \draw (5) -- node {$\Tens{\COCONTR_X}{\COCONTR_X}$} (4);
    \draw (12) -- node {$\COWEAK_X$} (11);
    \draw (11) -- node {$\WEAK_X$} (13);
    \draw (12) -- node [swap] {$\Id_\One$} (13);
  \end{tikzpicture}
\end{center}

Moreover, we also require the following commutations (in the
dereliction/cocontraction and codereliction/contraction diagrams, we omit the
isos $\Treeiso\ITens{\TreeB\TreeZ\TreeO}{\TreeO}$ and
$\Treeiso\ITens{\TreeB\TreeO\TreeZ}{\TreeO}$ for the sake of readability).

\begin{center}
     \begin{tikzpicture}[->, >=stealth]
    \node (1) {$X$};
    \node (1r) [right of=1, node distance=16mm] {};
    \node (2) [below of=1r, node distance=6mm]{$\Excl X$};
    \node (3) [below of=1, node distance=12mm]{$\One$};
    \node (11) [below of=1, node distance=20mm] {$X$};
    \node (11r) [right of=11, node distance=16mm] {};
    \node (12) [below of=11r, node distance=6mm]{$\Excl X$};
    \node (13) [below of=11, node distance=12mm]{$\One$};
    \node (21) [right of=1, node distance=56mm] {$X$};
    \node (21r) [right of=21, node distance=16mm] {};
    \node (22) [below of=21r, node distance=6mm]{$\Excl X$};
    \node (23) [below of=21, node distance=12mm]{$\Tens{\Excl X}{\Excl X}$};
    \node (31) [right of=11, node distance=56mm] {$X$};
    \node (31r) [right of=31, node distance=16mm] {};
    \node (32) [below of=31r, node distance=6mm]{$\Excl X$};
    \node (33) [below of=31, node distance=12mm]{$\Tens{\Excl X}{\Excl X}$};
    \tikzstyle{every node}=[midway,auto,font=\scriptsize]
    \draw (1) -- node {$\CODER_X$} (2);
    \draw (2) -- node {$\WEAK_X$} (3);
    \draw (1) -- node [swap] {$0$} (3);
    \draw (12) -- node [swap] {$\DER_X$} (11);
    \draw (13) -- node [swap] {$\COWEAK_X$} (12);
    \draw (13) -- node {$0$} (11);
    \draw (21) -- node {$\CODER_X$} (22);
    \draw (22) -- node {$\CONTR_X$} (23);
    \draw (21) -- node [swap] {$\Tens{\Coder X}{\Coweak X}+\Tens{\Coweak
        X}{\Coder X}$} (23);
    \draw (32) -- node [swap] {$\DER_X$} (31);
    \draw (33) -- node [swap] {$\COCONTR_X$} (32);
    \draw (33) -- node {$\Tens{\Der X}{\Weak X}+\Tens{\Weak X}{\Der X}$} (31);
  \end{tikzpicture}
\end{center}

and
\begin{center}
     \begin{tikzpicture}[->, >=stealth]
    \node (1) {$X$};
    \node (1r) [right of=1, node distance=16mm] {};
    \node (2) [below of=1r, node distance=6mm]{$\Excl X$};
    \node (3) [below of=1, node distance=12mm]{$X$};
    \tikzstyle{every node}=[midway,auto,font=\scriptsize]
    \draw (1) -- node {$\CODER_X$} (2);
    \draw (2) -- node {$\DER_X$} (3);
    \draw (1) -- node [swap] {$\Id_X$} (3);
  \end{tikzpicture}
\end{center}

\paragraph{The \emph{why not} modality.}
We define $\Int X=\Orth{(\Excl{(\Orth X)})}$ and we extend this operation to a
functor $\Groupoid\cL\to\Groupoid\cL$ in the same way (using the contravariant
functoriality of $\Orth{(\_)}$). We define 
\[
\Weak X'=\Orth{\Weak{\Orth X}}:\Bot\to\Int X\,.
\]
Since $\Contr{\Orth X}:\Excl{(\Orth X)}\to\Tens{\Excl{(\Orth X)}}{\Excl{(\Orth
    X)}}$, we have $\Orth{\Contr{\Orth X}}:\Orth{\Tensp{\Excl{(\Orth
      X)}}{\Excl{(\Orth X)}}}\to\Int X$. But $\Biorthiso_{\Excl{(\Orth
    X)}}:\Excl{(\Orth X)}\to\Orth{(\Int X)}$, hence
$\Orth{\Tensp{\Biorthiso_{\Excl{(\Orth X)}}}{\Biorthiso_{\Excl{(\Orth
        X)}}}}:\Par{\Int X}{\Int X}\to\Orth{\Tensp{\Excl{(\Orth
      X)}}{\Excl{(\Orth X)}}}$ and we set
\[
\Contr X'=\Orth{\Contr{\Orth X}}\Complin\Orth{\Tensp{\Biorthiso_{\Excl{(\Orth
        X)}}}{\Biorthiso_{\Excl{(\Orth X)}}}}\in\cL(\Par{\Int X}{\Int X},\Int
X)\,.
\]
Then it can be shown that $(\Int X,\Weak X',\Contr X')$ is a commutative
$\IPar$-monoid (that is, a monoid in the monoidal category $(\cL,\IPar)$). Of
course, $\Weak X'$ and $\Contr X'$ are natural transformations.

Last, we have $\Der{\Orth X}:\Excl{(\Orth X)}\to\Orth X$ and hence
$\Orth{(\Der{\Orth X})}:\Biorth X\to\Int X$, so we can define the natural
morphism
\[
\Der X'=\Orth{(\Der{\Orth X})}\Complin\Biorthiso_X:X\to\Int X\,.
\]

\paragraph{Interpreting \DILLZ{} derivations.}
We extend the interpretation of derivations presented in
Section~\ref{par:MLL-der-interp} to the fragment \DILLZ{} presented in
Section~\ref{sec:DILLZ}. 

We first have to extend the interpretation of formulas --~this is done in the
obvious way~-- and the definition of the De Morgan isomorphisms.  We have
$\Typesem{\Orth{(\Excl A)}}=\Int{\Typesem{\Orth A}}$ and $\Orth{\Typesem{\Excl
    A}}=\Orth{(\Excl{\Typesem A})}$. By inductive hypothesis, we have the iso
$\Demorgan_A:\Typesem{\Orth A}\to\Orth{\Typesem A}$, hence
$\Int{\Demorgan_A}:\Typesem{\Orth{(\Excl A)}}=\Typesem{\Int{(\Orth
    A)}}\to\Int{(\Orth{\Typesem A})}=\Orth{\Excl{(\Biorth{\Typesem A})}}$ and
since we have $\Orth{(\Excl{\Biorthiso_{\Typesem
      A}})}:\Orth{\Excl{(\Biorth{\Typesem A})}}\to\Orth{(\Excl{\Typesem A})}$,
we set
\[
\Demorgan_{\Excl A}=
\Orth{(\Excl{\Biorthiso_{\Typesem A}})}\Complin\Int{\Demorgan_A}\in\Groupoid\cL(\Typesem{\Orth{(\Excl A)}},\Orth{\Typesem{\Excl A}})\,.
\]

We have $\Typesem{\Orth{(\Int A)}}=\Excl{\Typesem{\Orth A}}$ and
$\Orth{\Typesem{\Int A}}=\Biorth{(\Excl{(\Orth{\Typesem A})})}$ so we set
\begin{equation*}
  \Demorgan_{\Int A}=\Biorthiso_{\Excl{(\Orth{\Typesem A})}}\Complin
  \Excl{\Demorgan_A}\in\Groupoid\cL(\Typesem{\Orth{(\Int A)}},
  \Orth{\Typesem{\Int A}})
\end{equation*}

Let $\pi$ be a derivation of $\Logic {\Phi}p\Gamma$, where $\Gamma=(\List
A1n)$.

Assume first that $\Gamma=(\Delta,\Int A)$, $p=\Net{\Vect s,\WEAK}{\Vect c}$
and that $\pi$ is the following derivation, denoting with $\lambda$ the
derivation of the premise:
\begin{center}
  \AxiomC{$\Logic{\Phi}{\Net{\Vect s}{\Vect c}}{\Gamma}$}
  \RightLabel{\quad\WEAKENING}
  \UnaryInfC{$\Logic{\Phi}{\Net{\Vect s,\WEAK}{\Vect c}}{\Gamma,\Int A}$}
  \DisplayProof
\end{center}
By inductive hypothesis we have
$\Dersem\lambda\in\Vectors\cL{\Typesem{\Gamma}}$. Let $\tau\in\Trees n$, we
have $\Dersem\lambda_\tau\in
\cL(\One,\TreeExt\IPar\tau(\Typesem{\Gamma}))$. 
We have $\Weak{\Typesem A}'\in\cL(\Bot,\Int{\Typesem A})$ and hence
\[
\Par{\Dersem\lambda_\tau}{\Weak{\Typesem A}'}
\in\cL(\Par\One\Bot,\TreeExt{\IPar}{\TreeB\tau\TreeO}{(\Typesem{\Gamma,\Int
    A})})
\]
so that we can set
\begin{equation*}
  \Dersem\pi_{\theta}=
  \Treeiso\IPar{\TreeB{\tau}{\TreeO}}\theta
  (\Par{\Dersem\lambda_\tau}{\Weak{\Typesem A}'})\Complin
  \Treeiso\IPar{\TreeO}{\TreeB{\TreeO}{\TreeZ}}
\end{equation*}
for any $\theta\in\Trees n$. 
The fact that the family $\Dersem\pi$ defined in that way does not depend on
the choice of $\tau$ results from the fact that
$\Dersem\lambda\in\Vectors\cL{\Typesem\Gamma}$.

Assume that $\Gamma=(\Delta,\Int A)$, $p=\Net{\Vect s,\CONTR(t_1,t_2)}{\Vect
  c}$ and that $\pi$ is the following derivation, denoting with $\lambda$ the
derivation of the premise:
\begin{center}
  \AxiomC{$\Logic{}{\Net{\Vect s,t_1,t_2}{\Vect c}}{\Delta,\Int A,\Int A}$}
  \RightLabel{\quad\CONTRACTION}
  \UnaryInfC{$\Logic{}{\Net{\Vect s,\CONTR(t_1,t_2)}{\Vect c}}{\Delta,\Int A}$}
  \DisplayProof
\end{center}
We have $\Contr{\Typesem A}'\in\cL(\Par{\Typesem{\Int A}}{\Typesem{\Int
    A}},\Typesem{\Int A})$. By inductive hypothesis
$\Dersem\lambda\in\Vectors\cL{\Typesem{\Delta,\Int A,\Int A}}$. 
Let
$\tau\in\Trees n$ where $n$ is the length of $\Delta$. 
We have
\[
\Dersem\lambda_{\TreeB\tau{\TreeB\TreeO\TreeO}}
\in\cL(\One,\Par {(\TreeExt\IPar\tau(\Dersem\Delta))}{\Parp{\Dersem{\Int
      A}}{\Dersem{\Int A}}})
\] 
and hence, given $\theta\in\Trees{n+1}$, we set
\begin{equation*}
  \Dersem\pi_\theta=
  \Treeiso\IPar{\TreeB{\tau}{\TreeO}}{\theta} 
  \Complin
  \Parp{\TreeExt\IPar\tau(\Dersem\Delta)}{\Contr{\Typesem A}'}
  \Complin
  \Dersem\lambda_{\TreeB\tau{\TreeB\TreeO\TreeO}}
\end{equation*}
defining in that way 
$\Dersem\pi\in\Vectors\cL{\Typesem{\Delta,\Int A}}$.

Assume that $\Gamma=(\Excl A)$, $p=\Net{\COWEAK}{}$ and that $\pi$ is the
following derivation
\begin{center}
  \AxiomC{}
  \RightLabel{\quad\COWEAKENING}
  \UnaryInfC{$\Logic{}{\Net{\WEAK}{}}{\Excl A}$}
  \DisplayProof
\end{center}
then, for $\theta\in\Trees 1$ we set
$\Dersem\pi_\theta=\Treeiso{\IPar}{\TreeO}{\theta}(\Typesem{\Excl A})
\Complin\Coweak{\Typesem A}$ defining in that way an element $\Dersem\pi$ of
$\Vectors\cL{\Typesem{\Excl A}}$.

Assume that $\Gamma=(\Delta,\Lambda,\Excl A)$, $p=\Net{\Vect s,\Vect
  t,\COCONTR(u,v)}{\Vect c,\Vect d}$ and that $\pi$ is the following derivation
\begin{center}
  \AxiomC{$\Logic{\Phi}{\Net{\Vect s,u}{\Vect c}}{\Delta,\Excl A}$}
  \AxiomC{$\Logic{\Phi}{\Net{\Vect t,v}{\Vect d}}{\Lambda,\Excl A,}$}
  \RightLabel{\quad\COCONTRACTION}
  \BinaryInfC{$\Logic{\Phi}{\Net{\Vect s,\Vect t,\COCONTR(u,v)}{\Vect
        c,\Vect d}}{\Delta,\Lambda,\Excl A}$} \DisplayProof
\end{center}
and we denote with $\lambda$ and $\rho$ the derivations of the two premises. By
inductive hypothesis, we have
$\Dersem\lambda\in\Vectors\cL{\Typesem{\Delta},\Typesem{\Excl A}}$ and
$\Dersem\rho\in\Vectors\cL{\Typesem{\Lambda},\Typesem{\Excl A}}$. We have
$\Tensvect{\Dersem\lambda}{\Dersem\rho}
\in\Vectors\cL{\Typesem\Delta,\Typesem\Lambda,\Tens{\Typesem{\Excl
      A}}{\Typesem{\Excl A}}}$.

Let $m$ be the length of $\Delta$ and $n$ be the length of $\Lambda$.  Let
$\tau\in\Trees{m+n}$, we have
$\Tensvect{\Dersem\lambda}{\Dersem\rho}_{\TreeB{\tau}{\TreeO}}
\in\cL(\One,\Par{\TreeExtp\IPar\tau{\Typesem\Delta,\Typesem\Lambda}}
{(\Tensp{\Typesem{\Excl A}} {\Typesem{\Excl A}}})$.  Hence, given
$\theta\in\Trees{m+n+1}$ we set
\begin{equation*}
  \Dersem\pi_\theta=
  \Treeiso\IPar{\TreeB{\tau}{\TreeO}}{\theta}
  \Complin
  \Parp{\TreeExtp\IPar\tau{\Typesem\Delta,\Typesem\Lambda}}{\Cocontr
    X}\Complin
  \Tensvect{\Dersem\lambda}{\Dersem\rho}_{\TreeB{\tau}{\TreeO}}\,.
\end{equation*}
so that
$\Dersem\pi\in\Vectors\cL{\Typesem{\Delta},\Typesem{\Lambda},\Typesem{\Excl
    A}}$, and this definition does not depend on the choice of $\tau$.

Assume that $\Gamma=(\Delta,\Int A)$, $p=\Net{\Vect s,\DER(s)}{\Vect c}$ and
that $\pi$ is the following derivation
\begin{center}
  \AxiomC{$\Logic{\Phi}{\Net{\Vect s,s}{\Vect c}}{\Delta,A}$}
  \RightLabel{\quad\DERELICTION}
  \UnaryInfC{$\Logic{\Phi}{\Net{\Vect s,\DER(s)}{\Vect c}}{\Delta,\Int A}$}
  \DisplayProof
\end{center}
Let $\lambda$ be the derivation of the premise, so that
$\Dersem\lambda\in\Vectors\cL{\Typesem\Delta,\Typesem A}$.

We have $\Der{\Typesem A}':\Typesem A\to\Typesem{\Int A}$. Let $n$ be the length
of $\Delta$, let $\tau\in\Trees n$ and let $\theta\in\Trees{n+1}$. We set
\begin{equation*}
  \Dersem\pi_\theta
  =\Treeiso\IPar{\TreeB{\tau}{\TreeO}}{\theta}
  \Complin
  \Parp{\TreeExtp\IPar\tau{\Typesem{\Delta}}}{\Der{\Typesem A}'}\Complin
  \Dersem\lambda_{\TreeB\tau\TreeO}
\end{equation*}
and we define in that way an element $\pi$ of
$\Vectors\cL{\Typesem\Delta,\Typesem{\Int A}}$ which does not depend on the
choice of $\tau$.

Assume that $\Gamma=(\Delta,\Excl A)$, $p=\Net{\Vect s,\CODER(s)}{\Vect c}$ and
that $\pi$ is the following derivation
\begin{center}
  \AxiomC{$\Logic{\Phi}{\Net{\Vect s,s}{\Vect c}}{\Delta,A}$}
  \RightLabel{\quad\CODERELICTION}
  \UnaryInfC{$\Logic{\Phi}{\Net{\Vect s,\CODER(s)}{\Vect c}}{\Delta,\Excl A}$}
  \DisplayProof
\end{center}
Let $\lambda$ be the derivation of the premise, so that
$\Dersem\lambda\in\Vectors\cL{\Typesem\Delta,\Typesem A}$.  We have
$\Coder{\Typesem A}:\Typesem A\to\Typesem{\Excl A}$. Let $n$ be the length of
$\Delta$, let $\tau\in\Trees n$ and let $\theta\in\Trees{n+1}$. We set
\begin{equation*}
  \Dersem\pi_\theta
  =\Treeiso\IPar{\TreeB{\tau}{\TreeO}}{\theta}
  \Complin
  \Parp{\TreeExtp\IPar\tau{\Typesem{\Delta}}}{\Coder{\Typesem A}}\Complin
  \Dersem\lambda_{\TreeB\tau\TreeO}
\end{equation*}
and we define in that way an element $\pi$ of
$\Vectors\cL{\Typesem\Delta,\Typesem{\Excl A}}$ which does not depend on the
choice of $\tau$.

Last assume that $p=\sum_{i=1}^n\mu_ip_i$, that $\pi$ is the following
derivation
\begin{center}
    \AxiomC{$\Logic{\Phi}{p_i}{\Gamma}\quad\forall i\in\{1,\dots,n\}$}
   \RightLabel{\quad\SUMRULE}
   \UnaryInfC{$\Logic{\Phi}{\sum_{i=1}^n\mu_ip_i}\Gamma$}
   \DisplayProof
\end{center}
and that $\lambda_i$ is the derivation of the $i$-th premise in this
derivation. Then by inductive hypothesis we have
$\Dersem{\lambda_i}\in\Vectors\cL{\Typesem\Delta}$ and we set of course
\[
\Dersem\pi=\sum_{i=1}^n\mu_i\Dersem{\lambda_i}\,.
\]

One can prove for this extended interpretation the same results as for the
\MLL{} fragment.
\begin{theorem}
  Let $\pi$ and $\pi'$ be derivations of $\Logic{\Phi}p\Gamma$. Then
  $\Dersem\pi=\Dersem{\pi'}$. 
\end{theorem}
Again, we set $\Netsem p=\Dersem\pi$ where $\pi$ is a derivation of
$\Logic{\Phi}p\Gamma$.

\begin{theorem}
  Assume that $\Logic{\Phi}p\Gamma$, $\Logic{\Phi}{p'}\Gamma$ and that
  $p\Rel\Red p'$. Then $\Netsem p=\Netsem{p'}$.
\end{theorem}

\subsection{Functorial exponential}
\label{sec:finctorial-exp}

Let $\cL$ be a preadditive *-autonomous category with an exponential
structure. A \emph{promotion} operation on $\cL$ is given by an extension of
the functor $\Excl{\_}$ to all morphisms of $\cL$ and by a lax symmetric
monoidal comonad structure on the ``$\oc$'' operation which satisfies
additional conditions. More precisely:
\begin{itemize}
\item For each $f\in\cL(X,Y)$ we are given a morphism $\Excl f\in\cL(\Excl
  X,\Excl Y)$ and the correspondence $f\mapsto\Excl f$ is functorial. This
  mapping $f\mapsto\Excl f$ extends the action of $\Excl\_$ on isomorphisms.
\item The morphisms $\Der X$, $\Coder X$, $\Weak X$, $\Coweak X$, $\Contr X$
  and $\Cocontr X$ are natural with respect to this extended functor.
\item There is a natural transformation $\Digg X:\Excl X\to\Excl{\Excl X}$
  which turns $(\Excl X,\Der X,\Digg X)$ into a comonad.
\item There is a morphism $\ExpMonZ:\One\to\Excl\One$ and a natural
  transformation\footnote{These morphisms are not required to be isos, whence
    the adjective ``lax'' for the monoidal structure.}
  $\ExpMonB_{X,Y}:\Tens{\Excl X}{\Excl Y}\to\Excl{\Tensp XY}$ which satisfy the
  following commutations

\begin{equation}\label{eq:excl-mon-unit-left}
    \begin{tikzpicture}[->, >=stealth]
    \node (1) {$\Tens\One{\Excl X}$};
    \node (2) [right of=1, node distance=24mm] {$\Tens{\Excl\One}{\Excl X}$};
    \node (3) [right of=2, node distance=24mm]{$\Excl{(\Tens\One X)}$};
    \node (4) [below of=3, node distance=12mm]{$\Excl X$};
    \tikzstyle{every node}=[midway,auto,font=\scriptsize]
    \draw (1) -- node {$\Tens{\ExpMonZ}{\Excl X}$} (2);
    \draw (2) -- node {$\ExpMonB_{\One,X}$} (3);
    \draw (3) -- node 
      {$\Excl{\Treeiso\ITens{\TreeB\TreeZ\TreeO}{\TreeO}}$} (4);
    \draw (1) -- node [swap] 
      {$\Treeiso\ITens{\TreeB\TreeZ\TreeO}{\TreeO}$} (4);
  \end{tikzpicture}
\end{equation}


\begin{equation}\label{eq:excl-mon-unit-right}
    \begin{tikzpicture}[->, >=stealth]
    \node (1) {$\Tens{\Excl X}\One$};
    \node (2) [right of=1, node distance=24mm] {$\Tens{\Excl X}{\Excl\One}$};
    \node (3) [right of=2, node distance=24mm]{$\Excl{(\Tens X\One)}$};
    \node (4) [below of=3, node distance=12mm]{$\Excl X$};
    \tikzstyle{every node}=[midway,auto,font=\scriptsize]
    \draw (1) -- node {$\Tens{\Excl X}{\ExpMonZ}$} (2);
    \draw (2) -- node {$\ExpMonB_{X,\One}$} (3);
    \draw (3) -- node 
      {$\Excl{\Treeiso\ITens{\TreeB\TreeO\TreeZ}{\TreeO}}$} (4);
    \draw (1) -- node [swap] 
      {$\Treeiso\ITens{\TreeB\TreeO\TreeZ}{\TreeO}$} (4);
  \end{tikzpicture}
\end{equation}


\begin{equation}\label{eq:excl-mon-ass}
    \begin{tikzpicture}[->, >=stealth]
    \node (1) {$\Tens{\Tensp{\Excl{X}}{\Excl Y}}{\Excl Z}$};
    \node (2) [right of=1, node distance=38mm] 
      {$\Tens{\Excl X}{\Tensp{\Excl Y}{\Excl Z}}$};
    \node (3) [right of=2, node distance=38mm]
      {$\Tens{\Excl X}{\Excl{\Tensp YZ}}$};
    \node (4) [below of=1, node distance=12mm]
      {$\Tens{\Excl{\Tensp XY}}{\Excl Z}$};
    \node (5) [below of=2, node distance=12mm]
      {$\Excl{\Tensp{\Tensp{X}{Y}}{Z}}$};
    \node (6) [below of=3, node distance=12mm]
      {$\Excl{\Tensp{X}{\Tensp{Y}{Z}}}$};
    \tikzstyle{every node}=[midway,auto,font=\scriptsize]
    \draw (1) -- node 
      {$\Treeiso\ITens{\TreeB{\TreeB\TreeO\TreeO}{\TreeO}}
          {\TreeB{\TreeO}{\TreeB\TreeO\TreeO}}$} (2);
    \draw (2) -- node 
      {$\Tens{\Excl X}{\ExpMonB_{Y,Z}}$} (3);
    \draw (3) -- node 
      {$\ExpMonB_{X,\Tens YZ} $} (6);
    \draw (1) -- node [swap] 
      {$\Tens{\ExpMonB_{X,Y}}{\Excl Z} $} (4);
    \draw (4) -- node {$\ExpMonB_{\Tens XY,Z} $} (5);
    \draw (5) -- node {$\Excl{\Treeiso\ITens{\TreeB{\TreeB\TreeO\TreeO}{\TreeO}}
          {\TreeB{\TreeO}{\TreeB\TreeO\TreeO}}} $} (6);
  \end{tikzpicture}
\end{equation}


\begin{equation}\label{eq:excl-mon-sym}
    \begin{tikzpicture}[->, >=stealth]
    \node (1) {$\Tens{\Excl{X}}{\Excl{Y}}$};
    \node (2) [right of=1, node distance=28mm] 
      {$\Tens{\Excl{Y}}{\Excl{X}}$};
    \node (4) [below of=1, node distance=12mm]
      {$\Excl{\Tensp XY} $};
    \node (5) [below of=2, node distance=12mm]
      {$\Excl{\Tensp YX} $};
    \tikzstyle{every node}=[midway,auto,font=\scriptsize]
    \draw (1) -- node 
      {$\Tsym_{\Excl X,\Excl Y} $} (2);
    \draw (2) -- node 
      {$\ExpMonB_{Y,X} $} (5);
    \draw (1) -- node [swap] 
      {$\ExpMonB_{X,Y} $} (4);
    \draw (4) -- node {$\Excl{\Tsym_{X,Y}} $} (5);
  \end{tikzpicture}
\end{equation}

\item The following diagrams commute

\begin{equation*}
    \begin{tikzpicture}[->, >=stealth]
    \node (1) {$\Tens{\Excl{X}}{\Excl{Y}}$};
    \node (2) [right of=1, node distance=24mm] 
      {$\Exclp{\Tens XY} $};
    \node (5) [below of=2, node distance=12mm]
      {${\Tens XY} $};
    \node (11) [right of=1, node distance=52mm] {$\One$};
    \node (12) [right of=11, node distance=16mm] 
      {$\Excl{\One} $};
    \node (15) [below of=12, node distance=12mm]
      {${\One} $};
    \tikzstyle{every node}=[midway,auto,font=\scriptsize]
    \draw (1) -- node 
      {$\ExpMonB_{X,Y} $} (2);
    \draw (2) -- node 
      {$\Der{\Tens XY} $} (5);
    \draw (1) -- node [swap] 
      {$\Tens{\Der X}{\Der Y} $} (5);
    \draw (11) -- node 
      {$\ExpMonZ $} (12);
    \draw (12) -- node 
      {$\Der{\One} $} (15);
    \draw (11) -- node [swap] 
      {$\One $} (15);
  \end{tikzpicture}
\end{equation*}


\begin{equation*}
    \begin{tikzpicture}[->, >=stealth]
    \node (1) {$\Tens{\Excl{X}}{\Excl{Y}}$};
    \node (2) [right of=1, node distance=56mm] 
      {$\Exclp{\Tens XY} $};
    \node (3) [below of=1, node distance=12mm]
      {$\Tens{\Excl{\Excl X}}{\Excl{\Excl Y}}$};
    \node (4) [right of=3, node distance=28mm] 
      {$\Excl{\Tensp{\Excl X}{\Excl Y}} $};
    \node (5) [below of=2, node distance=12mm]
      {$\Excl{\Excl{\Tensp XY}} $};
    \node (11) [right of=1, node distance=78mm] {$\One$};
    \node (12) [right of=11, node distance=16mm] 
      {$\Excl{\One} $};
    \node (13) [below of=11, node distance=12mm] {$\Excl\One$};
    \node (15) [below of=12, node distance=12mm]
      {$\Excl{\Excl\One} $};
    \tikzstyle{every node}=[midway,auto,font=\scriptsize]
    \draw (1) -- node 
      {$\ExpMonB_{X,Y} $} (2);
    \draw (2) -- node 
      {$\Digg{\Tens XY} $} (5);
    \draw (1) -- node [swap] 
      {$\Tens{\Digg X}{\Digg Y} $} (3);
    \draw (3) -- node {$\ExpMonB_{\Excl X,\Excl Y} $} (4);
    \draw (4) -- node {$\Excl{\ExpMonB_{X,Y}} $} (5);
    \draw (11) -- node 
      {$\ExpMonZ $} (12);
    \draw (12) -- node 
      {$\Digg{\One} $} (15);
    \draw (11) -- node [swap] 
      {$\ExpMonZ $} (13);
    \draw (13) -- node 
      {$\Excl\ExpMonZ $} (15);
  \end{tikzpicture}
\end{equation*}

\end{itemize}
When these conditions hold, one says that $(\ExpMonZ,\ExpMonB)$ is a lax
symmetric monoidal structure on the comonad $(\oc,\Der{},\Digg{})$. 

\paragraph{Monoidality and structural morphisms.}
This monoidal structure must also be compatible with the structural
constructions.

\begin{equation*}
    \begin{tikzpicture}[->, >=stealth]
    \node (1) {$\Tens{\Excl{X}}{\Excl{Y}}$};
    \node (2) [right of=1, node distance=24mm] 
      {$\Exclp{\Tens XY} $};
    \node (3) [below of=1, node distance=12mm]
      {$\Tens\One\One$};
    \node (5) [below of=2, node distance=12mm]
      {$\One $};
    \node (11) [right of=1, node distance=52mm] {$\One$};
    \node (12) [right of=11, node distance=16mm] 
      {$\Excl{\One} $};
    \node (15) [below of=12, node distance=12mm]
      {${\One} $};
    \tikzstyle{every node}=[midway,auto,font=\scriptsize]
    \draw (1) -- node 
      {$\ExpMonB_{X,Y} $} (2);
    \draw (2) -- node 
      {$\Weak{\Tens XY} $} (5);
    \draw (1) -- node [swap] 
      {$\Tens{\Weak X}{\Weak Y} $} (3);
    \draw (3) -- node  
      {$\Treeiso\ITens{\TreeB\TreeZ\TreeZ}\TreeZ $} (5);
    \draw (11) -- node 
      {$\ExpMonZ $} (12);
    \draw (12) -- node 
      {$\Weak{\One} $} (15);
    \draw (11) -- node [swap] 
      {$\One $} (15);
  \end{tikzpicture}
\end{equation*}


\begin{equation*}
    \begin{tikzpicture}[->, >=stealth]
    \node (1) {$\Tens{\Excl{X}}{\Excl{Y}}$};
    \node (2) [right of=1, node distance=40mm] 
      {$\Tens{\Tensp{\Excl{X}}{\Excl{X}}}{\Tensp{\Excl{Y}}{\Excl{Y}}} $};
    \node (3) [below of=2, node distance=12mm]
      {$\Tens{\Tensp{\Excl{X}}{\Excl{Y}}}{\Tensp{\Excl{X}}{\Excl{Y}}} $};
    \node (4) [below of=1, node distance=24mm]
      {$\Excl{\Tensp XY}$};
    \node (5) [below of=3, node distance=12mm]
      {$\Tens{\Excl{\Tensp XY}}{\Excl{\Tensp XY}} $};
    \node (1c) [right of=3, node distance=32mm] {};
    \node (11) [above of=1c, node distance=6mm] {$\One$};
    \node (12) [right of=11, node distance=16mm] 
      {$\Tens\One\One $};
    \node (13) [below of=11, node distance=12mm]
      {$\Excl\One $};
    \node (14) [below of=12, node distance=12mm]
      {$\Tens{\Excl\One}{\Excl\One}$}; 
    \tikzstyle{every node}=[midway,auto,font=\scriptsize]
    \draw (1) -- node {$\Tens{\Contr X}{\Contr Y}$} (2);
    \draw (1) -- node [swap]
      {$\ExpMonB_{X,Y} $} (4);
    \draw (2) -- node 
      {$\Symiso\ITens{\phi}{\sigma} $} (3);
    \draw (3) -- node {$\Tens{\ExpMonB_{X,Y}}{\ExpMonB_{X,Y}} $} (5);
    \draw (4) -- node {$\Contr{\Tens XY}$} (5);
    \draw (11) -- node {$\Treeiso\ITens\TreeZ{\TreeB\TreeZ\TreeZ} $} (12);
    \draw (12) -- node {$\Tens\ExpMonZ\ExpMonZ $} (14);
    \draw (11) -- node [swap] {$\ExpMonZ $} (13);
    \draw (13) -- node {$\Contr\One$} (14);
  \end{tikzpicture}
\end{equation*}

where $\phi=(1,3,2,4)\in\Symgrp 4$ and
$\sigma=\TreeB{\TreeB{\TreeO}{\TreeO}}{\TreeB{\TreeO}{\TreeO}}$.

\paragraph{Monoidality and costructural morphisms.}
\label{par:monoidality-costruct}
We need the following diagrams to commute in order to validate the reduction
rules of \DILL.


\begin{equation*}
    \begin{tikzpicture}[->, >=stealth]
    \node (1) {$\Tens{\One}{\Excl{Y}}$};
    \node (2) [right of=1, node distance=32mm] 
      {$\Tens{\Excl{X}}{\Excl{Y}} $};
    \node (3) [below of=1, node distance=12mm]
      {$\Tens\One\One $};
    \node (4) [below of=3, node distance=12mm]
      {$\One$};
    \node (5) [below of=2, node distance=24mm]
      {$\Excl{\Tensp XY} $};
    \node (1c) [right of=3, node distance=68mm] {};
    \node (11) [above of=1c, node distance=18mm] 
      {$\Tens {\Tensp{\Excl X}{\Excl X}}{\Excl Y} $};
    \node (12) [right of=11, node distance=36mm] 
      {$\Tens{\Excl X}{\Excl Y}$};
    \node (13) [below of=11, node distance=12mm]
      {$\Tens{\Tensp{\Excl X}{\Excl X}}{\Tensp{\Excl Y}{\Excl Y}} $};
    \node (14) [below of=13, node distance=12mm]
      {$\Tens{\Tensp{\Excl X}{\Excl Y}}{\Tensp{\Excl X}{\Excl Y}} $}; 
    \node (15) [below of=14, node distance=12mm]
      {$\Tens{\Excl{\Tensp XY}}{\Excl{\Tensp XY}} $}; 
    \node (16) [below of=12, node distance=36mm] {$\Excl{\Tensp XY}$};
    \tikzstyle{every node}=[midway,auto,font=\scriptsize]
    \draw (1) -- node {$\Tens{\Coweak X}{\Excl Y}$} (2);
    \draw (1) -- node [swap] {$\Tens\One{\Weak Y} $} (3);
    \draw (3) -- node [swap] 
      {$\Treeiso\ITens{\TreeB{\TreeZ}{\TreeZ}}\TreeZ $} (4);
    \draw (4) -- node {$\Coweak{\Tens XY} $} (5);
    \draw (2) -- node {$\ExpMonB_{X,Y}$} (5);
    \draw (11) -- node {$\Tens{\Cocontr X}{\Excl Y} $} (12);
    \draw (12) -- node {$\ExpMonB_{X,Y}$} (16);
    \draw (11) -- node [swap] 
      {$\Tens{\Tensp{\Excl X}{\Excl X}}{\Contr Y} $} (13);
    \draw (13) -- node [swap] {$\Symiso\ITens{\phi}{\sigma}$} (14);
    \draw (14) -- node [swap] {$\Tens{\ExpMonB_{X,Y}}{\ExpMonB_{X,Y}}$} (15);
    \draw (15) -- node {$\Cocontr{\Tens XY}$} (16);
  \end{tikzpicture}
\end{equation*}

where $\phi=(1,3,2,4)\in\Symgrp 4$ and
$\sigma=\TreeB{\TreeB{\TreeO}{\TreeO}}{\TreeB{\TreeO}{\TreeO}}$.


\begin{equation*}
    \begin{tikzpicture}[->, >=stealth]
    \node (1) {$\Tens{X}{\Excl{Y}}$};
    \node (2) [right of=1, node distance=32mm] 
      {$\Tens{\Excl{X}}{\Excl{Y}} $};
    \node (4) [below of=1, node distance=12mm]
      {$\Tens XY$};
    \node (5) [below of=2, node distance=12mm]
      {$\Excl{\Tensp XY} $};
    \tikzstyle{every node}=[midway,auto,font=\scriptsize]
    \draw (1) -- node {$\Tens{\Coder X}{\Excl Y}$} (2);
    \draw (1) -- node [swap] {$\Tens X{\Der Y} $} (4);
    \draw (4) -- node {$\Coder{\Tens XY} $} (5);
    \draw (2) -- node {$\ExpMonB_{X,Y}$} (5);
  \end{tikzpicture}
\end{equation*}

\paragraph{Digging and structural morphisms.} We assume that $\Digg X$ is a
comonoid morphism from $(\Excl X,\Weak X,\Contr X)$ to $(\Excl{\Excl
  X},\Weak{\Excl X},\Contr{\Excl X})$, in other words, the following diagrams
commute.

\begin{equation*}
    \begin{tikzpicture}[->, >=stealth]
    \node (1) {$\Excl X$};
    \node (2) [right of=1, node distance=18mm] 
      {$\Excl{\Excl X} $};
    \node (3) [below of=2, node distance=12mm] {$\One$};
    \node (11) [right of=2, node distance=24mm] {$\Excl X$};
    \node (12) [right of=11, node distance=28mm] {$\Excl{\Excl X}$};
    \node (13) [below of=11, node distance=12mm] {$\Tens{\Excl X}{\Excl X} $};
    \node (14) [below of=12, node distance=12mm] 
      {$\Tens{\Excl{\Excl X}}{\Excl{\Excl X}} $};
    \tikzstyle{every node}=[midway,auto,font=\scriptsize]
    \draw (1) -- node {$\Digg X$} (2);
    \draw (2) -- node {$\Weak{\Excl X} $} (3);
    \draw (1) -- node [swap] {$\Weak X $} (3);
    \draw (11) -- node {$\Digg X$} (12);
    \draw (11) -- node [swap] {$\Contr X$} (13);
    \draw (12) -- node {$\Contr{\Excl X}$} (14);
    \draw (13) -- node {$\Tens{\Digg X}{\Digg X} $} (14);
  \end{tikzpicture}
\end{equation*}

\paragraph{Digging and costructural morphisms.}
\label{par:digging-costruct} 
It is not required that $\Digg X$ be a monoid morphism from $(\Excl X,\Coweak
X,\Cocontr X)$ to $(\Excl{\Excl X},\Coweak{\Excl X},\Cocontr{\Excl X})$, but
the following diagrams must commute.

\begin{equation*}
    \begin{tikzpicture}[->, >=stealth]
    \node (1) {$\One$};
    \node (2) [right of=1, node distance=18mm] 
      {${\Excl X} $};
    \node (3) [below of=1, node distance=12mm] {$\Excl\One$};
    \node (4) [below of=2, node distance=12mm] {$\Excl{\Excl X}$};
    \node (11) [right of=2, node distance=24mm] {$\Tens{\Excl X}{\Excl X}$};
    \node (12) [right of=11, node distance=22mm] {${\Excl X}$};
    \node (13) [right of=12, node distance=18mm] {$\Excl{\Excl X} $};
    \node (14) [below of=11, node distance=12mm] 
      {$\Tens{\Excl{\Excl X}}{\Excl{\Excl X}} $};
    \node (15) [below of=13, node distance=12mm] 
      {$\Excl{\Tensp{\Excl{X}}{\Excl{X}}} $};
    \tikzstyle{every node}=[midway,auto,font=\scriptsize]
    \draw (1) -- node {$\Coweak X$} (2);
    \draw (2) -- node {${\Digg X} $} (4);
    \draw (1) -- node [swap] {$\ExpMonZ $} (3);
    \draw (3) -- node [swap] {$\Excl{\Coweak X} $} (4);
    \draw (11) -- node {$\Cocontr X$} (12);
    \draw (12) -- node {$\Digg X$} (13);
    \draw (11) -- node [swap] {$\Tens{\Digg X}{\Digg X}$} (14);
    \draw (14) -- node {$\ExpMonB_{\Excl X,\Excl X} $} (15);
    \draw (15) -- node [swap] {$\Excl{\Cocontr X} $} (13);
  \end{tikzpicture}
\end{equation*}

In the same spirit, we need a last diagram to commute, which describes the
interaction between codereliction and digging.

\begin{equation*}
    \begin{tikzpicture}[->, >=stealth]
    \node (1) {$X$};
    \node (2) [right of=1, node distance=26mm] {${\Excl X}$};
    \node (3) [right of=2, node distance=28mm] {$\Excl{\Excl X} $};
    \node (4) [below of=1, node distance=12mm] 
      {$\Tens\One X $};
    \node (5) [below of=2, node distance=12mm] 
      {$\Tens{\Excl{X}}{\Excl{X}} $};
    \node (6) [below of=3, node distance=12mm] 
      {$\Tens{\Excl{\Excl X}}{\Excl{\Excl X}}$};
    \tikzstyle{every node}=[midway,auto,font=\scriptsize]
    \draw (1) -- node {$\Coder X$} (2);
    \draw (2) -- node {$\Digg X$} (3);
    \draw (1) -- node [swap] 
      {$\Treeiso{\ITens}{\TreeO}{\TreeB{\TreeZ}{\TreeO}} $} (4);
    \draw (4) -- node {$\Tens{\Coweak X}{\Coder X} $} (5);
    \draw (5) -- node {$\Tens{\Digg X}{\Coder{\Excl X}} $} (6);
    \draw (6) -- node [swap] {$\Excl{\Cocontr X} $} (3);
  \end{tikzpicture}
\end{equation*}

\paragraph{Preadditive structure and functorial exponential.} 
\label{par:add-funct-exp}
Our last requirement justifies the term ``exponential'' since it expresses that
sums are turned into products by this functorial operation.

\begin{equation*}
    \begin{tikzpicture}[->, >=stealth]
    \node (1) {$\Excl X$};
    \node (1c) [right of=1, node distance=8mm] {};
    \node (2) [right of=1, node distance=16mm] 
      {${\Excl X} $};
    \node (3) [below of=1c, node distance=12mm] {$\One$};
    \node (11) [right of=2, node distance=24mm] {$\Excl X$};
    \node (12) [right of=11, node distance=26mm] {${\Excl Y}$};
    \node (13) [below of=11, node distance=12mm] {$\Tens{\Excl X}{\Excl X} $};
    \node (14) [below of=12, node distance=12mm] {$\Tens{\Excl Y}{\Excl Y} $};
    \tikzstyle{every node}=[midway,auto,font=\scriptsize]
    \draw (1) -- node {$\Excl 0$} (2);
    \draw (1) -- node [swap] {${\Weak X} $} (3);
    \draw (3) -- node [swap] {$\Coweak X $} (2);
    \draw (11) -- node {$\Excl{(f+g)} $} (12);
    \draw (11) -- node [swap] {$\Contr X$} (13);
    \draw (14) -- node [swap] {$\Cocontr X$} (12);
    \draw (13) -- node {$\Tens{\Excl f}{\Excl g} $} (14);
  \end{tikzpicture}
\end{equation*}

\begin{remark}
  There is another option in the categorical axiomatization of models of Linear
  Logic that we briefly describe as follows.
  \begin{itemize}
  \item One requires the linear category $\cL$ to be cartesian, with a terminal
    object $\Top$ and a cartesian product usually denoted as $\With{X_1}{X_2}$,
    projections $\Proj i\in\cL(\With{X_1}{X_2},X_i)$ and pairing
    $\Pair{f_1}{f_2}\in\cL(Y,\With{X_1}{X_2})$ for $f_i\in\cL(Y,X_i)$. This
    provides in particular $\cL$ with another symmetric monoidal structure.
  \item As above, one require the functor $\Excl\_$ to be a comonad. But we
    equip it now with a symmetric monoidal structure $(\SeelyZ,\SeelyB)$ from
    the monoidal category $(\cL,\IWith)$ to the monoidal category
    $(\cL,\ITens)$. This means in particular that
    $\SeelyZ\in\cL(\One,\Excl\Top)$ and
    $\SeelyB_{X_1,X_2}\in\cL(\Tens{\Excl{X_1}}{\Excl{X_2}},
    \Excl{(\With{X_1}{X_2})})$ are isos. These isos are often called
    \emph{Seely isos} in the literature, though Girard already stressed their
    significance in~\cite{Girard87}, admittedly not in the general
    categorical setting of monoidal comonads. An additional commutation is
    required, which describes the interaction between $\SeelyB$ and $\Digg{}$.
  \end{itemize}
  Using this structure, the comonad $(\Excl\_,\Der{},\Digg{})$ can be equipped
  with a lax symmetric monoidal structure $(\ExpMonZ,\ExpMonB)$. Again, our
  main reference for these notions and constructions is~\cite{Mellies09}. In
  this setting, the structural natural transformations $\Weak X$ and $\Contr X$
  can be defined and it is well known that the Kleisli category $\cL_\oc$ of
  the comonad $\Excl\_$ is cartesian closed.

  If we require the category $\cL$ to be preadditive in the sense of
  Section~\ref{sec:additive-lin-cat}, it is easy to see that $\Top$ is also an
  initial object and that $\IWith$ is also a coproduct. Using this fact, the
  natural transformations $\Coweak X$ and $\Cocontr X$ can also be defined.

  To describe a model of $\DILL$ in this setting, one has to require these
  Seely monoidality isomorphisms to satisfy some commutations with the
  $\Coder{}$ natural transformation.

  Here, we prefer a description which does not use cartesian products because
  it is closer to the basic constructions of the syntax of proof-structures and
  makes the presentation of the semantics conceptually simpler and more
  canonical, to our taste at least.
\end{remark}

\paragraph{Generalized monoidality, contraction and digging.}
\label{par:gen-monoidality}
Just as the monoidal structure of a monoidal category, the monoidal structure
of $\Excl\_$ can be parameterized by monoidal trees.  Let $n\in\Nat$ and let
$\tau\in\Trees n$. Given a family of objects $\Vect X=(\List X1n)$ of $\cL$, we
define $\ExpMonT\tau_{\Vect X}:\TreeExt\ITens\tau(\Excl{\Vect X})
\to\Excl{\TreeExt\ITens\tau(\Vect X)}$ by induction on $\tau$ as follows: 
\begin{align*}
 \ExpMonT\TreeZ &=\ExpMonZ \\
 \ExpMonT\TreeO_X &=\Id_{\Excl X} \\
 \ExpMonT{\TreeB\sigma\tau}_{\Vect X,\Vect
  Y} &=\ExpMonB_{\TreeExt\ITens\sigma(\Vect X),\TreeExt\ITens\tau(\Vect
  Y)}\Complin\Tensp{\ExpMonT\sigma_{\Vect X}}{\ExpMonT\tau_{\Vect Y}}\,.
\end{align*}

Given $\sigma,\tau\in\Trees n$ and $\phi\in\Symgrp n$, one can prove that the
following diagrams commute

\begin{equation*}
    \begin{tikzpicture}[->, >=stealth]
    \node (1) {$\TreeExt\ITens\sigma(\Excl{\Vect X}) $};
    \node (2) [right of=1, node distance=28mm] 
      {$\TreeExt\ITens\tau(\Excl{\Vect X}) $};
    \node (3) [below of=1, node distance=12mm] 
      {$\Excl{\TreeExt\ITens\sigma(\Vect X)} $};
    \node (4) [below of=2, node distance=12mm]
      {$\Excl{\TreeExt\ITens\tau(\Vect X)} $};
    \node (11) [right of=2, node distance=24mm] 
      {$\TreeExt\ITens\sigma(\Excl{\Vect X}) $};
    \node (12) [right of=11, node distance=32mm] 
      {$\TreeExt\ITens\sigma(\Symfunc\phi(\Excl{\Vect X})) $};
    \node (13) [below of=11, node distance=12mm] 
      {$\Excl{\TreeExt\ITens\sigma(\Vect X)} $};
    \node (14) [below of=12, node distance=12mm] 
      {$\Excl{\TreeExt\ITens\sigma(\Symfunc\phi(\Vect X))} $};
    \tikzstyle{every node}=[midway,auto,font=\scriptsize]
    \draw (1) -- node {$\Treeiso\ITens\sigma\tau(\Excl{\Vect X}) $} (2);
    \draw (2) -- node {$\ExpMonT\tau_{\Vect X} $} (4);
    \draw (1) -- node [swap] 
      {$\ExpMonT\sigma_{\Vect X} $} (3);
    \draw (3) -- node {$\Excl{\Treeiso\ITens\sigma\tau(\Excl{\Vect X})} $} (4);
    \draw (11) -- node {$\Symiso\ITens\phi\sigma(\Excl{\Vect X}) $} (12);
    \draw (12) -- node {$\ExpMonT\sigma_{\Symfunc\phi(\Vect X)} $} (14);
    \draw (11) -- node [swap] 
      {$\ExpMonT\sigma_{\Vect X} $} (13);
    \draw (13) -- node {$\Excl{\Symiso\ITens\phi\sigma(\Vect X)} $} (14);
  \end{tikzpicture}
\end{equation*}


\begin{equation*}
    \begin{tikzpicture}[->, >=stealth]
    \node (1) {$\TreeExt\ITens\sigma(\Excl{\Vect X}) $};
    \node (2) [right of=1, node distance=28mm] 
      {$\TreeExt\ITens\sigma{\Vect\One}=\TreeExt\ITens{\sigma_0} $};
    \node (3) [below of=1, node distance=12mm] 
      {$\Excl{\TreeExt\ITens\sigma(\Vect X)} $};
    \node (4) [below of=2, node distance=12mm]
      {$\One=\TreeExt\ITens\TreeZ{} $};
    \node (11) [right of=2, node distance=24mm] 
      {$\TreeExt\ITens\sigma(\Excl{\Vect X}) $};
    \node (12u) [right of=11, node distance=26mm] {}; 
    \node (12) [below of=12u, node distance=6mm]
      {$\TreeExt\ITens\sigma{\Vect X} $};
    \node (13) [below of=11, node distance=12mm] 
      {$\Excl{\TreeExt\ITens\sigma(\Vect X)} $};
    \tikzstyle{every node}=[midway,auto,font=\scriptsize]
    \draw (1) -- node {$\TreeExt\ITens\sigma(\WEAK_{\Vect X}) $} (2);
    \draw (2) -- node {$\Treeiso\ITens{\sigma_0}{\TreeZ} $} (4);
    \draw (1) -- node [swap] 
      {$\ExpMonT\sigma_{\Vect X} $} (3);
    \draw (3) -- node {$\WEAK_{\TreeExt\ITens\sigma(\Vect X)} $} (4);
    \draw (11) -- node {$\TreeExt\ITens\sigma(\DER_{\Vect X}) $} (12);
    \draw (13) -- node [swap]  {$\DER_{\TreeExt\ITens\sigma(\Vect X)} $} (12);
    \draw (11) -- node [swap] 
     {$\ExpMonT\sigma_{\Vect X} $} (13);
  \end{tikzpicture}
\end{equation*}
where $\Vect\One$ is the sequence $(\One,\dots,\One)$ ($n$ elements) and
$\sigma_0=\Subst\sigma\TreeZ\TreeO\in\Trees 0$ (the tree obtained from $\sigma$
by replacing each occurrence of $\TreeO$ by $\TreeZ$). 

Before stating the next commutation, we define a generalized form of
contraction $\Contr{\Vect X}^\sigma:\TreeExt\ITens\sigma{\Excl{\Vect
    X}}\to\TreeExt\ITens{\TreeB\sigma\sigma}{(\Excl{\Vect X},\Excl{\Vect X})}$
as the following composition of morphisms:
\begin{equation*}
    \begin{tikzpicture}[->, >=stealth]
    \node (1) {$\TreeExt\ITens\sigma\Excl{\Vect X} $};
    \node (2) [right of=1, node distance=22mm] 
      {$\TreeExt\ITens{\sigma_2}{\Excl{\Vect{X'}}} $};
    \node (3) [right of=2, node distance=28mm] 
      {$\TreeExt\ITens{\sigma_2}{(\Excl{\Vect X},\Excl{\Vect X})} $};
    \node (4) [right of=3, node distance=36mm] 
      {$\TreeExt\ITens{\TreeB\sigma\sigma}{(\Excl{\Vect X},\Excl{\Vect X})} $};
    \tikzstyle{every node}=[midway,auto,font=\scriptsize]
    \draw (1) -- node {$\TreeExt\ITens\sigma{\Contr{\Vect X}} $} (2);
    \draw (2) -- node {$\Symiso\ITens\phi\sigma $} (3);
    \draw (3) -- node {$\Treeiso\ITens{\sigma_2}{\TreeB\sigma\sigma} $} (4);
  \end{tikzpicture}
\end{equation*}
where $\Vect{X'}=(X_1,X_1,X_2,X_2,\dots,X_n,X_n)$,
$\sigma_2=\Subst\sigma{\TreeB\TreeO\TreeO}{\TreeO}$ and $\phi\in\Symgrp{2n}$ is
defined by $\phi(2i+1)=i+1$ and $\phi(2i+2)=i+n+1$ for
$i\in\{0,\dots,n-1\}$. With these notations, one can prove that

\begin{equation*}
    \begin{tikzpicture}[->, >=stealth]
    \node (1) {$\TreeExt\ITens\sigma\Excl{\Vect X} $};
    \node (2) [right of=1, node distance=48mm] 
      {$\TreeExt\ITens{\TreeB\sigma\sigma}{(\Excl{\Vect X},\Excl{\Vect X})}
        =\Tens{(\TreeExt\ITens\sigma\Excl{\Vect X})}
              {(\TreeExt\ITens\sigma\Excl{\Vect X})} $};
    \node (3) [below of=1, node distance=12mm] 
      {$\Excl{\TreeExt\ITens\sigma\Vect X} $};
    \node (4) [below of=2, node distance=12mm] 
      {$\Tens{(\Excl{\TreeExt\ITens\sigma\Vect X})}
        {(\Excl{\TreeExt\ITens\sigma\Vect X})} $};
    \tikzstyle{every node}=[midway,auto,font=\scriptsize]
    \draw (1) -- node {$\Contr{\Vect X}^\sigma $} (2);
    \draw (1) -- node [swap] {$\ExpMonT\sigma_{\Vect X} $} (3);
    \draw (2) -- node 
      {$\Tens{\ExpMonT\sigma_{\Vect X}}{\ExpMonT\sigma_{\Vect X}} $} (4);
    \draw (3) -- node {$\CONTR_{\TreeExt\ITens\sigma(\Vect X)} $} (4);
  \end{tikzpicture}
\end{equation*}


We also define a generalized version of digging
$\DiggT\sigma{\Vect X}:\TreeExt\ITens\sigma{\Excl{\Vect X}}
\to\Excl{\TreeExt\ITens\sigma{\Excl{\Vect X}}}$ as the following composition of
morphisms:
\begin{equation*}
    \begin{tikzpicture}[->, >=stealth]
    \node (1) {$\TreeExt\ITens\sigma\Excl{\Vect X} $};
    \node (2) [right of=1, node distance=22mm] 
      {$\TreeExt\ITens\sigma{\Excl{\Excl{\Vect X}}} $};
    \node (3) [right of=2, node distance=22mm] 
      {$\Excl{\TreeExt\ITens\sigma{\Excl{\Vect X}}} $};
    \tikzstyle{every node}=[midway,auto,font=\scriptsize]
    \draw (1) -- node {$\TreeExt\ITens\sigma{\Digg{\Vect X}} $} (2);
    \draw (2) -- node {$\ExpMonT\sigma_{\Excl{\Vect X}} $} (3);
  \end{tikzpicture}
\end{equation*}
With this notation, one can prove that
\begin{equation*}
    \begin{tikzpicture}[->, >=stealth]
    \node (1) {$\TreeExt\ITens\sigma\Excl{\Vect X} $};
    \node (2) [right of=1, node distance=22mm] 
      {$\Excl{\TreeExt\ITens\sigma{\Excl{\Vect X}}} $};
    \node (3) [below of=1, node distance=12mm] 
      {$\Excl{\TreeExt\ITens\sigma\Vect X} $};
    \node (4) [below of=2, node distance=12mm] 
      {$\Excl{\Excl{\TreeExt\ITens\sigma{\Vect X}}} $};
    \tikzstyle{every node}=[midway,auto,font=\scriptsize]
    \draw (1) -- node {$\DiggT\sigma{\Vect X} $} (2);
    \draw (2) -- node {$\Excl{\ExpMonT\sigma_{\Vect X}} $} (4);
    \draw (1) -- node [swap] {$\ExpMonT\sigma_{\Vect X} $} (3);
    \draw (3) -- node {$\Digg{\TreeExt\ITens\sigma\Vect X} $} (4);
  \end{tikzpicture}
\end{equation*}

We have $\DiggT{\TreeZ}{}=\ExpMonZ$, $\DiggT{\TreeO}{X}=\Digg X$, and observe
that the following generalizations of the comonad laws hold. The two
commutations involving digging and dereliction generalize to:
\begin{equation*}
    \begin{tikzpicture}[->, >=stealth]
    \node (1) {$\TreeExt\ITens\sigma\Excl{\Vect X} $};
    \node (2) [below of=1, node distance=14mm] 
      {$\Excl{\TreeExt\ITens\sigma{\Excl{\Vect X}}} $};
    \node (2l) [left of=2, node distance=24mm] 
      {$\TreeExt\ITens\sigma{\Excl{\Vect X}} $};
    \node (2r) [right of=2, node distance=24mm] 
      {$\Excl{\TreeExt\ITens\sigma{\Vect X}} $};
    \tikzstyle{every node}=[midway,auto,font=\scriptsize]
    \draw (1) -- node {$\DiggT\sigma{\Vect X} $} (2);
    \draw (1) -- node [swap] {$\TreeExt\ITens\sigma{\Excl{\Vect X}} $} (2l);
    \draw (1) -- node {$\ExpMonT\sigma_{\Vect X} $} (2r);
    \draw (2) -- node {$\Der{\TreeExt\ITens\sigma{\Excl{\Vect X}}} $} (2l);
    \draw (2) -- node [swap] 
      {$\Excl{\TreeExt\ITens\sigma{\Der{\Vect X}}} $} (2r);
  \end{tikzpicture}
\end{equation*}
The square diagram involving digging generalizes as follows.
Let $\Vect Y=(\List Y1m)$ be another list of objects and let $\tau\in\Trees
m$. One can prove that
\begin{equation*}
    \begin{tikzpicture}[->, >=stealth]
    \node (1) {$\TreeExt\ITens\sigma\Excl{\Vect X} $};
    \node (2) [right of=1, node distance=22mm] 
      {$\Excl{\TreeExt\ITens\sigma{\Excl{\Vect X}}} $};
    \node (3) [below of=1, node distance=12mm] 
      {$\Excl{\TreeExt\ITens\sigma{\Excl{\Vect X}}} $};
    \node (4) [below of=2, node distance=12mm] 
      {$\Excl{\Excl{\TreeExt\ITens\sigma{\Excl{\Vect X}}}} $};
    \tikzstyle{every node}=[midway,auto,font=\scriptsize]
    \draw (1) -- node {$\DiggT\sigma{\Vect X} $} (2);
    \draw (2) -- node {$\Excl{\DiggT\sigma{\Vect X}} $} (4);
    \draw (1) -- node [swap] {$\DiggT\sigma{\Vect X} $} (3);
    \draw (3) -- node {$\Digg{\TreeExt\ITens\sigma{\Excl{\Vect X}}} $} (4);
  \end{tikzpicture}
\end{equation*}
and then one can generalize this property as follows
\begin{equation}\label{eq:gen-monad-digg-digg-bis}
    \begin{tikzpicture}[->, >=stealth]
    \node (1) {$\Tens{(\TreeExt\ITens\sigma{\Excl{\Vect X}})}
           {(\TreeExt\ITens\tau{\Excl{\Vect Y}})} $};
    \node (2) [right of=1, node distance=48mm] 
      {$\Excl{(\Tens{(\TreeExt\ITens\sigma{\Excl{\Vect X}})}
                  {(\TreeExt\ITens\tau{\Excl{\Vect Y}})})} $};
    \node (3) [below of=1, node distance=12mm] 
      {$\Tens{\Excl{(\TreeExt\ITens\sigma{\Excl{\Vect X}})}}
           {(\TreeExt\ITens\tau{\Excl{\Vect Y}})} $};
    \node (4) [below of=2, node distance=12mm] 
      {$\Excl{(\Tens{\Excl{(\TreeExt\ITens\sigma{\Excl{\Vect X}})}}
                  {(\TreeExt\ITens\tau{\Excl{\Vect Y}})})} $};
    \tikzstyle{every node}=[midway,auto,font=\scriptsize]
    \draw (1) -- node {$\DiggT{\TreeB\sigma\tau}{\Vect X,\Vect Y} $} (2);
    \draw (2) -- node {$\Exclp{\Tens{\DiggT\sigma{\Vect X}}
                           {(\TreeExt\ITens\tau{\Excl{\Vect Y}})}} $} (4);
    \draw (1) -- node [swap] {$\Tens{\DiggT\sigma{\Vect X}}
                    {(\TreeExt\ITens\tau{\Excl{\Vect Y}})} $} (3);
    \draw (3) -- node {$\DiggT{\TreeB{\TreeO}{\tau}}
                     {\TreeExt\ITens\sigma{\Excl{\Vect X}},\Vect Y} $} (4);
  \end{tikzpicture}
\end{equation}

%
   
\paragraph{Generalized promotion and structural constructions.}
\label{par:gen-prom-struct}
Let $f:\TreeExt\ITens\sigma{\Excl{\Vect X}}\to Y$, we define the
\emph{generalized promotion} $\Prom f:\TreeExt\ITens\sigma{\Excl{\Vect
    X}}\to\Excl Y$ by $\Prom f=\Excl f\Complin\DiggT\sigma{\Vect X}$. Using the 
commutations of Section~\ref{par:gen-monoidality}, one can prove that this
construction obeys the following commutations.
\begin{equation*}
    \begin{tikzpicture}[->, >=stealth]
    \node (1) {$\TreeExt\ITens\sigma\Excl{\Vect X} $};
    \node (2) [right of=1, node distance=28mm] {$\Excl Y $};
    \node (3) [below of=1, node distance=12mm] 
      {$\TreeExt\ITens\sigma{\Vect\One}=\TreeExt\ITens{\sigma_0} $};
    \node (4) [below of=2, node distance=12mm] 
      {$\TreeExt\ITens\TreeZ=\One $};
    \tikzstyle{every node}=[midway,auto,font=\scriptsize]
    \draw (1) -- node {$\Prom f $} (2);
    \draw (2) -- node {$\Weak Y $} (4);
    \draw (1) -- node [swap] {$\TreeExt\ITens\sigma{\Weak{\Vect X}} $} (3);
    \draw (3) -- node {$\Treeiso\ITens{\sigma_0}{\TreeZ} $} (4);
  \end{tikzpicture}
\end{equation*}
with the same notations as before.

With these notations, we have
\begin{equation*}
    \begin{tikzpicture}[->, >=stealth]
    \node (1) {$\TreeExt\ITens\sigma\Excl{\Vect X} $};
    \node (2) [right of=1, node distance=32mm] {$\Excl Y $};
    \node (3) [below of=1, node distance=12mm] 
      {$\TreeExt\ITens{\TreeB\sigma\sigma}{(\Excl{\Vect X},\Excl{\Vect X})} $};
    \node (4) [below of=2, node distance=12mm] 
      {$\Tens{\Excl Y}{\Excl Y} $};
    \tikzstyle{every node}=[midway,auto,font=\scriptsize]
    \draw (1) -- node {$\Prom f $} (2);
    \draw (2) -- node {$\Contr Y $} (4);
    \draw (1) -- node [swap] {$\Contr{\Vect X}^\sigma $} (3);
    \draw (3) -- node {$\Tens{\Prom f}{\Prom f} $} (4);
  \end{tikzpicture}
\end{equation*}

The next two diagrams deal with the interaction between generalized promotion
and dereliction (resp.~digging).
\begin{equation*}
    \begin{tikzpicture}[->, >=stealth]
    \node (1) {$\TreeExt\ITens\sigma\Excl{\Vect X} $};
    \node (2) [right of=1, node distance=26mm] {$\Excl Y $};
    \node (3) [below of=2, node distance=12mm] 
      {$Y $};
    \tikzstyle{every node}=[midway,auto,font=\scriptsize]
    \draw (1) -- node {$\Prom f $} (2);
    \draw (2) -- node {$\Der Y $} (3);
    \draw (1) -- node [swap] {$f $} (3);
  \end{tikzpicture}
\end{equation*}
\begin{equation*}
    \begin{tikzpicture}[->, >=stealth]
    \node (1) {$\Tens{(\TreeExt\ITens\sigma{\Excl{\Vect
            X}})}{(\TreeExt\ITens\tau{\Excl{\Vect Y}})} $};
    \node (2) [right of=1, node distance=52mm] 
      {$\Tens{\Excl Y}{(\TreeExt\ITens\tau{\Excl{\Vect Y}})} $};
    \node (3) [below of=1, node distance=12mm] 
      {$\Excl{(\Tens{(\TreeExt\ITens\sigma{\Excl{\Vect
            X}})}{(\TreeExt\ITens\tau{\Excl{\Vect Y}})})} $};
    \node (4) [below of=2, node distance=12mm] 
      {$\Excl{(\Tens{\Excl Y}{(\TreeExt\ITens\tau{\Excl{\Vect Y}})})} $};
    \tikzstyle{every node}=[midway,auto,font=\scriptsize]
    \draw (1) -- node 
      {$\Tens{\Prom f}{(\TreeExt\ITens\tau{\Excl{\Vect Y}})} $} (2);
    \draw (2) -- node {$\DiggT{\TreeB{\TreeO}{\tau}}{Y,\Vect Y} $} (4);
    \draw (1) -- node [swap] {$\DiggT{\TreeB\sigma\tau}{\Vect X,\Vect Y}$} (3);
    \draw (3) -- node 
      {$\Exclp{\Tens{\Prom f}{(\TreeExt\ITens\tau{\Excl{\Vect Y}})}} $} (4);
  \end{tikzpicture}
\end{equation*}
The second diagram follows easily from~(\ref{eq:gen-monad-digg-digg-bis}) and
allows one to prove the following property. Let
$f:\TreeExt\ITens\sigma{\Vect{\Excl X}}\to Y$ and $g:\Tens{\Excl
  Y}{(\TreeExt\ITens\tau{\Vect{\Excl Y}})}\to Z$ so that $\Prom
f:\TreeExt\ITens\sigma{\Vect{\Excl X}}\to\Excl Y$ and $\Prom g:\Tens{\Excl
  Y}{(\TreeExt\ITens\tau{\Vect{\Excl Y}})}\to\Excl Z$; one has
\begin{equation*}
  \Prom g\Complin\Tensp{\Prom f}{(\TreeExt\ITens\tau{\Excl{\Vect Y}})}
  =\Prom{(g\Complin\Tensp{\Prom f}{(\TreeExt\ITens\tau{\Excl{\Vect Y}})})}:
  \Tens{(\TreeExt\ITens\sigma{\Vect{\Excl
    X}})}{(\TreeExt\ITens\tau{\Vect{\Excl Y}})}\to\Excl Z
\end{equation*}

\begin{remark}
  We actually need a more general version of this property, where $\Prom f$ is
  not necessarily in leftmost position in the $\ITens$ tree. It is also easy to
  obtain, but notations are more heavy. We use the same kind of convention in
  the sequel but remember that the corresponding properties are easy to
  generalize.
\end{remark}

\paragraph{Generalized promotion and costructural constructions.} 
\label{par:gen-prom-costruct}
Let $f:\Tens{\Excl X}{(\TreeExt\ITens\sigma{\Excl{\Vect X}})}\to Y$. Observe
that $f\Complin\Tensp{\Coweak X}{(\TreeExt\ITens\sigma{\Excl{\Vect
      X}})}\Complin\Treeiso\ITens\sigma{\TreeB\TreeZ\sigma}_{\Vect{\Excl
    X}}:\TreeExt\ITens\sigma{\Excl{\Vect X}}\to Y$. The following equation
holds:
\begin{equation*}
  \Prom f\Complin\Tensp{\Coweak X}{(\TreeExt\ITens\sigma{\Excl{\Vect
      X}})}\Complin\Treeiso\ITens\sigma{\TreeB\TreeZ\sigma}_{\Vect{\Excl
    X}}=\Prom{(f\Complin\Tensp{\Coweak X}{(\TreeExt\ITens\sigma{\Excl{\Vect
      X}})}\Complin\Treeiso\ITens\sigma{\TreeB\TreeZ\sigma}_{\Vect{\Excl
    X}})}
\end{equation*}
Similarly, we have $f\Complin\Tensp{\Cocontr
  X}{(\TreeExt\ITens\sigma{\Excl{\Vect
      X}})}:
\Tens{\Tensp{\Excl X}{\Excl X}}{(\TreeExt\ITens\sigma{\Excl{\Vect X}})}\to Y$
and the following equation holds:
\begin{equation*}
  \Prom f\Complin\Tensp{\Cocontr
    X}{(\TreeExt\ITens\sigma{\Excl{\Vect
        X}})}=
  \Prom {(f\Complin\Tensp{\Cocontr
    X}{(\TreeExt\ITens\sigma{\Excl{\Vect
        X}})})}
\end{equation*}
This results from the commutations of Sections~\ref{par:monoidality-costruct}
and~\ref{par:digging-costruct}.

\paragraph{Generalized promotion and codereliction (also known as \emph{chain
    rule}).} 
\label{par:gen-prom-coder}
Let $f:\Tens{\Excl X}{(\TreeExt\ITens\sigma{\Excl{\Vect X}})}\to Y$. We set
\[
f_0=f\Complin\Tensp{\Coweak X}{(\TreeExt\ITens\sigma{\Excl{\Vect
      X}})}\Complin\Treeiso\ITens\sigma{\TreeB\TreeZ\sigma}
:\TreeExt\ITens\sigma{\Excl{\Vect X}}\to Y
\]
Then we have
\begin{equation*}
    \begin{tikzpicture}[->, >=stealth]
    \node (1) {$\Tens{X}{(\TreeExt\ITens\sigma{\Excl{\Vect X}})} $};
    \node (2) [right of=1, node distance=42mm] 
      {$\Tens{\Excl X}{(\TreeExt\ITens\sigma{\Excl{\Vect X}})} $};
    \node (3) [right of=2, node distance=28mm] {$\Excl Y $};
    \node (4) [below of=1, node distance=12mm] 
      {$\Tens{\Excl X}{\Tensp{(\TreeExt\ITens\sigma{\Excl{\Vect X}})}
                         {(\TreeExt\ITens\sigma{\Excl{\Vect X}}})} $};
    \node (5) [below of=4, node distance=12mm] 
      {$\Tens{\Tensp{\Excl X}
                {(\TreeExt\ITens\sigma{\Excl{\Vect X}}})}
         {(\TreeExt\ITens\sigma{\Excl{\Vect X}})} $};
    \node (6) [below of=2, node distance=24mm]
      {$\Tens Y{\Excl Y} $};
    \node (7)  [below of=3, node distance=24mm]
      {$\Tens{\Excl Y}{\Excl Y} $};
    \tikzstyle{every node}=[midway,auto,font=\scriptsize]
    \draw (1) -- node 
      {$\Tens{\Coder X}{(\TreeExt\ITens\sigma{\Excl{\Vect X}})} $} (2);
    \draw (2) -- node {$\Prom f $} (3);
    \draw (1) -- node [swap] {$\Tens{\Coder X}{\Contr{\Vect X}^\sigma} $} (4);
    \draw (4) -- node [swap] 
      {$\Treeiso\ITens{\TreeB{\TreeO}{\TreeB\TreeO\TreeO}}
                           {\TreeB{\TreeB\TreeO\TreeO}{\TreeO}} $} (5);
    \draw (5) -- node {$\Tens f{\Prom{f_0}} $} (6);
    \draw (6) -- node {$\Tens{\Coder Y}{\Excl Y} $} (7);
    \draw (7) -- node [swap] {$\Cocontr Y $} (3);
  \end{tikzpicture}
\end{equation*}
This results from the commutations of Sections~\ref{par:monoidality-costruct}
and~\ref{par:digging-costruct}.

\paragraph{Interpreting $\DILL$ derivations.}
For the sake of readability, we assume here that the De Morgan isomorphisms
(see~\ref{par:MLL-der-interp}) are identities, so that $\Typesem{\Orth
  A}=\Orth{\Typesem A}$ for each formula $A$. The general definition of the
semantics can be obtained by inserting De Morgan isomorphisms at the correct
places in the forthcoming expressions.

Let $P$ be a net of arity $n+1$ and let $p_i=\Net{\Vect{t_i},t_i}{\Vect{c_i}}$
  for $i=1,\dots,n$. Consider the following derivation $\pi$, where we denote
  as $\lambda,\List\rho 1n$ the derivations of the premises.
\begin{center}
  \AxiomC{$\Logic{\Phi}{P}
    {\Int{\Orth{A_1}},\dots,\Int{\Orth{A_n}},B}$}
  \AxiomC{$\Logic{\Phi}{p_1}
    {\Gamma_1,\Excl{A_1}}\ 
    \cdots\ \Logic{\Phi_n}{p_n}{\Gamma_n,\Excl{A_n}}$}
  \BinaryInfC{$\Logic{\Phi}
    {\Net{\Vect{t_1},\dots,\Vect{t_n},\PROM n{P}(\List t1n)}
    {\Vect{c_1},\dots,\Vect{c_n}}}{\Gamma_1,\dots,\Gamma_n,\Excl B}$}
  \DisplayProof
\end{center}
By inductive hypothesis, we have $\Dersem\lambda\in\Vectors\cL{\Orth{(\Excl{\Typesem{A_1}})},\dots,\Orth{(\Excl{\Typesem{A_n}})},\Typesem
  B}$ so
that, picking an element $\sigma$ of $\Trees n$ we have
\begin{align*}
\Dersem\lambda_{\TreeB\sigma\TreeO} 
& \in\cL(\One,\Par{\TreeExt\IPar{\sigma}(\Orth{(\Excl{\Typesem{A_1}})},
\dots,\Orth{(\Excl{\Typesem{A_n}})})}{\Typesem
B})\\
&\quad\quad=\cL(\One,\Limpl{\TreeExt\ITens\sigma(\Excl{\Typesem{A_1}},
\dots,\Excl{\Typesem{A_n}})}{\Typesem B})
\end{align*}
and hence
\[\Prom{(\Funinv\Curlin(\Dersem\lambda_{\TreeB\sigma\TreeO})
\Complin\Treeiso\ITens{\sigma}{\TreeB\TreeZ\sigma})}
\in\cL(\TreeExt\ITens\sigma(\Excl{\Typesem{A_1}},
\dots,\Excl{\Typesem{A_n}}),\Excl{\Typesem B})\,.
\]

For $i=1,\dots,n$, we have
$\Dersem{\rho_i}\in\Vectors{\cL}{\Typesem{\Gamma_i},\Excl{\Typesem{A_i}}}$. Let
$l_i$ be the length of $\Gamma_i$, and let us choose
$\tau_i\in\Trees{l_i}$. 
We have
$\Dersem{\rho_i}_{\TreeB{\tau_i}\TreeO}\in\cL(\One,\Par{\TreeExt\IPar{\tau_i}{(\Typesem{\Gamma_i})}}{\Excl{\Typesem{A_i}}})$
and hence, setting
\[
r_i=\Funinv\Curlin(\Dersem{\rho_i}_{\TreeB{\tau_i}\TreeO})\Complin
\Treeiso\ITens{\tau_i}{\TreeB\TreeZ{\tau_i}}\in\cL(\TreeExt{\ITens}{\tau_i}{\Orth{(\Typesem{\Gamma_i})}},\Excl{\Typesem{A_i}})
\]
we have $\TreeExt\ITens\sigma{(\Vect
  r)}\in\cL(\TreeExt\ITens\theta{(\Orth{\Typesem\Delta})},
\TreeExt\ITens\sigma{(\Excl{\Typesem{A_1}},
\dots,\Excl{\Typesem{A_n}})})$ where
\begin{align*}
  \Delta &=\Gamma_1,\dots,\Gamma_n\\
  \theta &=\sigma(\List\tau 1n)
\end{align*}
where $\sigma(\List\tau 1n)$ (for $\sigma\in\Trees n$ and $\tau_i\in\Trees{n_i}$
for $i=1,\dots,k$) is the element of $\Trees{n_1+\cdots+n_k}$ defined
inductively by
\begin{align*}
  \TreeZ()&=\TreeZ\\
  \TreeO(\tau)&=\tau\\
  \TreeB{\sigma}{\sigma'}(\List\tau
  1n)&=\TreeB{\sigma(\List\tau1k)}{\sigma'(\List\tau{k+1}n)}\\
  &\hspace{4em}\text{where }\sigma\in\Trees k,\ \sigma'\in\Trees{n-k}
\end{align*}
We have therefore
\[\Prom{(\Funinv\Curlin(\Dersem\lambda_{\TreeB\sigma\TreeO})
\Complin\Treeiso\ITens{\sigma}{\TreeB\TreeZ\sigma})}
\Complin\TreeExt\ITens\sigma{(\Vect r)}\in
\cL(\TreeExt\ITens\theta{(\Orth{\Typesem\Delta})},\Typesem B)
\]
We set
\[
\Dersem\pi_\theta=
\Curlin(\Prom{(\Funinv\Curlin(\Dersem\lambda_{\TreeB\sigma\TreeO})
\Complin\Treeiso\ITens{\sigma}{\TreeB\TreeZ\sigma})}
\Complin\TreeExt\ITens\sigma{(\Vect r)}\Complin
\Treeiso\ITens{\TreeB\TreeZ\theta}\theta)
\in\cL(\One,\TreeExt\IPar{\TreeB\theta\TreeO}(\Typesem{\Delta,\Excl B}))
\]
and this gives us a definition of
$\Typesem\pi\in\Vectors\cL{\Typesem{\Delta,\Excl B}}$ which does not depend on
the choice of $\theta$.

\begin{theorem}
  Let $\pi$ and $\pi'$ be derivations of $\Logic{\Phi}p\Gamma$. Then
  $\Dersem\pi=\Dersem{\pi'}$. 
\end{theorem}
Again, we set $\Netsem p=\Dersem\pi$ where $\pi$ is a derivation of
$\Logic{\Phi}p\Gamma$.

\begin{theorem}
  Assume that $\Logic{\Phi}p\Gamma$, $\Logic{\Phi}{p'}\Gamma$ and that
  $p\Rel\Red p'$. Then $\Netsem p=\Netsem{p'}$.
\end{theorem}

The proofs of these results are tedious inductions, using the commutations
described in paragraphs~\ref{par:gen-prom-struct},
\ref{par:gen-prom-costruct} and~\ref{par:gen-prom-coder}.


\subsection{The differential $\lambda$-calculus}
Various $\lambda$-calculi have been proposed, as possible extensions of the
ordinary $\lambda$-calculus with constructions corresponding to the above
differential and costructural rules of differential \LL{}. We record
here briefly our original syntax of~\cite{EhrhardRegnier02}, simplified by Vaux
in~\cite{Vaux05}\footnote{Alternative syntaxes have been proposed, which are
  formally closer to Boudol's calculus with multiplicities or with resources
  and are therefore often called \emph{resource $\lambda$-calculi}}.

A \emph{simple term} is either
\begin{itemize}
\item a variable $x$,
\item or an ordinary application $\App MR$ where $M$ is a simple terms and $R$
  is a term,
\item or an abstraction $\Abs xM$ where $x$ is a variable and $M$ is a simple
  term,
\item or a differential application $\Dapp MN$ where $M$ and $N$ are simple
  terms.
\end{itemize}
A \emph{term} is a finite linear combination of simple terms, with coefficients
in $\Field$. Substitution of a term $R$ for a variable $x$ in a simple term
$M$, denoted as $\Subst MRx$ is defined as usual, whereas differential (or
linear) substitution of a simple term for a variable in another simple term,
denoted as $\Dsubst MNx$, is defined as follows:
\begin{align*}
  \Dsubst ytx&=
  \begin{cases}
    t & \text{if $x=y$}\\
    0 & \text{otherwise}
  \end{cases}\\
  \Dsubst{\Abs yM}Nx&=\Abs y{\Dsubst MNx}\\
  \Dsubst{\Dapp MN}{P}{x}&=\Dapp{\left(\Dsubst
      MPx\right)}{N}+\Dapp{M}{\left(\Dsubst
      NPx\right)}\\
  \Dsubst{\App MR}{N}{x}&= \App{\Dsubst MNx}{R}+\App{\Dapp M{\left(\Dsubst
        RNx\right)}}{R}
\end{align*}
All constructions are linear, except for ordinary application which is not
linear in the argument. This means that when we write e.g.~$\App{M_1+M_2}{R}$,
what we actually intend is $\App{M_1}{R}+\App{M_2}{R}$. Similarly, substitution
$\Subst MRx$ is linear in $M$ and not in $R$, whereas differential substitution
$\Dsubst MNx$ is linear in both $M$ and $N$. There are two reduction rules:
\begin{align*}
  \App{\Abs xM}{R}&\Rel\Beta\Subst MRx\\
  \Dapp{(\Abs xM)}{N}&\Rel\Dbeta \Abs x{\left(\Dsubst MNx\right)}
\end{align*}
which have of course to be closed under arbitrary contexts. The resulting
calculus can be proved to be Church-Rosser using fairly standard techniques
(Tait - Martin-L\"of), to have good normalization properties in the typed case
etc, see~\cite{EhrhardRegnier02,Vaux05}. To be more precise, Church-Rosser
holds only up to the least congruence on terms which identifies $\Dapp{(\Dapp
  M{N_1})}{N_2}$ and $\Dapp{(\Dapp M{N_2})}{N_1}$, a syntactic version of
Schwarz Lemma: terms are always considered up to this congruence called below
\emph{symmetry of derivatives}.

\subsubsection{Resource calculus.} Differential application can be iterated:
given simple terms $M,\List N1n$, we define $\Dappm nM{\List
  N1n}=\Dapp{(\cdots\Dapp M{N_1}\cdots)}{N_n}$; the order on the terms $\List
N1n$ does not matter, by symmetry of derivatives.  The (general) resource
calculus is another syntax for the differential $\lambda$-calculus, in which
the combination $\App{\Dappm nM{\List N1n}}{R}$ is considered as one single
operation denoted e.g.~as $M[\List N1n,R^\infty]$ where the superscript
$\infty$ is here to remind that $R$ can be arbitrarily duplicated during
reduction, unlike the $N_i$'s. This presentation of the calculus, studied in
particular by Tranquilli and Pagani, and also used for instance
in~\cite{BucciarelliCarraroEhrhardManzonetto11}, has very good properties as
well. It is formally close to Boudol's $\lambda$-calculus with multiplicities
such as presented in~\cite{BoudolCurienLavatelli99}, with the difference that
the operational semantics of Boudol's calculus is given as a rewriting strategy
whereas in the differential version of the resource $\lambda$-calculus, redexes
can be reduced everywhere in terms. The price to pay is that reduction becomes
non-deterministic in the sense that it can produce formal sums of terms.

\subsubsection{The finite resource
  calculus.}\label{sec:finite-resource-calculus}
If, in the resource calculus above, one restricts one's attention to the terms
where all applications are of the form
\[
M[\List N1n,0^\infty]
\]
which corresponds to the differential term $\App{\Dapp{(\cdots\Dapp
    M{N_1}\cdots)}{N_n}}0$, then one gets a calculus which is stable under
reduction and where all terms are strongly normalizing. This calculus, called
the \emph{finite resource calculus}, can be presented as follows.
\begin{itemize}
\item Any variable $x$ is a term.
\item If $x$ is a variable and $s$ is a simple term then $\Abs xs$ is a simple
  term.
\item If $S$ is a finite multiset (also called \emph{bunch} in the sequel) of
  simple terms then $\Linapp sS$ is a simple term. Intuitively, this term
  stands for the application $s[s_1,\dots,s_n,0^\infty]$ of the resource
  calculus, where $\Mset{\List s1n}=S$.
\end{itemize}
A term is a (possibly infinite\footnote{When considering infinite linear
  combinations, one has to deal with the possibility of unbounded coefficients
  appearing during the reduction. One option is to accept infinite
  coefficients, but it is also possible to prevent this phenomenon to occur by
  topological means as explained in~\cite{Ehrhard10b}.}) linear combination of
finite terms. This syntax is extended from simple terms to general terms by
linearity. For instance, the term $\Linapp x{\Mset{y+z,y+z}}$ stands for
$\Linapp x{\Mset{y,y}}+2\Linapp x{\Mset{y,z}}+\Linapp x{\Mset{z,z}}$.

In the finite resource calculus, it is natural to perform several
$\Dbeta$-reductions in one step, and one gets
\begin{equation*}
  \Linapp{\Abs xs}{S}\Rel\Dbeta
  \begin{cases}
    \sum_{f\in\Symgrp n}\Substbis{s}{s_{f(1)}/x_1,\dots,s_{f(n)}/x_n}&
    \text{if $\Deg sx=n$}\\
    0 & \text{otherwise}
  \end{cases}
\end{equation*}
where $d=\Deg xs$ is the number of occurrences of $x$ in $s$ (which is a simple
term), $x_1,\dots,x_d$ are the occurrences of $x$ in $s$ and the multiset $S$
is $\Mset{\List s1n}$.

Again, this calculus enjoys confluence, and also strong normalization (even in
the untyped case). It can be used for hereditarily Taylor expanding
$\lambda$-terms as explained
in~\cite{EhrhardRegnier06a,EhrhardRegnier06b,Ehrhard10b}.

Taylor expansion consists in hereditarily replacing, in a differential
$\lambda$-term, any ordinary application $\App MN$ by the infinite sum
\[
\sum_{n=0}^\infty\frac 1{\Factor n}\App{\Dappm nM{N,\dots,N}}0\,.
\]
More precisely, it is a transformation $M\mapsto\Tay M$ from resource terms to
finite resource terms which is defined as follows:
\begin{align*}
  \Tay x &= x\\
  \Tay{(\Abs xM)} &= \Abs x{(\Tay M)}\\
  \Tay{M[\List N1n,R^\infty]} &= \sum_{p=0}^\infty\frac{1}{\Factor p}
  \Linapp{\Tay M}{\Mset{\Tay{N_1},\dots,\Tay{N_n},\Tay R,\dots,\Tay R}}
\end{align*}
so that the Taylor expansion of a resource term is a generally infinite linear
combination of finite resource terms. In the definition above, we use the
extension by linearity of the syntax of finite resource terms to arbitrary
(possibly infinite) linear combinations. The coefficients belong to the
considered semi-ring $\Field$ where division by positive natural numbers must
be possible.

In~\cite{EhrhardRegnier06a,EhrhardRegnier06b} we studied the behavior of this
expansion with respect to differential $\beta$-reduction in the case where the
expanded terms come from the $\lambda$-calculus (that is, do not contain
differential applications; this is the \emph{uniform} case), and we exhibited
tight connections between this operation and Krivine's machine, an
implementation of linear head reduction. 

There is a simple translation from resource terms (or differential terms) to
$\DILL$ proof-nets. When restricted to the finite resource calculus, this
translation ranges in $\DILLZ$. This translation extends Girard's Translation
from the $\lambda$-calculus to \LL{} proof-nets.

\section{More on exponential structures}
\label{sec:more-exp-struct}
We address here two aspects of exponential structures: we study a
simple condition expressing that a map whose derivative is uniformly equal to
$0$ must be constant, and we propose an axiomatization of antiderivatives in
this categorical setting.

So we assume to be given a preadditive *-autonomous category $\cL$ equipped
with an exponential structure in the sense of Section~\ref{sec:exp-struct}, and
we use the same notations as in this section.

For the sake of notational simplicity, we do as if $\ITens$ were strictly
associative and $\One$ were strictly neutral for $\ITens$. In other words, we
do not mention the isos $\lambda$, $\rho$ and $\alpha$ in our computations,
just as if they were identities (see Section~\ref{sec:monoidal-struct}). 
Given an object $X$ and $n\in\Nat$ of $\cL$, we use $\Tpower Xn$ for the $n$th
tensor power of $X$: $\Tpower X0=\One$ and $\Tpower X{(n+1)}=\Tens{\Tpower
  Xn}{X}$.  

Given an object $X$ of $\cL$, we define a morphism $\Coderc X\in\cL(\Tens{\Excl
X}{X},\Excl X)$ as the following composition of morphisms
\begin{equation*}
  \begin{tikzpicture}[->, >=stealth]
    \node (1) {$\Tens{\Excl X}{X} $};
    \node (2) [right of=1, node distance=28mm] 
      {$\Tens{\Excl X}{\Excl X} $};
    \node (3) [right of=2, node distance=16mm]
      {$\Excl X $}; 
    \tikzstyle{every node}=[midway,auto,font=\scriptsize]
    \draw (1) -- node {$\Tens{\Excl X}{\Coder X} $} (2);
    \draw (2) -- node {$\Cocontr X $} (3);
  \end{tikzpicture}  
\end{equation*}

More generally, we define $\Codercm Xn\in\cL(\Tens{\Excl X}{\Tpower Xn},\Excl
X)$:
\begin{align*}
  \Codercm X0 &= \Id_{\Excl X}\\
  \Codercm X{n+1} &= \Coderc X\Complin\Tensp{\Codercm Xn}{X}
\end{align*}
Last we set 
\begin{equation*}
\Coderm Xn=\Codercm Xn\Complin\Tensp{\Coweak X}{\Tpower Xn}\in\cL(\Tpower
Xn,\Excl X)\,.
\end{equation*}

We define dually $\Derc X\in\cL(\Excl X,\Tens{\Excl X}{X})$ as $\Derc
X=\Tensp{\Excl X}{\Der X}\Complin\Contr X$ and $\Dercm Xn\in\cL(\Excl
X,\Tens{\Excl X}{\Tpower Xn})$ by $\Dercm X0=\Id_{\Excl X}$ and $\Dercm
X{n+1}=\Tensp{\Dercm Xn}X\Complin\Derc X$. And we set
\begin{equation*}
  \Derm Xn=\Tensp{\Weak X}{\Tpower Xn}\Complin\Dercm Xn
  \in\cL(\Excl X,\Tpower Xn)\,.  
\end{equation*}

Observe that we have $\Coderm X1=\Coder X$ and $\Derm X1=\Der X$.

Consider now some $f\in\cL(\Excl X,Y)$, to be intuitively seen as a non linear
map from $X$ to $Y$ (if $\Excl\_$ were assumed to be a comonad as in
Section~\ref{sec:finctorial-exp}, then $f$ would be a morphism in the Kleisli
category $\cL_\oc$). Such a morphism will sometimes be called a ``regular
function'' in the sequel, but keep in mind that it is not even a function in
general. For instance, if $X$ and $Y$ are vector spaces (with a topological
structure, in the infinite dimensional case), such a regular function could
typically be a smooth or an analytic function.

With these notations, $f\Complin\Coweak X\in\cL(\One,Y)$ should be
understood as the point of $Y$ obtained by applying the regular function $f$ to
$0$. Similarly, $f\Complin\Cocontr X\in\cL(\Tens{\Excl X}{\Excl X},Y)$ should
be understood as the regular function $g:X\times X\to Y$ defined by
$g(x_1,x_2)=f(x_1+x_2)$. 

Dually, given $y\in\cL(\One,Y)$ considered as a point of $Y$, then
$y\Complin\Weak X\in\cL(\Excl X,Y)$ should be understood as the constant
regular function which takes $y$ as unique value. If $g\in\Complin(\Tens{\Excl
  X}{\Excl X},Y)$, to be considered as a regular function with two parameters
$X\times X\to Y$, then $f=g\Complin\Contr X\in\cL(\Excl X,Y)$ should be
understood as the regular function $X\to Y$ given by $f(x)=g(x,x)$. Given
$g\in\cL(X,Y)$, to be considered as a linear function from $X$ to $Y$,
$f=g\Complin\Der X\in\cL(\Excl X,Y)$ is $g$, considered now as a regular
function from $X$ to $Y$.

The basic idea of $\DILL$ is that $f'=f\Complin\Coderc X\in\cL(\Tens{\Excl
  X}{X},Y)$ (that is $f'\in\cL(\Excl X,\Limpl XY)$, up to linear curryfication)
represents the \emph{derivative} of $f$.

Remember indeed that if $f:X\to Y$ is a smooth function from a vector
space $X$ to a vector space $Y$, the derivative of $f$ is a function $f'$ from
$X$ to the space $\Limpl XY$ of (continuous) linear functions from $X$ to $Y$:
$f'$ maps $x\in X$ to a linear map $f'(x):X\to Y$, the differential (or
Jacobian) of $f$ at point $x$, which maps $u\in X$ to $f'(x)\cdot u$.

In particular, $f\Complin\Coder X=f\Complin\Coderc X\Complin\Tensp{\Coweak
  X}{X}\in\cL(X,Y)$ corresponds to $f'(0)$, the differential of $f$ at $0$.

More generally, $f\Complin\Codercm Xn\in\cL(\Tens{\Excl X}{\Tpower Xn},Y)$
represents the $n$th derivative of $f$, which is a regular function from $X$
to the space of $n$-linear functions from $X^n$ to $Y$ (these $n$ linear
functions are actually symmetric, a property called ``Schwarz Lemma'' and
is axiomatized here by the commutativity of the algebra structure of $\Excl X$).

The axioms of an exponential structure express that this categorical definition
of differentiation satisfies the usual laws of differentiation. Let us give a
simple example. Consider $f\in\cL(\Tens{\Excl X}{\Excl X}, Y)$, to be seen as a
regular function $f(x_1,x_2)$ depending on two parameters $x_1$ and $x_2$ in
$X$. Remember that $g=f\Complin\Contr X\in\cL(\Excl X,Y)$ represents the map
depending on one parameter $x$ in $X$ given by $g(x)=f(x,x)$.

Using the axioms of exponential structures, one checks easily that
\begin{equation*}
  \Contr X\Complin\Coderc X=((\Tens{\Excl X}{\Coderc X})
  +\Tensp{\Coderc X}{\Excl X}\Complin\Tensp{\Excl X}{\sigma})
  \Complin(\Tens{\Contr X}{X})
\end{equation*}
which, by composing with $f$ and using standard algebraic notations, gives
\[
g'(x)\cdot u=f'_1(x,x)\cdot u+f'_2(x,x)\cdot u
\] 
where we use $f'_i$ for the $i$th partial derivative of $f$ and ``$\cdot$'' for
the linear application of the differential. This is Leibniz law.

Similarly, the easily proven equation $\Weak X\Complin\Coderc X=0$ expresses
that the derivative of a constant map is equal to $0$.

It is a nice exercise to interpret similarly the dual equations
\begin{align*}
  \Derc X\Complin\Cocontr X&=\Tensp{\Cocontr X}{X}\Complin(\Tensp{\Excl
    X}{\Derc X}+\Tensp{\Excl X}{\sigma}\Complin\Tensp{\Derc X}{\Excl X})\\
  \Derc X\Complin\Coweak X&=0
\end{align*}
Consider $g\in\cL(\Tens{\Excl X}{X},Y)$, to be considered as a regular function
$X\times X\to Y$ which is linear in its second parameter. Then
$f=g\Complin\Derc X\in\cL(\Excl X,Y)$ is the regular function $X\to Y$ given by
$g(x)=f(x,x)$. The second equation corresponds to the fact that $g(0,0)=0$,
and the first one, to the fact that
$g(x_1+x_2,x_1+x_2)=g(x_1+x_2,x_1)+g(x_1+x_2,x_2)$.

\subsection{Taylor exponential structures.}
\label{sec:Taylor-structure}
Let $\cL$ be an 
exponential structure and let $f\in\cL(\Excl X,Y)$, to be considered as a
``regular function'' from $X$ to $Y$. The condition $f\Compl\Coderc X=0$ means
intuitively that the derivative of $f$ is uniformly equal to $0$, and hence,
according to standard intuitions on differentiation, $f$ should be a constant
map. In other words we should have $f=f\Compl\Coweak X\Compl\Weak X$.

This property can be stated in a more general way as follows: let
$f_1,f_2:\Excl X\to Y$, then
\begin{equation*}
  f_1\Compl\Coderc X=f_2\Compl\Coderc X\Implies
  f_1+(f_2\Compl\Coweak X\Compl\Weak X)=f_2+(f_1\Compl\Coweak X\Compl\Weak X) 
\end{equation*}
and does not seem to be derivable from the other axioms of exponential
structures. The converse implication is easy to prove.

\begin{remark}
There is a dual condition which reads as
follows: if $f_1,f_2:Y\to\Excl X$, then
\begin{equation*}
  \Derc X\Compl f_1=\Derc X\Compl f_2\Implies
  f_1+(\Coweak X\Compl\Weak X\Compl f_2)=f_2+(\Coweak X\Compl\Weak X\Compl f_1) 
\end{equation*}
The intuition is that, given $f:Y\to\Excl X$, to be considered as a generalized
point of $\Excl X$, if $\Derc X\Compl f=0$, then the range of $f$ is included
in the subspace of $\Excl X$ generated by the unit of the bialgebra $\Excl
X$. In other words, this condition means that the kernel of $\Derc X$ is
generated by this unit.
\end{remark}

We say that the exponential structure $\cL$ is Taylor if it satisfies the first
condition. 

For $n\in\Nat$, let $\Taym Xn\in\cC(\Excl X,\Excl X)$ be defined by
\begin{equation*}
  \Taym Xn=\sum_{i=0}^n\frac 1{\Factor i}\,\Coderm Xi\Compl\Derm Xi\,.
\end{equation*}
\begin{lemma}\label{lemma:Taylor-coderc-commute}
  For any $n>0$, we have
  \begin{equation*}
    \Taym Xn\Compl\Coderc X=\Coderc X\Compl(\Tens{\Taym X{n-1}}{\Id_X})\,.
  \end{equation*}
\end{lemma}
\Beginproof
This results from
\begin{equation*}
  \Coderm Xn\Compl\Derm Xn\Compl\Coderc X
  =n\Coderc X\Compl(\Tens{(\Coderm X{n-1}\Compl\Derm X{n-1})}{\Id_X})
\end{equation*}
which comes from the basic equations of exponential structures.
\Endproof

Remember that a commutative monoid $M$ is cancellative if, in $M$, one has
$u+v=u'+v\Implies u=u'$.

\begin{proposition}\label{prop:Taylor-expo-polynomial}
  Assume that $\cL$ is Taylor and that each homset $\cL(X,Y)$ is a cancellative
  monoid. Let $n\in\Nat$ and let $f_1,f_2:\Excl X\to Y$. If $f_1\Compl\Codercm
  X{n+1}=f_2\Compl\Codercm X{n+1}$ then $f_1+(f_2\Compl\Taym
  Xn)=f_2+(f_1\Compl\Taym Xn)$.
\end{proposition}
In particular, if $f\Compl\Codercm X{n+1}=0$ (that is, the $(n+1)$-th
derivative of $f$ is uniformly equal to $0$), then $f=f\Compl\Taym Xn$, meaning
that $f$ is equal to its Taylor expansion of rank $n$.
\Beginproof
By induction on $n$. For $n=0$, this is simply the hypothesis that $\cL$ is
Taylor. Assume now that $f_1\Compl\Codercm X{n+2}=f_2\Compl\Codercm X{n+2}$ and
let us prove that $f_1+(f_2\Compl\Taym X{n+1})=f_2+(f_1\Compl\Taym X{n+1})$.

We have $f_1\Compl\Coderc X\Compl(\Tens{\Codercm
  X{n+1}}{\Id_X})=f_2\Compl\Coderc X\Compl(\Tens{\Codercm X{n+1}}{\Id_X})$. By
monoidal closeness, we have $\Curlin{(f_1\Compl\Coderc X)}\Compl\Codercm
X{n+1}=\Curlin{(f_2\Compl\Coderc X)}\Compl\Codercm X{n+1}$ and hence, by
inductive hypothesis, we have
\begin{equation*}
  \Curlin{(f_1\Compl\Coderc X)}+(\Curlin{(f_2\Compl\Coderc X)}\Compl\Taym Xn)
  =\Curlin{(f_2\Compl\Coderc X)}+(\Curlin{(f_1\Compl\Coderc X)}\Compl\Taym Xn)
\end{equation*}
that is
\begin{equation*}
  \Curlin{(f_1\Compl\Coderc X)}+
    (\Curlinp{f_2\Compl\Coderc X\Compl(\Tens{\Taym Xn}{\Id_X})})
    =\Curlinp{f_2\Compl\Coderc X}+
    (\Curlinp{f_1\Compl\Coderc X\Compl(\Tens{\Taym Xn}{\Id_X})})
\end{equation*}
and hence
\begin{equation*}
  (f_1\Compl\Coderc X)+(f_2\Compl\Coderc X\Compl(\Tens{\Taym Xn}{\Id_X}))
  =(f_2\Compl\Coderc X)+(f_1\Compl\Coderc X\Compl(\Tens{\Taym Xn}{\Id_X}))
\end{equation*}
so applying Lemma~\ref{lemma:Taylor-coderc-commute}, we get
\begin{equation*}
  (f_1+f_2\Taym X{n+1})\Compl\Coderc X=(f_2+f_1\Taym X{n+1})\Compl\Coderc X\,.
\end{equation*}
Applying the hypothesis that $\cL$ is Taylor, we get
\begin{equation*}
  f_1+f_2\Taym X{n+1}+(f_2+f_1\Taym X{n+1})\Compl\Coweak X\Compl\Weak X
  =f_2+f_1\Taym X{n+1}+(f_1+f_2\Taym X{n+1})\Compl\Coweak X\Compl\Weak X
\end{equation*}
and since $\Taym X{n+1}\Compl\Coweak X\Compl\Weak X=\Coweak X\Compl\Weak X$ we
get $f_1+f_2\Taym X{n+1}+(f_1+f_2)\Compl\Coweak X\Compl\Weak X=f_2+f_1\Taym
X{n+1}+(f_2+f_1)\Compl\Coweak X\Compl\Weak X$ and so, applying the
cancellativeness hypothesis, we get finally
\begin{equation*}
  f_1+f_2\Taym X{n+1}=f_2+f_1\Taym X{n+1}
\end{equation*}
as required.
\Endproof

\subsubsection{The category of polynomials.}
We say that $f\in\cL(\Excl X,Y)$ is \emph{polynomial} if there exists
$n\in\Nat$ such that $f\Compl\Codercm X{n+1}=0$, and we call \emph{degree} of
$f$ the least such $n$. The morphism $\Der X\in\cC(\Excl X,X)$ is polynomial of
degree $1$.

Let $f\in\cL(\Excl X,Y)$ and $g\in\cL(\Excl Y,Z)$ be polynomial of degree $m$
and $n$ respectively. We define the composition $g\Comp f\in\cL(\Excl X,Z)$ as
follows
\begin{equation*}
  g\Comp f=\sum_{i=0}^n \frac 1{\Factor i}\,g\Compl\Coderm Yi\Compl\Tensexp
  fi\Compl\Contrm Xi\,.
\end{equation*}
where $\Tensexp fi=\overbrace{f\ITens\cdots\ITens f}^{i\times}$.

Since $\Der X\Compl\Coderm Xi=0$ for $i\not=1$, we get $\Der X\Comp g=g$. Next
observe that $g\Comp\Der X=g\Compl\Taym Xn=g$ by
Proposition~\ref{prop:Taylor-expo-polynomial}. One can prove that $g\Comp f$ is
polynomial of degree $\leq mn$, that is $(g\Comp f)\Compl\Codercm X{mn}=0$ by a
straightforward (though boring) categorical computation using the basic axioms
of exponential structures. Using the same axioms, one shows that this notion
of composition is associative, so that we have defined a category of polynomial
morphisms.

\subsubsection{Weak functoriality and the category of polynomials.}
We do not require this operation $X\mapsto\Excl X$ to be functorial, but some
weak form of functoriality can be derived from the above categorical
axioms. Let $f\in\cL(X,Y)$. By induction on $n$, we define a family of
morphisms $f^n:\Excl X\to\Excl X$ as follows: $f^0=\Coweak X\Compl\Weak X$ and
\begin{equation*}
  f^{n+1}=\Coderc X\Compl(\Tens{f^n}{f})\Compl\Derc X\,.
\end{equation*}

\begin{proposition}
  Let $f\in\cL(X,Y)$ and $g\in\cL(Y,Z)$ and let $n,p\in\Nat$. Then
  \begin{equation*}
    g^p\Compl f^n=
    \begin{cases}
      \Factor n(g\Compl f)^n & \text{if $n=p$}\\
      0 & \text{otherwise.}
    \end{cases}
  \end{equation*}
\end{proposition}
\Beginproof
Simple calculation using the diagram commutations which define an exponential
structure.
\Endproof

So for each $n\in\Nat$ we can define $\Partexcl nf=\sum_{q=0}^n\frac{1}{\Factor
  q}f^q:\Excl X\to\Excl X$, and we have $\Partexcl ng\Complin\Partexcl
nf=\Partexcl n{(g\Complin f)}$. So $f\mapsto\Partexcl n{f}$ is a quasifunctor,
but not a functor as it does not map $\Id_X$ to $\Id_{\Excl X}$, but to an
idempotent morphism $\rho^n_X:\Excl X\to\Excl X$.

In some concrete models, this sequence $(\Partexcl nf)_{n\in\Nat}$ can be said
to be convergent, in a sense which depends of course on the model. The limit is
then denoted as $\Excl f$ and the operation defined in that way turns out often
to be a true functor, defining a functorial exponential in the sense of
Section~\ref{sec:finctorial-exp}.

\subsection{Computing antiderivatives.}\label{sec:primitives}
We say that an exponential structure $\cL$ \emph{has antiderivatives} if the
morphism
\[
J_X=\Id_X+(\Coderc X\Compl\Derc X):\Excl X\to\Excl X
\] 
is an isomorphism. We explain why.

We assume to be given an exponential structure $\cL$ which has antiderivatives
in that sense and we set $\Primmor X=\Funinv{J_X}$.

In the sequel, we use the following notation
\begin{equation*}
  \psi_X=(\Tens{\Coderc X}{\Id_X})\Compl\sigma_{23}
  \Compl(\Tens{\Derc X}{\Id_X}):\Tens{\Excl X}{X}\to\Tens{\Excl X}{X}
\end{equation*}
where $\sigma_{23}=\Tens{\Excl X}{\sigma}$ is an automorphism on $\Excl
X\ITens X\ITens X$, because this morphism $\psi_X$ will show up quite
often. Observe in particular that
\begin{equation*}
  \Derc X\Compl\Coderc X=\Id_{\Tens{\Excl X}{X}}+\psi_X\,.
\end{equation*}

\begin{lemma}\label{lemma:primmor-tens-comm}
  The following commutation holds
  \begin{equation}\label{eq:prim-tens-commut}
    (\Tens{\Primmor X}{\Id_X})\Compl\psi_X
    =\psi_X\Compl(\Tens{\Primmor X}{\Id_X})
  \end{equation}
\end{lemma}
\Beginproof
Since $\Primmor X=\Funinv{(\Id_{\Excl X}+(\Coderc X\Compl\Derc X))}$, we have
$\Tens{\Primmor X}{\Id_{X}}=\Funinv\phi$ where $\phi=\Id_{\Tens{\Excl
    X}{X}} +(\Tens{(\Coderc X\Compl\Derc X)}{\Id_X})$ by functoriality of
$\ITens$. To prove~\Eqref{eq:prim-tens-commut}, it suffices therefore to prove
that $\phi$ commutes with $\psi_X$. For this, it suffices to show that
\[
(\Tens{(\Coderc X\Compl\Derc X)}{\Id_X})\Complin\psi_X
=\psi_X\Complin(\Tens{(\Coderc X\Compl\Derc X)}{\Id_X})\,.
\] 
We have
\begin{equation*}
  (\Tens{(\Coderc X\Compl\Derc X)}{\Id_X})\Compl\psi_X
  =(\Tens{(\Coderc X\Compl\Derc X\Compl\Coderc
    X)}{\Id_X})\Compl\sigma_{23}\Compl(\Tens{\Derc X}{\Id_X})
\end{equation*}
but remember that $\Derc X\Compl\Coderc X=\Id_{\Tens{\Excl X}{X}}+\psi_X$, and
hence
\begin{equation*}
  (\Tens{(\Coderc X\Compl\Derc X)}{\Id_X})\Compl\psi_X
  =\psi_X+(\Tens{(\Coderc X\Compl\psi_X)}{\Id_X})
  \Compl\sigma_{23}\Compl(\Tens{\Derc X}{\Id_X})
\end{equation*}
But $\Coderc X\Compl(\Tens{\Coderc X}{\Id_X})\Compl\sigma_{23}=\Coderc
X\Compl(\Tens{\Coderc X}{\Id_X})$ by commutativity of the bialgebra $\Excl X$
and by definition of $\Coderc X$. Therefore $\Coderc X\Compl\psi_X=\Coderc
X\Compl(\Tens{(\Coderc X\Compl\Derc X)}{\Id_X})$. So we can write
\begin{align*}
  &(\Tens{(\Coderc X\Compl\Derc X)}{\Id_X})\Compl\psi_X\\
  &\quad\quad=\psi_X+(\Tens{\Coderc X}{\Id_X})
  \Compl(\Tens{(\Coderc X\Compl\Derc X)}{\Tens{\Id_X}{\Id_X}})
  \Compl\sigma_{23}\Compl(\Tens{\Derc X}{\Id_X})\\
  &\quad\quad=\psi_X+(\Tens{\Coderc X}{\Id_X})
  \Compl(\Tens{(\Coderc X\Compl\Derc X)}{\sigma})
  \Compl(\Tens{\Derc X}{\Id_X})
\end{align*}
A similar, and completely symmetric computation, using this time the
cocommutativity of the bialgebra $\Excl X$, leads to
\begin{equation*}
  \psi_X\Compl(\Tens{(\Coderc X\Compl\Derc X)}{\Id_X})
  =\psi_X+(\Tens{\Coderc X}{\Id_X})
  \Compl(\Tens{(\Coderc X\Compl\Derc X)}{\sigma})
  \Compl(\Tens{\Derc X}{\Id_X})
\end{equation*}
and we are done.
\Endproof

We can now prove a completely categorical version of the following proposition
which is the key step in the usual proof of Poincar\'e's Lemma.

\begin{proposition}\label{prop:poincare-categorique}
  Let $f:\Tens{\Excl X}{X}\to Y$ be such that the differential
  $f\Compl(\Tens{\Derc X}{\Id_X}):\Excl X\ITens X\ITens X\to Y$ satisfies
\begin{equation*}
  f\Compl(\Tens{\Derc X}{\Id_X})\Compl\sigma_{23}
  =f\Compl(\Tens{\Derc X}{\Id_X})\,.
\end{equation*}
  Then there exists $g:\Excl X\to Y$ such that $g\Compl\Coderc X=f$; in other
  words, $g$ is an ``antiderivative'' of $f$.
\end{proposition}
\Beginproof
One sets 
\begin{equation}\label{eq:poincare-primitive-cat}
  g=f\Compl(\Tens{\Primmor X}{\Id_X})\Compl\Derc X\,.
\end{equation}
Then we have
\begin{align*}
  g\Compl\Coderc X
  &= f\Compl(\Tens{\Primmor X}{\Id_X})\Compl\Derc X\Compl\Coderc X\\
  &= f\Compl(\Tens{\Primmor X}{\Id_X})\Compl(\Id_{\Tens{\Excl X}{X}}+\psi_X)\\
  &= f\Compl(\Tens{\Primmor X}{\Id_X})+f\Compl(\Tens{\Primmor
    X}{\Id_X})\Compl\psi_X\\
  &= f\Compl(\Tens{\Primmor X}{\Id_X})+f\Compl\psi_X\Compl(\Tens{\Primmor
    X}{\Id_X})
\end{align*}
by Lemma~\ref{lemma:primmor-tens-comm}. But
\begin{align*}
  f\Compl\psi_X
  &= f\Compl(\Tens{\Coderc X}{\Id_X})\Compl\sigma_{23}
  \Compl(\Tens{\Derc X}{\Id_X})\\
  &= f\Compl(\Tens{\Coderc X}{\Id_X})
  \Compl(\Tens{\Derc X}{\Id_X})\quad\text{by our hypothesis on $f$}\\
  &= f\Compl(\Tens{(\Coderc X\Compl\Derc X)}{\Id_X})\,.
\end{align*}
So we get
\begin{align*}
  g\Compl\Coderc X
  &= f\Compl(\Tens{\Primmor X}{\Id_X})+
  f\Compl(\Tens{(\Coderc X\Compl\Derc X\Compl\Primmor X)}{\Id_X})\\
  &= f\Compl ((\Tens{(\Id_{\Excl X}+(\Coderc X\Compl\Derc X))\Compl\Primmor
    X)}{\Id_X})\\
  &= f
\end{align*}
since $\Primmor X$ is the inverse of $\Id_{\Excl X}+(\Coderc X\Compl\Derc X)$.
\Endproof

\subsubsection{Comments.}
Let us give some intuition about our axiom that $J_X$ has an inverse.  Given
$f:\Excl X\to Y$ seen as a ``regular function'' from $X$ to $Y$, we explain why
the morphism $f\Compl\Primmor X:\Excl X\to Y$ should be understood as
representing the regular function $g$ defined by
\begin{equation*}
  g(x)=\int_0^1 f(tx)\,dt
\end{equation*}
assuming of course that this integral makes sense. With this interpretation,
$g\Compl\Coderc X:\Tens{\Excl X}{X}\to Y$ represents the differential
$\Diffsymb g$ of $g$, a regular function $X\times X\to Y$ which maps $(x,y)$ to
$\Diffsymb g(x)\cdot y$ and is linear in $y$. Then, applying the ordinary rules
of differential calculus, and the fact that differentiation commutes with
integration, we get
\begin{equation*}
  \Diffsymb g(x)\cdot y=\int_0^1t(\Diffsymb f(tx)\cdot y)dt\,.
\end{equation*}
The morphism $h=g\Compl\Coderc X\Compl\Derc X:\Excl X\to Y$ corresponds to the
regular function from $X$ to $Y$ such that
\begin{align*}
  h(x)&=\int_0^1(\Diffsymb f(tx)\cdot(tx))dt\\
  &=\int_0^1t(\Diffsymb f(tx)\cdot x)dt\\
  &=\int_0^1 t\frac{df(tx)}{dt}dt\\
  &=f(x)-\int_0^1f(tx)\,dt\,,  
\end{align*}
integrating by parts. In other words, we have seen that
\begin{equation*}
  f\Compl\Primmor X\Compl\Coderc X\Compl\Derc X=f-(f\Compl\Primmor X)
\end{equation*}
that is
\begin{equation*}
  f\Compl\Primmor X\Compl(\Id_{\Excl X}+(\Coderc X\Compl\Derc X))=f
\end{equation*}
this is why our first axiom on $\Primmor X$ is that $\Primmor
X\Compl(\Id_{\Excl X}+(\Coderc X\Compl\Derc X))=\Id_{\Excl X}$. To explain why
we also require $(\Id_{\Excl X}+(\Coderc X\Compl\Derc X))\Compl\Primmor
X=\Id_{\Excl X}$, observe that 
\begin{equation*}
l=f\Compl\Coderc X\Compl\Derc X:\Excl X\to Y  
\end{equation*}
corresponds to the regular function defined by $l(x)=\Diffsymb f(x)\cdot x$
and hence $l\Compl\Primmor X$ corresponds to the regular function $m:X\to Y$
given by
\begin{equation*}
  m(x)=\int_0^1(\Diffsymb f(tx)\cdot(tx))dt=h(x)
\end{equation*}
by linearity of the differential. So we have 
\begin{equation*}
  f\Compl\Coderc X\Compl\Derc X\Compl\Primmor X=f-(f\Compl\Primmor X)
\end{equation*}
that is 
\begin{equation*}
  f\Compl(\Id_{\Excl X}+(\Coderc X\Compl\Derc X))\Compl\Primmor X=f\,.
\end{equation*}

A remarkable and quite natural feature of this axiomatization of
antiderivatives is the fact that it is actually a mere \emph{property} of the
exponential structure, and not an additional structure: it must be such that
$\Id_{\Excl X}+(\Coderc X\Compl\Derc X)$ has an inverse.

\begin{remark}
  The definition~\Eqref{eq:poincare-primitive-cat} of the antiderivative $g$ of
  $f$ in the proof above reads as follows, if we use this intuitive
  interpretation of $\Primmor X$:
  \begin{equation}\label{eq:poincare-primitive_int}
    g(x)=\int_0^1f(tx)\cdot x\,dt
  \end{equation}
  which is exactly its definition, in the standard proof of Poincar\'e's
  Lemma. The proof of Proposition~\ref{prop:poincare-categorique} is a
  rephrasing of the standard proof, which uses an integration by parts. 
\end{remark}

\subsubsection{The fundamental theorem of calculus.}
This is the statement according to which one can use antiderivatives for
computing integrals: if $f,g:\Real\to\Real$ are such that $g'=f$, then
$\int_a^bf(t)\,dt=g(b)-g(a)$. In the present setting, it boils down to a simple
categorical equation.
\begin{proposition}
  Let $\cL$ be an exponential structure which has antiderivatives and is
  Taylor. Then
  \begin{equation*}
    \Coderc X\Compl(\Tens{\Primmor X}{\Id_X})\Compl\Derc X+\Coweak
    X\Compl\Weak X=\Id_{\Excl X}\,.
  \end{equation*}
\end{proposition}
\Beginproof
Let $f_1=\Coderc X\Compl(\Tens{\Primmor X}{\Id_X})\Compl\Derc X:\Excl X\to\Excl
X$ and let $f_2=\Id_{\Excl X}$. 
We have
  \begin{align*}
    f_1\Compl\Coderc X
    &=\Coderc X\Compl(\Tens{\Primmor X}{\Id_X})\Compl\Derc X\Compl\Coderc X\\
    &=\Coderc X\Compl(\Tens{\Primmor X}{\Id_X})+\Coderc X\Compl(\Tens{\Primmor
      X}{\Id_X})\Compl(\Tens{\Coderc
      X}{\Id_X})\Compl\sigma_{23}\Compl(\Tens{\Derc X}{\Id_X})\\
    &=\Coderc X\Compl(\Tens{\Primmor X}{\Id_X})+\Coderc X\Compl(\Tens{\Coderc
      X}{\Id_X})\Compl\sigma_{23}\Compl(\Tens{\Derc
      X}{\Id_X})\Compl(\Tens{\Primmor X}{\Id_X})\\
    &\quad\quad\quad\quad\text{by Lemma~\ref{lemma:primmor-tens-comm}}\\
    &=\Coderc X\Compl(\Tens{\Primmor X}{\Id_X})+\Coderc X\Compl(\Tens{\Coderc
      X}{\Id_X})\Compl(\Tens{\Derc
      X}{\Id_X})\Compl(\Tens{\Primmor X}{\Id_X})\\
    &\quad\quad\quad\quad\text{by commutativity of cocontraction}\\
    &=\Coderc X\Compl(\Tens{(\Id_{\Excl
        X}+(\Coderc X\Compl\Derc X))}{\Id_X})\Compl(\Tens{\Primmor X}{\Id_X})\\
    &=\Coderc X\quad\quad\text{since $\Primmor X=\Funinv{(\Id_{\Excl
        X}+(\Coderc X\Compl\Derc X))}$}\\
    &=f_2\Compl\Coderc X\,.
  \end{align*}
  Since $\cL$ is Taylor and since $f_1\Compl\Coweak X\Compl\Weak X=0$, we have
  therefore $f_1+(f_2\Compl\Coweak X\Compl\Weak X)=f_2+(f_1\Compl\Coweak
  X\Compl\Weak X)$, which is exactly the announced equation.
\Endproof
\begin{remark}
  We give now an intuitive interpretation of this property. Let $f:\Excl X\to
  Y$, considered as a regular function from $X$ to $Y$. Then $f\Compl\Coderc
  X\Compl(\Tens{\Primmor X}{\Id_X})\Compl\Derc X:\Excl X\to Y$ represents the
  regular function $g:X\to Y$ given by
  \begin{equation*}
    g(x)=\int_0^1\Diffsymb f(tx)\cdot x\,dt=\int_0^1\frac{df(tx)}{dt}\,dt
  \end{equation*}
  so that $g(x)=f(x)-f(0)$ by the Fundamental Theorem of Calculus. In other
  words $g(x)+f(0)=f(x)$, that is $f\Compl(\Coderc X\Compl(\Tens{\Primmor
    X}{\Id_X})\Compl\Derc X+\Coweak X\Compl\Weak X)=f$.
\end{remark}

\subsection{Computing antiderivatives in the resource calculus}

We can consider finite linear combinations of finite resource terms
(see~\ref{sec:finite-resource-calculus}) as polynomials, and with this respect,
it seems natural to formally compute the antiderivative of such a term, as one
does for polynomials. This is the purpose of this short section. We use
$\Sterms$ for the set of simple resource terms and $\Fmod\Field S$ for the free
$\Field$-module generated by the set $S$.

As with ordinary polynomials, we define first the antiderivative of a
monomial, that is, of a simple resource term. Remember that, for ordinary one
variable polynomials, the antiderivative of $X^d$ is $\frac 1{d+1}X^{d+1}$;
the definition is completely similar here.  Let $t\in\Sterms$ be a simple
resource term and let $x$ be a variable. We set
\begin{equation*}
  \Primint x(t)=\frac 1{\Deg xt+1}t\,.
\end{equation*}
We extend this operation by linearity to all elements $u\in\Fmod\Field\Sterms$,
that is we set $\Primint x(u)=\sum_{t\in\Sterms}u_t\Primint x(t)$. 

For $d\in\Nat$, let $\Stermsh xd=\{t\in\Sterms\St\Deg xt=d\}$ be the set of all
simple resource terms of degree $d$ in $x$. The elements of
$\Fmod\Field{\Stermsh xd}$ are said to be homogeneous of degree $d$ in $x$.

With these notations, we can write
\begin{equation*}
  \Primint x(u)=\sum_{d=0}^\infty\frac 1{d+1}\sum_{t\in\Stermsh xd}u_t t
\end{equation*}
Intuitively, $\Primint x(u)$ stands for the integral $\int_0^1u(\tau x)\,d\tau$
which is the basic ingredient in the proof above of Poincar\'e's Lemma.

Let $u\in\Fmod\Field\Sterms$ which is linear in the variable $h$, in other
words $u\in\Fmod\Field{\Stermsh h1}$. Let $h'$ be a variable which does not
occur free in $u$, we assume that
\begin{equation*}
  \Derp ux{h'}=\Derp{\Subst u{h'}h}xh
\end{equation*}
which is our symmetry hypothesis on $u$. In other words, for any $d\in\Nat$, we have
\begin{equation}\label{eq:symetrie-homogene}
  \sum_{t\in\Stermsh xd}u_t\Derp tx{h'}
  =\sum_{t\in\Stermsh xd}u_t\Derp{\Subst t{h'}h}xh\,.
\end{equation}

Mimicking~\Eqref{eq:poincare-primitive_int}, we set
\begin{equation*}
  v=\Subst{\Primint x(u)}xh
\end{equation*}
and we prove that
\begin{equation}\label{eq:primitive-resource}
  \Derp vxh=u\,.
\end{equation}
Choose $h'$ as above, we prove that $\Derp vx{h'}=\Subst u{h'}h$ which of course
implies~\Eqref{eq:primitive-resource}. We have, keeping in mind that $h$ has
exactly one free occurrence in each $t\in\Sterms$ such that $u_t\not=0$
\begin{align*}
  \Derp vx{h'}
  &= \sum_{d=0}^\infty\frac 1{d+1}\sum_{t\in\Stermsh xd}u_t\Derp{\Subst
    txh}x{h'}\\
  &= \sum_{d=0}^\infty\frac 1{d+1}\sum_{t\in\Stermsh xd}u_t
  \left(\Subst{\Derp tx{h'}}xh+\Subst t{h'}h\right)\\
  &= \sum_{d=0}^\infty\frac 1{d+1}\sum_{t\in\Stermsh xd}u_t
  \left(\Subst{\left(\Derp{\Subst t{h'}h}xh\right)}xh+\Subst t{h'}h\right)
  \quad\text{by~\Eqref{eq:symetrie-homogene}}\\
  &= \sum_{d=0}^\infty\frac 1{d+1}\sum_{t\in\Stermsh xd}u_t
  \left(\Derp{\Subst t{h'}h}xx+\Subst t{h'}h\right)\\
  &= \sum_{d=0}^\infty\frac 1{d+1}\sum_{t\in\Stermsh xd}u_t
  \left(d\,\Subst t{h'}h+\Subst t{h'}h\right)
  \quad\text{since $\Derp sxx=d\,s$ for all $s\in\Stermsh xd$}\\
  &= \Subst u{h'}h
\end{align*}
and we are done.

\section{Concrete models}
We want now to give concrete examples of categorical models of \DILL.

\subsection{Products, coproducts and the Seely
  isomorphisms}\label{sec:modele-DILL}
In Section~\ref{sec:finctorial-exp}, we introduced the functorial version of
the exponential without mentioning the Seely isomorphisms: as explained in
Paragraph~\ref{par:add-funct-exp}, this choice is very natural when presenting
the denotational interpretation of proof-nets. But when describing the
structure of concrete models, as we want to do now, it is more natural to
assume that the linear category $\cL$ is cartesian and that the $\Excl\_$
comonad is equipped with a strong symmetric monoidal structure from $\IWith$ to
$\ITens$.

So we assume to be given a preadditive *-autonomous category $\cL$ equipped with
an exponential structure (Section~\ref{sec:exp-struct}) where $\Excl\_$ is a
monoidal comonad satisfying the conditions of Section~\ref{sec:finctorial-exp}.

We assume moreover that $\cL$ is cartesian, with terminal object $\Top$,
cartesian product $\IWith$, projections
$\Proj i\in\cL(\With{X_1}{X_2},X_i)$. Because $\cL$ is preadditive, this implies
that $\Top$ is also an initial object, and that $\With{X_1}{X_2}$ (together
with suitably defined injections) is also the coproduct of $X_1$ and $X_2$. In
other words, $\cL$ is an additive monoidal category.

We assume to be given an isomorphism $\SeelyZ\in\cL(\One,\Excl\Top)$ and a
natural isomorphism
$\SeelyB_{X_1,X_2}
\in\cL(\Tens{\Excl{X_1}}{\Excl{X_2}},\Excl{(\With{X_1}{X_2})})$ which endow
the functor $\Excl\_$ with a monoidal structure. 
This means that diagrams similar to~\Eqref{eq:excl-mon-unit-left},
\Eqref{eq:excl-mon-unit-right}, \Eqref{eq:excl-mon-ass} and
\Eqref{eq:excl-mon-sym} hold.

We also require the following diagram to commute
\begin{equation*}
    \begin{tikzpicture}[->, >=stealth]
    \node (1) {$\Tens{\Excl X}{\Excl Y} $};
    \node (2) [right of=1, node distance=28mm] 
      {$\Excl{(\With XY)} $};
    \node (3) [below of=2, node distance=12mm] 
      {$\Excl{\Excl{(\With XY)}} $};
    \node (4) [below of=1, node distance=24mm] 
      {$\Tens{\Excl{\Excl X}}{\Excl{\Excl Y}} $};
    \node (5) [below of=3, node distance=12mm]
      {$\Excl{(\With{\Excl X}{\Excl Y})} $};
    \tikzstyle{every node}=[midway,auto,font=\scriptsize]
    \draw (1) -- node {$\SeelyB_{X,Y} $} (2);
    \draw (1) -- node [swap] {$\Tens{\Digg X}{\Digg Y} $} (4);
    \draw (2) -- node {$\Digg{\With XY} $} (3);
    \draw (3) -- node {$\Excl{\Pair{\Excl{\Proj 1}}{\Excl{\Proj 2}}} $} (5);
    \draw (4) -- node {$\SeelyB_{\Excl X,\Excl Y} $} (5);
  \end{tikzpicture}
\end{equation*}

  Of course, there is a connection between these two monoidal structures on
  $\Excl\_$. The morphism $\ExpMonZ$ is the following composition of morphisms:
\begin{equation*}
    \begin{tikzpicture}[->, >=stealth]
    \node (1) {$\One $};
    \node (2) [right of=1, node distance=12mm] {$\Excl\Top $};
    \node (3) [right of=2, node distance=12mm] {$\Excl{\Excl\Top} $};
    \node (4) [right of=3, node distance=20mm] {$\Excl\One $};
    \tikzstyle{every node}=[midway,auto,font=\scriptsize]
    \draw (1) -- node {$\SeelyZ $} (2);
    \draw (2) -- node {$\Digg\Top $} (3);
    \draw (3) -- node {$\Excl{(\Funinv{(\SeelyZ_\One)})} $} (4);
  \end{tikzpicture}
\end{equation*}

and $\ExpMonB_{X,Y}$ is
\begin{equation*}
    \begin{tikzpicture}[->, >=stealth]
    \node (1) {$\Tens{\Excl X}{\Excl Y} $};
    \node (2) [right of=1, node distance=26mm] {$\Excl{(\With XY)} $};
    \node (3) [right of=2, node distance=26mm] {$\Excl{\Excl{(\With XY)}} $};
    \node (4) [right of=3, node distance=32mm] 
      {$\Excl{(\Tens{\Excl X}{\Excl Y})} $};
    \node (5) [right of=4, node distance=32mm] 
      {$\Excl{(\Tens XY)} $};
    \tikzstyle{every node}=[midway,auto,font=\scriptsize]
    \draw (1) -- node {$\SeelyB_{X,Y} $} (2);
    \draw (2) -- node {$\Digg{\With XY} $} (3);
    \draw (3) -- node {$\Excl{(\Funinv{\SeelyB_{X,Y}})} $} (4);
    \draw (4) -- node {$\Excl{(\Tens{\Der X}{\Der Y})} $} (5);
  \end{tikzpicture}
\end{equation*}

The bi-algebraic structure of $\Excl X$ presented in
Section~\ref{sec:exp-struct} is also related to this Seely monoidal
structure. 

For the coalgebraic part, let $\Diag X\in\cL(X,\With XX)$ be the
diagonal morphism associated with the cartesian product of $X$ with
itself. Then we have 
\[
\Contr X=\Funinv{\SeelyB_{X,X}}\Compl\Excl{\Diag X}:\Excl
X\to\Tens{\Excl X}{\Excl X}
\]
Similarly we set 
\[
\Weak X=\SeelyZ_\One\Compl\Excl{\tau_X}
\]
where $\tau_X:X\to \Top$ is the unique morphism to the terminal object. The
algebraic part satisfies similar conditions, using the codiagonal $\Codiag
X:\With XX\to X$ and the morphism $\tau'_X:\Top\to X$.

\subsection{Relational semantics}\label{sec:relational}

We introduce now the simplest $*$-autonomous category equipped with an
exponential structure: the category of sets and relations. For this model, we
assume that $\Field=\{0,1\}$ with addition defined by $1+1=1$.

Let $\REL$ be the category whose objects are sets and where
$\REL(X,Y)=\Part{X\times Y}$, identities being the diagonal relations and
composition being defined as follows: if $R\in\REL(X,Y)$ and $S\in\REL(Y,Z)$
then
\[
S\Compl R=\{(a,c)\in X\times Z\St\exists b\in Y\ (a,b)\in R\
\text{and}\ (b,c)\in S\}\,.
\]
Let $x\subseteq X$, we set $Rx=\{b\in Y\St\exists a\in x\ (a,b)\in R\}\subseteq
Y$ which is the direct image of $x$ by $R$. We also define $\Transp
R=\{(b,a)\in Y\times X\St (a,b)\in Y\}$ which is the transpose of $R$. Given
$x\subseteq X$ and $y'\subseteq Y$, we have
\begin{equation}\label{eq:rel-linapp}
  (Rx)\cap y'=\Setpr 2(R\cap(x\times y'))\quad\text{and}\quad
  (\Transp Ry')\cap x=\Setpr 1(R\cap (x\times y'))
\end{equation}
where $\Setpr 1$ and $\Setpr 2$ are the two projections of the cartesian
product in the category $\SET$ of sets and functions (the ordinary
cartesian product ``$\times$'').

Observe that an isomorphism in $\REL$ is a relation which is a bijection.

The symmetric monoidal structure is given by the tensor product $\Tens
XY=X\times Y$ and the unit $\One$ an arbitrary singleton. The neutrality,
associativity and symmetry isomorphisms are defined as the obvious
corresponding bijections (for instance, the symmetry isomorphism
$\sigma_{X,Y}\in\REL(\Tens XY,\Tens YX)$ is given by $\sigma(a,b)=(b,a)$). This
symmetric monoidal category is closed, with linear function space given by
$\Limpl XY=X\times Y$, the natural bijection between $\REL(\Tens ZX,Y)$ and
$\REL(Z,\Limpl XY)$ being induced by the cartesian product associativity
isomorphism. Last, one takes for $\Bot$ an arbitrary singleton, and this turns
$\REL$ into a $*$-autonomous category. One denotes as $\Star$ the unique
element of $\One$ and $\Bot$.

This category is additive, with cartesian product $\With{X_1}{X_2}$ of $X_1$
and $X_2$ defined as $\{1\}\times X_1\cup\{2\}\times X_2$ with projections
$\Proj i=\{((i,a),a)\St a\in X_i\}$ (for $i=1,2$), and terminal object
$\Top=\emptyset$. Then the commutative monoid structure on homsets $\REL(X,Y)$
is defined by $0=\emptyset$ and $f+g=f\cup g$ and the action of $\Field$ on
morphisms is defined by $0\,f=0$ and $1\,f=f$ (there are no other
possibilities).

$\REL$ is also a Seely category (see Section~\ref{sec:modele-DILL}), for a
comonad $\Excl\_$ defined as follows:
\begin{itemize}
\item $\Excl X$ is the set of all finite multisets of elements of $X$;
\item if $R\in\REL(X,Y)$, then we set $\Excl R=\{(\Mset{\List a1n},\Mset{\List
    b1n})\St n\in\Nat\text{ and }\forall i\,(a_i,b_i)\in R\}$;
\item $\Der X\in\REL(\Excl X,X)$ is $\Der X=\{(\Mset a,a)\St a\in X\}$;
\item $\Digg X=\{(m_1+\cdots+m_n,\Mset{\List m1n})\St n\in\Nat\text{ and }\List
  m1n\in\Excl X\}$.
\end{itemize}
The monoidality isomorphism $\SeelyB_{X,Y}\in\REL(\Tens{\Excl X}{\Excl
  Y},\Excl{(\With XY)})$ is the bijection which maps $(\Mset{\List
  a1l},\Mset{\List b1r})$ to
$\Mset{(1,a_1),\dots,(1,a_l),(2,b_1),\dots,(2,b_r)}$. 

Last, we also provide a codereliction natural transformation $\Coder
X\in\REL(X,\Excl X)$ which is simply given by $\Coder X=\{(a,\Mset a)\St a\in
X\}$.

With these definitions, it is easy to see that $\Contr X=\{(l+r,(l,r))\St
l,r\in\Excl X\}$, $\Weak X=\{(\Mset{},\Star)\}$, $\Cocontr X=\{((l,r),l+r)\St
l,r\in\Excl X\}$ and $\Coweak X=\{(\Star,\Mset{})\}$. 
The required diagrams are easily seen to commute.

\paragraph{Antiderivatives.}
This exponential structure is bicommutative and can easily seen to be Taylor in
the sense of Section~\ref{sec:Taylor-structure}. Moreover, it has
antiderivatives in the sense of Section~\ref{sec:primitives}, simply because
the morphism $J_X=\Id_X+(\Coderc X\Compl\Derc X):\Excl X\to\Excl X$ coincides
here with the identity. Indeed $\Derc X=\{(l+\Mset a,(l,a))\St l\in\Excl
X\text{ and }a\in X\}$, $\Coderc X=\{((l,a),l+\Mset a)\St l\in\Excl X\text{ and
}a\in X\}$ and therefore $\Coderc X\Compl\Derc X=\{(l,l)\St l\in\Excl X\text{
  and }\Card l>0\}$.

Concretely, saying that a morphism $f\in\REL(\Tens{\Excl X}{X},Y)$ satisfies
the symmetry condition of Proposition~\ref{prop:poincare-categorique} simply
means that, given $m\in\Mfin X$, $a,a'\in X$ and $b\in Y$, one has $((m+\Mset
a,a'),b)\in f\Equiv((m+\Mset{a'},a),b)\in f$. In that case, the antiderivative
$g\in\REL(\Excl X,Y)$ given by that proposition is simply
\begin{equation*}
  g=\{(m+\Mset a,b)\St ((m,a),b)\in f\}\,.
\end{equation*}



\subsection{Finiteness spaces}\label{sec:finiteness-spaces}
This model can be seen as an enrichment of the model of sets and relations of
Section~\ref{sec:relational}. It can also be described as a category of
topological vector spaces and linear continuous maps. From now on, $\Field$
denotes an arbitrary field which is always endowed with the discrete topology.

\subsection{Linearly topologized vector spaces (ltvs)}
Let $E$ be a $\Field$-vector space.
A \emph{linear topology} on $E$ is a topology $\lambda$ such that there is a
filter $\cL$ of linear subspaces of $E$ with the following property: a subset
$U$ of $E$ is $\lambda$-open iff for any $x\in U$ there exists $V\in\cL$ such
that $x+V\subseteq U$. One says that such a filter $\cL$ \emph{generates} the
topology $\cL$. A $\Field$-ltvs is a $\Field$-vector space equipped with a
linear topology. Observe that $E$ is Hausdorff iff $\bigcap\cL=\{0\}$ (for some,
and hence any, generating filter $\cL$); from now on we assume always that this
is the case.

\begin{proposition}
  Let $E$ be a $\Field$-ltvs.  Any linear subspace $U$ of $E$ which is a
  neighborhood of $0$ is both open and closed. So $E$ is totally disconnected
  (the only subsets of $E$ which are connected are the empty set and the one
  point sets).
\end{proposition}
\Beginproof
Let $\cL$ be a generating filter for the topology of $E$.  First, let $x\in U$
and let $V\in\cL$ be such that $V\subseteq U$ (such a $V$ exists because $U$ is
a neghborhood of $0$), then we have $x+V\subseteq U$ since $U$ is a linear
subspace and hence $U$ is open. Next let $x\in E\setminus U$. If $y\in U\cap
(x+U)$ then we have $y-x\in U$ and hence $x\in U$ since $y\in U$ and $U$ is a
linear subspace: contradiction. Therefore $U\cap(x+U)=\emptyset$ and $U$ is
closed since $U$ is open.
\Endproof

Any linear subspace which contains an open linear subspace is open.

\subsubsection{Cauchy completeness.}
A \emph{net} in $E$ is a family $(x(d))_{d\in D}$ of elements of $E$ indexed by
a directed set $D$. The net $(x(d))_{d\in D}$ \emph{converges} to $x\in E$ if,
for any neighborhood $U$ of $0$, there exists $d\in D$ such that $\forall e\in
D\ e\geq d\Implies x(e)-x\in U$. Because $E$ is Hausdorff, a net converges to at
most one point. As usual, one can check that a subset $U$ of $E$ is open iff,
for any net $(x(d))_{d\in D}$ which converges to a point $x\in U$, there exists
$d\in D$ such that $\forall e\in D\ e\geq d\Implies x(e)\in U$.

A net $(x(d))_{d\in D}$ is \emph{Cauchy} if, for any neighborhood $U$ of $0$,
there exists $d\in D$ such that $\forall e,e'\in D\ e,e'\geq d\Implies
x(e)-x({e'})\in U$. This latter statement is equivalent to $\forall e\in D\
e\geq d\Implies x(e)-x(d)\in U$.

One says that $E$ is \emph{complete} if any Cauchy net in $E$ converges.

\subsubsection{Linear boundedness.}
Let $E$ be an ltvs and let $U$ be an open linear subspace of $E$. Let
$\pi_U:E\to E/U$ be the canonical projection. This map is of course linear, and
its kernel is $U$ which is a neighborhood of $0$. This means that, endowing
$E/U$ with the discrete topology, $\pi_U$ is continuous. Hence the quotient
topology on $E/U$ is the discrete topology.

We say that a subspace $B$ of $E$ is \emph{linearly bounded} if $\pi_U(B)$ is
finite dimensional, for every linear open subspace $U$ of $E$. In other words,
for any linear open subspace $U$, there is a finite dimensional subspace $A$ of
$E$ such that $B\subseteq U+A$.

\begin{proposition}\label{prop:lbounded-properties}
  Any finite dimensional subspace of an ltvs $E$ is linearly bounded. Let $B_1$
  and $B_2$ be subspaces of $E$. If $B_1\subseteq B_2$ and $B_2$ is linearly
  bounded, so is $B_1$. If $B_1$ and $B_2$ are linearly bounded, so is
  $B_1+B_2$.
\end{proposition}
A collection of subspaces of a vector space $F$ having these properties is
called a \emph{linear bornology} on $F$.

An ltvs $E$ is \emph{locally linearly bounded} if it has a linear open subspace
which is linearly bounded. 


\subsubsection{Linear and multilinear maps.}
Let $\List E1n$ and $F$ be $\Field$-ltvs's. An $n$-multilinear function
$f:E_1\times\cdots\times E_n\to F$ is \emph{hypocontinuous} if, for any
$i\in\{1,\dots,n\}$, any linear open subspace $V\subseteq F$ and any linearly
bounded subspaces $B_1\subseteq E_1$,\dots,$B_{i-1}\subseteq E_{i-1}$,
$B_{i+1}\subseteq E_{i+1}$,\dots,$B_{n}\subseteq E_{n}$, there exists an open
linear subspace $U\subseteq E_i$ such that $f(B_1\times\cdots\times
B_{i-1}\times U\times B_{i+1}\times\cdots\times B_n)\subseteq V$.

We denote by $\Limpl{(\List E1n)}{F}$ the $\Field$-vector space of all such
multilinear maps. Given linearly bounded subspaces $\List B1n$ of $\List E1n$
respectively and given a linear open subspace $V$ of $F$, we define
\begin{eqnarray*}
  \Ann(\List B1n,V)=\{f\in\Limpl{(\List E1n)}{F}\St f(B_1\times\cdots\times
  B_n)\subseteq V\}\,.
\end{eqnarray*}
This is a linear subspace of $\Limpl{(\List E1n)}{F}$ and by
Proposition~\ref{prop:lbounded-properties} these subspaces form a filter which
defines a linear topology on $\Limpl{(\List E1n)}{F}$ and this topology is
Hausdorff. Indeed, if $f\in\Limpl{(\List E1n)}{F}$ is $\not=0$, then take
$x_i\in E_i$ such that $f(\List x1n)\not=0$. Since $F$ is Hausdorff, there is a
linear neighborhood $V$ of $0$ in $F$ such that $f(\List x1n)\notin V$. Let
$B_i=\Field x_i$; this is a linearly bounded subspace of $E_i$ and
$f(B_1\times\cdots\times B_n)\not\subseteq V$.

In the case $n=1$ (and $E=E_1$), the corresponding maps $f:E\to F$ are simply
called linear, and they are continuous. The corresponding function space is
denoted as $\Limpl EF$. 

If $F=\Field$, the corresponding maps are called (multi)linear
(hypo)continuous forms. If furthermore $n=1$ the corresponding function space
is denoted as $E'$ and is called \emph{topological dual} of $E$.

\begin{proposition}\label{prop:multilinear-bounded}
  Let $f:E_1\times\cdots\times E_n\to F$ be multilinear and hypocontinuous and
  let $B_i\subseteq E_i$ be linearly bounded subspaces for $i=1,\dots,n$. Then
  $f(B_1\times\cdots\times B_n)$ is a linearly bounded subspace of $F$.
\end{proposition}
\Beginproof
Let $V$ be an open linear subspace of $F$. Let $U_1$ be an open linear subspace
of $E_1$ such that $f(U_1\times B_1\times\cdots\times B_n)\subseteq V$. Let
$A_1$ be a finite dimensional subspace of $E_1$ such that $B_1\subseteq
U_1+A_1$, we have $f(B_1\times\cdots\times B_n)\subseteq V+f(A_1\times
B_2\times\cdots\times B_n)$. Since $A_1$ is bounded, one can find similarly a
finite dimensional subspace $A_2$ of $E_2$ such that $f(A_1\times
B_2\times\cdots\times B_n)\subseteq V+f(A_1\times A_2\times
B_3\times\cdots\times B_n)$ and hence (since $V+V=V$) we get
$f(B_1\times\cdots\times B_n)\subseteq V+f(A_1\times A_2\times
B_3\times\cdots\times B_n)$. Continuing this process, we find finite
dimensional subspaces $A_i$ of $E_i$ for $i=1,\dots,n$ such that
$f(B_1\times\cdots\times B_n)\subseteq V+f(A_1\times\cdots\times A_n)$ and we
conclude that $f(B_1\times\cdots\times B_n)$ is linearly bounded since
$f(A_1\times\cdots\times A_n)$ is finite dimensional.
\Endproof

It is tempting to think that (multi)linear continuous maps could be
characterized as those which preserve linear boundedness. This cannot be the
case: think of a linear map $f:E\to F$ where $F$ is finite dimensional. Such a
map preserves linear boundedness (any subspace of $F$ is linearly bounded) but
has no reason to be continuous.

\subsection{Finiteness spaces and the related ltvs's}
We restrict now our attention to particular ltvs's which can be described in a
simple combinatorial way.

\subsubsection{Basic definitions.}
Let $I$ be a set. Given $\cF\subseteq\Part I$, we define
$\Orth\cF\subseteq\Part I$ by
\[
\Orth\cF=\{u'\subseteq I\St\forall u\in\cF\  u\cap u'\text{ is finite}\}\,.
\]
We have $\cF\subseteq\cG\Implies\Orth\cG\subseteq\Orth\cF$,
$\cF\subseteq\Biorth\cF$ and therefore $\Triorth\cF=\Orth\cF$.

A \emph{finiteness space} is a pair $X=(\Web X,\Fin X)$ where $\Web X$ is a set
and $\Fin X\subseteq{\Part{\Web X}}$ satisfies $\Fin X=\Biorth{\Fin X}$. The
following properties follow easily from the definition
\begin{itemize}
\item if $u\subseteq\Web X$ is finite then $u\in\Fin X$
\item if $u,v\in\Fin X$ then $u\cup v\in\Fin X$
\item if $u\subseteq v\in\Fin X$, then $u\in\Fin X$.
\end{itemize}
Let us prove for instance the second statement. Let $u'\in\Orth{\Fin X}$, then
$(u\cup v)\cap u'=(u\cap u')\cup(v\cap u')$ is finite since both sets $u\cap
u'$ and $v\cap u'$ are finite by our hypothesis that $u,v\in\Fin X$. Since this
holds for all $u'\in\Orth{\Fin X}$, we have $u\cup v\in\Biorth{\Fin X}=\Fin X$.

A \emph{strong isomorphism}\footnote{This would coincide with the categorical
  notion of isomorphism if we were using morphisms which are defined as
  relations. With linear continuous maps (between the associated ltvs's) as
  morphisms, the present notion of isomorphism is a particular case of the
  standard categorical one: we can have more linear homeomorphisms from
  $\Fmod\Field X$ to $\Fmod\Field Y$ than those which are generated by such
  finiteness-preserving bijections between webs.} between two finiteness spaces
$X$ and $Y$ is a bijection $\phi:\Web X\to\Web Y$ such that, for all
$u\subseteq\Web X$, one has $u\in\Fin X$ iff $\phi(u)\in\Fin Y$.

Let $X$ be a finiteness space. We define a $\Field$-vector space $\Fmod\Field
X$ as the set of all families $x\in\Field^{\Web X}$ such that the set $\Supp
x=\{a\in\Web X\St x_a\not=0\}$ belongs to $\Fin X$.

Given $u'\in\Orth{\Fin X}$, we define a linear subspace of $\Fmod\Field X$ by
\begin{equation*}
  \Neigh X{u'}=\{x\in\Fmod\Field X\St\Supp x\cap u'=\emptyset\}\,.
\end{equation*}
Observe first that $\forall u',v'\in\Orth{\Fin X}\quad u'\subseteq
v'\Equiv\Neigh X{v'}\subseteq\Neigh X{u'}$.

Since, given $u',v'\in\Orth{\Fin X}$, we have $\Neigh X{u'\cup v'}=\Neigh
X{u'}\cap\Neigh X{v'}$, the set $\{\Neigh X{u'}\St u'\in\Orth{\Fin X}\}$ is a
filter of linear subspaces of $\Fmod\Field X$. Moreover, observe that
$\bigcap_{u'\in\Orth{\Fin X}}\Neigh X{u'}=\{0\}$ (because $\forall a\in\Web X\
\{a\}\in\Orth{\Fin X}$), and therefore this filter defines an Hausdorff linear
topology on $\Fmod\Field X$, that we call the \emph{canonical topology} of
$\Fmod\Field X$.

\begin{proposition}
  For any finiteness space $X$, the ltvs $\Fmod\Field X$ is Cauchy-complete.
\end{proposition}
\Beginproof
Let $(x(d))_{d\in D}$ be a Cauchy net in $\Fmod\Field X$. Let $a\in\Web X$. By
taking $u'=\{a\}$ in the definition of a Cauchy net, we see that there exist
$x_a\in\Field$ and $d_a\in D$ such that $\forall e\geq d_a\ x(e)_a=x_a$. In
that way we have defined $x=(x_a)_{a\in\Web X}\in\Field^{\Web X}$

We prove first that
\begin{equation}\label{eq:complete-basic}
  \forall u'\in\Orth{\Fin X}\,\exists d\in D\,\forall e\geq d\,\forall a\in u'
  \quad
  x(e)_a=x_a\,.
\end{equation}
Let $u'\in\Orth{\Fin X}$. Let $d^0\in D$ be such that
$x(e)-x(d^0)\in\Neigh X{u'}$ for all $e\geq d^0$. Let $a\in u'$ and let $d_a\geq
d^0$ be such that $x(e)_a=x_a$ for all $e\geq d_a$. Let $e\geq d^0$. Let
$e'\geq e,d_a$. We have $x_a=x(e')_a$ since $e'\geq d_a$ and $x(e')_a=x(e)_a$
since $e,e'\geq d^0$ and $a\in u'$. It follows that $\forall a\in u'\
x_a=x(e)_a$.

From this we deduce now that $x\in\Fmod\Field X$. Let $u'\in\Orth{\Fin X}$. Let
$d\in D$ be such that $\forall e\geq d\,\forall a\in u' \ x(e)_a=x_a$. Then
$\Supp x\cap u'=\Supp{x(d)}\cap u'$ is finite, so $\Supp x\in\Biorth{\Fin
  X}=\Fin X$, that is $x\in\Fmod\Field X$.

Now Condition~(\ref{eq:complete-basic}) expresses exactly that $\lim_{d\in
  D}x(d)=x$ and hence the net $(x(d))_{d\in D}$ converges.
%
%
%
\Endproof

A natural question is whether the ltvs $\Fmod\Field X$, which is Hausdorff, is
always metrizable. We provide a necessary and sufficient condition under which
this is the case.

\begin{proposition}\label{prop:metrizable-charact}
  Let $X$ be a finiteness space. The ltvs $\Fmod\Field X$ is metrizable iff
  there exists a sequence $(u'_n)_{n\in\Nat}$ of elements of $\Orth{\Fin X}$
  which is monotone ($n\leq m\Implies u'_n\subseteq u'_m$) and such that
  $\forall u'\in\Orth{\Fin X}\,\exists n\in\Nat\ u'\subseteq u'_n$.
\end{proposition}
\Beginproof
Let first $(u'_n)_{n\in\Nat}$ be a sequence of elements of $\Orth{\Fin X}$
which satisfies the condition stated above. Given $x,y\in\Fmod\Field X$, we
define
\begin{equation*}
  d(x,y)=
  \begin{cases}
    0 & \text{if $x=y$}\\
    2^{-n} & \text{if $x\not=y$ and $n$ is the least integer}\\
    & \text{such that $u'_n\cap\Supp{x-y}\not=\emptyset$}\,.
  \end{cases}
\end{equation*}
Indeed, if $x\not=y$, then $\Supp{x-y}\not=\emptyset$ and hence, taking
$a\in\Supp{x-y}$, we can find $n\in\Nat$ such that $\{a\}\subseteq u'_n$. This
function $d$ is easily seen to be an ultrametric distance (that is
$d(x,z)\leq\max(d(x,y),d(y,z))$) and it generates the canonical
topology of $\Fmod\Field X$. Indeed we have
\begin{equation*}
  d(x,y)<2^{-n}\Iff x-y\in\Neigh X{u'_n}
\end{equation*}
(indeed, $d(x,y)<2^{-n}$ means that the least $m$ such that $x-y\notin\Neigh
X{u'_m}$ satisfies $m>n$) and hence $B_{2^{-n}}=\Neigh X{u'_n}$, where
$B_\epsilon$ is the open ball centered at $0$ and of radius $\epsilon$.

Conversely, assume that $\Fmod\Field X$ is metrizable and let $d$ be a distance
defining the canonical topology of $\Fmod\Field X$. For each $n\in\Nat$,
$B_{2^{-n}}$ is a neighborhood of $0$ and hence there exist $v'_n\in\Orth{\Fin
  X}$ such that $\Neigh X{v'_n}\subseteq B_{2^{-n}}$. Let
$u'_n=v'_0\cup\cdots\cup v'_n\in\Orth{\Fin X}$. Then
$\Neigh X{u'_n}\subseteq\Neigh X{v'_n}\subseteq B_{2^{-n}}$. Now let
$u'\in\Orth{\Fin X}$, then $\Neigh X{u'}$ is a neighborhood of $0$ and hence
there exists $n$ such that $B_{2^{-n}}\subseteq\Neigh X{u'}$, which implies
$\Neigh X{u'_n}\subseteq\Neigh X{u'}$ and hence $u'\subseteq u'_n$.
\Endproof

It follows that there are non metrizable ltvs associated with finiteness
spaces. We give in Proposition~\ref{lemma:non-metrizable} an example of this
situation which arises in the semantics of \LL{}, using exponential
constructions that will be introduced in
Section~\ref{subsubsec:contr-finiteness}.

\begin{proposition}\label{lemma:non-metrizable}
  The ltvs $\Fmod\Field{\Excl{\Int 1}}$ is not metrizable
\end{proposition}
\Beginproof
Let $X=\Excl{\Int 1}$, so that $\Web X=\Mfin\Nat$ and a subset $u$ for $\Web X$
belongs to $\Fin X$ iff $\exists n\in\Nat\ u\subseteq\Mfin{\{0,\dots,n\}}$. The
proof is a typical Cantor diagonal reasoning.  We assume towards a
contradiction that $\Fmod\Field X$ is metrizable, that is by
Proposition~\ref{prop:metrizable-charact}, we assume that there is a monotone
sequence $(u'_n)_{n\in\Nat}$ of elements of $\Orth{\Fin X}$ such that $\forall
u'\in\Orth{\Fin X}\,\exists n\in\Nat\ u'\subseteq u'_n$. Let $n\in\Nat$, we
have $\{p\Mset n\St p\in\Nat\}\in\Mfin{\{0,\dots,n\}}$ and hence
$u'_n\cap\{p\Mset n\St p\in\Nat\}$ is finite. Therefore we can find a function
$f:\Nat\to\Nat$ such that $\forall n\in\Nat\ f(n)\Mset n\notin u'_n$. Let
$u'=\{f(n)\Mset n\St n\in\Nat\}$. Then $u'\in\Orth{\Fin X}$ since, for any
$n\in\Nat$, $u'\cap\Mfin{\{0,\dots,n\}}=\{f(i)\Mset i\St i\in[0,n]\}$ is
finite. But for all $n\in\Nat$ we have $f(n)\Mset n\in u'\setminus u'_n$ and so
$u'\not\subseteq u'_n$.
\Endproof

We consider this as a very interesting phenomenon which seems to reveal a
relation between the topological complexity of the interpretation of a type
with its logical complexity (alternation of exponentials).

\subsubsection{Linearly bounded subspaces.}
Let $X$ be a finiteness space.  We are interested in characterizing the
linearly bounded subspaces of $\Fmod\Field X$.

Given $u\subseteq\Web X$, let $\Bnd Xu=\{x\in\Fmod\Field X\St\Supp x\subseteq
u\}$. This is a linear subspace of $\Fmod\Field X$.

Let $u\in\Fin X$. We prove that $\Bnd X{u}$ is linearly bounded. Let $u'$ in
$\Orth{\Fin X}$. Observe that $\Neigh X{u'}=\Bnd X{\Web X\setminus u'}$. We
have therefore $\Bnd X{u}\subseteq\Neigh X{u'}+ \Bnd X{u\cap u'}$, and since
$u\cap u'$ is finite, the space $\Bnd X{u\cap u'}$ is finite dimensional. Let
$U$ be an open subspace of $U$, let $u'\subseteq\Orth{\Fin X}$ be such that
$\Neigh X{u'}\subseteq U$. Then $\Bnd X{u}\subseteq U+ \Bnd X{u\cap u'}$. Hence
$\Bnd X{u}$ is linearly bounded. We show now that this condition is actually
sufficient.

\begin{proposition}\label{prop:fspace-bounded-char}
  A linear subspace $B$ of $\Fmod\Field X$ is linearly bounded iff there exists
  $u\in\Fin X$ such that $B\subseteq \Bnd X{u}$.
\end{proposition}
\Beginproof
Assume that $B$ is linearly bounded. Let $u=\bigcup_{x\in B}\Supp x$, so that
$B\subseteq \Bnd X{u}$, we prove that $u\in\Fin X$. Let $u'\in\Orth{\Fin
  X}$. Let $A$ be a finite dimensional subspace of $E$ such that
$B\subset\Neigh X{u'}+A$. Let $A_0$ be a finite generating subset of $A$ and let
$u_0=\bigcup_{y\in A_0}\Supp y\in\Fin X$. Then $x\in A\Implies\Supp x\subseteq
u_0$ (that is $A\subseteq \Bnd X{u_0}$).

Let $x\in B$, we write $x=x_1+x_2$ where $x_1\in\Neigh X{u'}$ and $x_2\in A$. We
have $\Supp x\subseteq\Supp{x_1}\cup\Supp{x_2}$ and hence $u'\cap\Supp
x\subseteq(u'\cap\Supp{x_1})\cup(u'\cap\Supp{x_2})\subseteq u'\cap u_0$ since
$u'\cap\Supp{x_1}=\emptyset$.  Since this holds for all $x\in B$, we have
$u'\cap u\subseteq u'\cap u_0$ so $u'\cap u$ is finite and hence $u\in\Fin X$
\Endproof

\begin{proposition}
  The ltvs $\Fmod\Field X$ is locally linearly bounded iff there exist
  $u\in\Fin X$ and $u'\in\Orth{\Fin X}$ such that $u\cup u'=\Web X$.
\end{proposition}
This is an obvious consequence of Proposition~\ref{prop:fspace-bounded-char}.

$\FINV$ is the category whose objects are the finiteness spaces and such
that $\FINV(X,Y)$ is the set of all continuous linear maps $\Fmod\Field
X\to\Fmod\Field Y$.

\subsubsection{Constructions of finiteness spaces.}
\label{subsubsec:contr-finiteness}
We give a number of constructions on finiteness spaces which allow one to
interpret differential \LL{}, starting with the most important one,
which is the linear function space.

The most striking features of these constructions can be summarized by the two
following statements.
\begin{itemize}
\item In spite of the fact that these constructions are algebraic in nature
  (tensor product, linear function space, topological dual etc), they are
  entirely performed on the webs of the finiteness spaces and do not involve
  the scalar coefficients. This means in particular that they do not depend on
  the choice of the field, and this is quite surprising.
\item So, these constructions are performed on the webs, but they do not really
  depend on them, in the following sense. Defining an \emph{intrinsic
    finiteness space} as a $\Field$-ltvs which is linearly homeomorphic to
  $\Fmod\Field X$ for \emph{some} finiteness space $X$, all these constructions
  can be transferred to the category of intrinsic finiteness spaces and
  continuous and linear maps.
\end{itemize}

Let $X$ and $Y$ be finiteness spaces. Let $\Limpl XY$ be the finiteness space
such that $\Web{\Limpl XY}=\Web X\times\Web Y$ and 
\begin{align*}
  \Fin{\Limpl XY} &= \Orth{\{u\times v'\St u\in\Fin X\ \text{and}\
    v'\in\Fin{\Orth
      Y}\}}\\
  &= \{w\subseteq\Web X\times\Web Y\St \forall u\in\Fin X\,\forall
  v'\in\Orth{\Fin Y}\ w\cap(u\times v')\ \text{is finite}\}
\end{align*}
Let $w\in\Fin{\Limpl XY}$, $u\in\Fin X$ and $v'\in\Orth{\Fin{X}}$. It follows
from~\Eqref{eq:rel-linapp} that $wu\in\Fin Y$ and that $\Transp
wv'\in\Orth{\Fin X}$.

Let $M\in\Fmod\Field{\Limpl XY}$. If $x\in\Fmod\Field X$ and $b\in\Web Y$, then
$\Transp{\Supp M}\{b\}\in\Orth{\Fin X}$ and hence the sum $\sum_{a\in\Web
  X}M_{a,b}x_a$ is finite. Therefore we can define $Mx\in\Field^{\Web Y}$ by
$Mx=(\sum_{a\in\Web X}M_{a,b}x_a)_{b\in\Web Y}$. Since $\Supp{Mx}\subseteq\Supp
M\Supp x$, we have $Mx\in\Fmod\Field Y$ and hence the function $\Funofmat M$
defined by $\Funofmat M(x)=Mx$ is a linear map $\Fmod\Field X\to\Fmod\Field
Y$. Moreover, $\Funofmat M$ is continuous. Indeed, for any $v'\in\Orth{\Fin Y}$
we have $\Neigh X{\Transp{\Supp M}v'}\subseteq\Funinv{\Funofmat M}(\Neigh
Y{v'})$ and hence $\Funinv{\Funofmat M}(\Neigh Y{v'})$ is open since
$\Transp{\Supp M}v'\in\Orth{\Fin X}$.

Given finiteness spaces $Z_1,Z_2$, we define immediately the finiteness space
$\Tens{Z_1}{Z_2}$ as $\Tens{Z_1}{Z_2}=\Orth{(\Limpl{Z_1}{\Orth{Z_2}})}$, so
that $\Web{\Tens{Z_1}{Z_2}}=\Web{Z_1}\times\Web{Z_2}$. One of the most pleasant
features of the theory of finiteness spaces is the following property
(see~\cite{Ehrhard00b}) which has been considerably generalized
in~\cite{TassonVaux10}.
\begin{proposition}\label{prop:fin-tensor-char}
  Let $w\subseteq\Web{Z_1}\times\Web{Z_2}$. One has $w\in\Fin{\Tens{Z_1}{Z_2}}$
  iff $\Setpr i(w)\in\Fin{Z_i}$ for $i=1,2$.
\end{proposition}

Coming back to linear function spaces, this means in particular that, given
$w\subseteq\Web X\times\Web Y$, one has $w\in\Orth{\Fin{\Limpl XY}}$ iff there
are $u\in\Fin X$ and $v'\in\Orth{\Fin Y}$ such that $w\subseteq u\times v'$,
from which we derive a simple characterization of the topology of linear
function spaces.

\begin{proposition}
  The function $\Funofmatm_{X,Y}:M\mapsto\Funofmat M$ is a linear homeomorphism
  from $\Fmod\Field{\Limpl XY}$ to $\Limpl{\Fmod\Field X}{\Fmod\Field Y}$,
  equipped with the topology of uniform convergence on linearly bounded
  subspaces.
\end{proposition}
\Beginproof
The proof that $\Funofmatm_{X,Y}$ is a linear isomorphism can be found
in~\cite{Ehrhard00b}. We prove that this linear isomorphism is an
homeomorphism. Let $B\subseteq\Fmod\Field X$ be a bounded subspace and
$V\subseteq\Fmod\Field Y$ is an open subspace. Let $u\in\Fin X$ be such that
$B\subseteq \Bnd X{u}$ and let $v'\in\Orth{\Fin{Y}}$ be such that $\Neigh
Y{v'}\subseteq V$. Then $u\times v'\in\Orth{\Fin{\Limpl XY}}$ and hence
$\Neigh{\Limpl XY}{u\times v'}\subseteq\Fmod\Field{\Limpl XY}$ is an open
subspace. Let $M\in\Neigh{\Limpl XY}{u\times v'}$, $x\in B$ and $b\in v'$, we
have $(Mx)_b=0$ since $\Supp x\subseteq u$, which shows that
$\Funofmatm_{X,Y}(M)(B)\subseteq V$ and hence $\Funofmatm_{X,Y}$ is
continuous. 

Let now $W\subseteq\Fmod\Field{\Limpl XY}$ be an open subspace. Let
$w\in\Orth{\Fin{\Limpl XY}}$ be such that $\Neigh{\Limpl XY}w\subseteq W$. By
Proposition~\ref{prop:fin-tensor-char}, there are $u\in\Fin X$ and
$v'\in\Orth{\Fin Y}$ such that $w\subseteq u\times v'$, and hence
$\Neigh{\Limpl XY}{u\times v'}\subseteq W$. Then, given $M\in\Neigh{\Limpl
  XY}{u\times v'}$, we have $\Funofmatm_{X,Y}(M)(\Bnd X{u})\subseteq\Neigh
Y{v'}$, which shows that $\Funofmatm_{X,Y}(W)$ is an open linear subspace of
$\Limpl{\Fmod\Field X}{\Fmod\Field Y}$.

We have seen that $\Funofmatm_{X,Y}$ is a continuous and open bijection and
hence it is an homeomorphism.
\Endproof

The tensor product $\Tens{X}{Y}$ defined above is characterized by a
standard universal property: it classifies the hypocontinuous bilinear maps. 

Given vectors $x\in\Fmod\Field{X}$ and $y\in\Fmod\Field{Y}$, then
$\Tens{x}{y}\in\Field^{\Web{\Tens{X}{Y}}}$ defined by $(\Tens
xy)_{(a,b)}=x_ay_b$ is clearly an element of $\Fmod\Field{\Tens XY}$ since
$\Supp{\Tens xy}=\Supp x\times\Supp y\in\Fin{\Tens XY}$. The map
\begin{align*}
  \tau:\Fmod\Field X\times\Fmod\Field Y&\to\Fmod\Field{\Tens XY}  \\
  (x,y)&\mapsto\Tens xy
\end{align*}
 is obviously
bilinear, let us check that it is hypocontinuous. 

Let $W$ be an open linear subspace of $\Fmod\Field{\Tens XY}$ and let
$w'\in\Orth{\Fin{\Tens XY}}$ be such that $\Neigh{\Tens XY}{w'}\subseteq
W$. Let $B\subseteq\Fmod\Field X$ be a linearly bounded subspace and let
$u\in\Fin X$ be such that $B\subseteq \Bnd X{u}$. Since
$w'\in\Fin{\Limpl{X}{\Orth Y}}$ we have $v'=w'u\in\Orth{\Fin Y}$. Let $x\in B$
and $y\in\Neigh{Y}{v'}$, we have $\Supp x\subseteq u$ and hence $\Setpr
2{(\Supp{\Tens xy}\cap{w'})}\subseteq\Setpr 2{((u\times\Supp y)\cap
  w')}=(w'u)\cap\Supp y=\emptyset$ by definition of $\Neigh Y{v'}$. Therefore
$\Tens xy\in\Neigh{\Tens XY}{w'}\subseteq W$. Symmetrically, taking a linearly
bounded subspace $C$ of $\Fmod\Field Y$, we show that there is an open linear
subspace $U$ of $\Fmod\Field X$ such that $\tau(U\times C)\subseteq W$. So the
map $\tau$ is bilinear and hypocontinuous.
\begin{proposition}
  Let $Z$ be a finiteness space and let $f:\Fmod\Field X\times\Fmod\Field
  Y\to\Fmod\Field Z$ be bilinear and hypocontinuous. There exists exactly
  one continuous linear map $\tilde f:\Fmod\Field{\Tens XY}\to\Fmod\Field Z$
  such that $f=\tilde f\Comp\tau$.
\end{proposition}
\Beginproof
We define a matrix $M\in\Field^{\Web X\times\Web Y\times\Web Z}$ by
$M_{a,b,c}=f(\Bcanon a,\Bcanon b)_c$ and we show first that $\Supp
M\in\Fin{\Limpl{\Tens XY}Z}$. 

So let $u\in\Fin X$, $v\in\Fin Y$ and $w'\in\Orth{\Fin Z}$; we must show that
$\Supp M\cap(u\times v\times w')$ is finite. Let $v'\in\Orth{\Fin Y}$ and $u'\in\Orth{\Fin X}$ be such
that 
\[
f(\Bnd X{u}\times\Neigh Y{v'})\subseteq\Neigh Z{w'} \text{ and } f(\Neigh
X{u'}\times \Bnd Y{v})\subseteq\Neigh Z{w'}\,.
\]
Let $(a,b,c)\in\Supp M\cap(u\times v\times w')$, since $f(\Bcanon a,\Bcanon
b)_c\notin\Neigh Z{w'}$ (by definition of $\Neigh Z{w'}$ and by our assumption
about $(a,b,c)$), we must have 
\[
(\Bcanon a,\Bcanon b)\notin\Bnd X u\times\Neigh
Y{v'}\text{ and }(\Bcanon a,\Bcanon b)\notin\Neigh X{u'}\times\Bnd Yv.
\]
But we know that $a\in u$ and $b\in v$, that is $\Bcanon a\in\Bnd Xu$ and
$\Bcanon b\in\Bnd Yv$. It follows that $\Bcanon a\notin\Neigh X{u'}$, that is
$a\in u'$, and similarly $b\in v'$. 

Since $\Bnd Xu$ and $\Bnd Yv$ are linearly bounded, so is $f(\Bnd Xu\times\Bnd
Yv)$ by Proposition~\ref{prop:multilinear-bounded} and hence there exists
$w\in\Fin Z$ such that 
\[
f(\Bnd Xu\times\Bnd Yv)\subseteq\Bnd Zw\,.
\]
Therefore $f(\Bcanon a,\Bcanon b)\in\Bnd Zw$ and hence $c\in w$. 

We have shown that 
\[\Supp M\cap(u\times v\times w')\subseteq(u\cap
u')\times(v\cap v')\times(w\cap w')
\]
and hence $\Supp M\cap(u\times v\times w')$ is finite, so
$M\in\Fmod\Field{\Limpl{\Tens XY}Z}$.

Let $\tilde f=\Funofmat M$, it is a linear and continuous map from
$\Fmod\Field{\Tens XY}$ to $\Fmod\Field Z$. We have $\tilde f(\Tens{\Bcanon
  a}{\Bcanon b})=f(\Bcanon a,\Bcanon b)$ for each $(a,b)\in\Web X\times\Web
Y$. Let $x\in\Fin X$ and $y\in\Fmod\Field Y$, by separate continuity of $f$
(which is a consequence of hypocontinuity) we
have 
\begin{align*}
  f(x,y)&=f(\sum_{a\in\Web X}x_a\Bcanon a,\sum_{b\in\Web Y}y_b\Bcanon b)\\
  &=\sum_{(a,b)\in\Web X\times\Web Y}x_ay_bf(\Bcanon a,\Bcanon b)\\
  &=\sum_{(a,b)\in\Web X\times\Web Y}x_ay_b\tilde f(\Tens{\Bcanon a}{\Bcanon
    b})\\
  &=\tilde f(\Tens xy)\text{\quad by continuity of $\Tilde f$}\,.
\end{align*}
Uniqueness of the continuous linear map $\tilde f$ results from the fact that
necessarily $\tilde f(\Tens{\Bcanon a}{\Bcanon b})=f(\Bcanon a,\Bcanon b)$.
\Endproof

Then one proves easily that the category $\FINV$ equipped with this tensor
product (whose neutral object is $\One$, which satisfies obviously
$\Fmod\Field\One=\Field$) is $*$-autonomous, the object of morphisms from $X$
to $Y$ being $\Limpl XY$ and the dualizing object being $\Bot=\One$ (indeed,
the finiteness spaces $\Limpl X\Bot$ and $\Orth X$ are obviously strongly
isomorphic).

This category is preadditive in the sense of Section~\ref{sec:additive-lin-cat}
since homsets $\FINV(X,Y)$ have an obvious structure of $\Field$-vector space
which is compatible with all the categorical operations introduced so far.

Countable products and coproducts are available as well. Let $(X_i)_{i\in I}$
be a countable family of finiteness spaces. The finiteness space
$X=\bigwith_{i\in I}X_i$ is given by $\Web X=\bigcup_{i\in I}\Web{X_i}$ and
$\Fin X=\{w\subseteq\Web X\St\forall i\in I\ w_i\in\Fin{X_i}\}$ where
$w_i=\{a\in\Web{X_i}\St (i,a)\in w\}$. It is easy to check that 
\[
\Orth{\Fin
  X}=\{w'\subseteq\Web X\St\forall i\in I\ w'_i\in\Orth{\Fin{X_i}}\text{ and
  $w'_i=\emptyset$ for almost all $i$}\}
\]
and it follows that $\Biorth{\Fin X}=\Fin X$. It is clear that
$\Fmod\Field{\bigwith_{i\in I}X_i}=\prod_{i\in I}\Fmod\Field{X_i}$ up to a
straightforward strong isomorphism and that $\bigwith_{i\in I}X_i$ together
with projections $\Proj j:\Fmod\Field{\bigwith_{i\in I}X_i}\to\Fmod\Field{X_j}$
defined in the obvious way, is the cartesian product of the $X_i$'s. 

Thanks to $*$-autonomy, the coproduct of the $X_i$'s is given by
$\bigoplus_{i\in I}X_i=\Orth{\left(\bigwith_{i\in I}\Orth{X_i}\right)}$ and
$\Fmod\Field{\bigoplus_{i\in I}X_i}\subseteq\prod_{i\in I}\Fmod\Field{X_i}$ is
the space of all families $(x_i)_{i\in I}$ of vectors such that $x_i=0$ for
almost all $i\in I$. Of course, the canonical linear topology on
$\Fmod\Field{\bigwith_{i\in I}X_i}$ is the product topology, but the canonical
topology on $\Fmod\Field{\bigoplus_{i\in I}X_i}$ is much finer: it is generated
by all products $\prod_{i\in I}V_i$ where $V_i$ is a linear neighborhood of $0$
in $\Fmod\Field{X_i}$. 

For finite families of objects, products and coproducts coincide.

Let $X$ be a finiteness space. We define $\Excl X$ by $\Web{\Excl X}=\Mfin{\Web
  X}$ and
\[
\Fin{\Excl X}=\{A\subseteq\Web{\Excl X}\St\bigcup_{m\in A}\Supp
m\in\Fin X\}
\] 
and it can be proved that indeed $\Fin{\Excl X}=\Biorth{\Fin{\Excl X}}$ (again,
see~\cite{TassonVaux10} for more general results of this kind). 

Given $x\in\Fmod\Field X$ and $m\in\Web{\Excl X}$, we set 
\[
x^m=\prod_{a\in\Web X}x_a^{m(a)}\in\Field
\] 
(this is a finite product since $\Supp m$ is a finite set), so that 
\[
\Prom
x=(x^m)_{m\in\Web{\Excl X}}\in\Fmod\Field{\Excl X}
\] 
by definition of $\Fin{\Excl X}$. Let $M\in\Fmod\Field{\Limpl{\Excl X}Y}$, it
is not hard to see that one defines a map $\Funkofmat M:\Fmod\Field
X\to\Fmod\Field Y$ by setting 
\[
\Funkofmat M(x)=\left(\sum_{m\in\Web{\Excl
      X}}M_{m,b}x^m\right)_{b\in\Web Y}
\]
all these sums are indeed finite, see~\cite{Ehrhard00b} for the details. When
the field $\Field$ is infinite, the map $M\mapsto\Funkofmat M$ is injective.

In~\cite{Ehrhard00b}, it is also proven that $\Excl\_$ is a functor. Given
$M\in\Fmod\Field{\Limpl XY}$ one defines $\Excl M\in\Fmod\Field{\Limpl{\Excl
    X}{\Excl Y}}$ by setting, for $m\in\Web{\Excl X}$ and $p\in\Web{\Excl Y}$,
\[
(\Excl M)_{m,p}=\sum_{r\in L(m,p)}\Multinom prM^r
\]
where 
\[
L(m,p)=\{r\in\Mfin{\Web X\times\Web Y}\St \sum_{b\in\Web
  Y}r(a,b)=m(a)\text{ and }\sum_{a\in\Web X}r(a,b)=p(b)\}
\]
(so that $r\in L(m,p)\Implies \Card m=\Card r=\Card p$) and 
\[
\Multinom
pr=\prod_{b\in\Web Y}\frac{\Factor{p(b)}}{\prod_{a\in\Web
    X}\Factor{r(a,b)}}\in\Natnz\,.
\]
is a generalized multinomial coefficient.  

This operation is functorial: $\Excl\Id=\Id$ and $\Excl M\Compmat\Excl
N=\Excl{(M\Compmat N)}$, and we also have 
\[
\Funkofmat{\Excl M}(x)=\Excl
M\Compmat\Prom x=\Prom{(\Compmat Mx)}\,.
\]
When $\Field$ is infinite, this latter equation completely characterizes $\Excl
M$, by injectivity of the operation $\Fun$ in that case. This functor has a
comonad structure, of which we recall here only the counit $\Der
X\in\Fmod\Field{\Limpl{\Excl X}{X}}$ given by $(\Der
X)_{m,a}=\Kronecker{m}{\Mset a}$.

The bijection $\Web{\Excl{(\With XY)}}\to\Web{\Tens{\Excl X}{\Excl Y}}$ which
maps the element $q\in\Web{\Excl{(\With XY)}}$ to the pair
$(m,p)\in\Web{\Tens{\Excl X}{\Excl Y}}$ defined by $m(a)=q(1,a)$ and
$p(b)=q(2,b)$ is a strong isomorphism of finiteness spaces. We also have a
strong isomorphism from $\Excl\Bot$ to $\One$. These strong isomorphisms induce
natural isomorphisms $\SeelyB_{X,Y}\in\FINV(\Tens{\Excl X}{\Excl
  Y},\Excl{(\With XY)})$ and $\SeelyZ\in\FINV(\One,\Excl\Top)$ which
endow the functor $\Excl\_$ with a monoidality structure from
$(\FINV,\IWith,\Top)$  to $(\FINV,\ITens,\One)$, satisfying
moreover the coherence diagram~\Eqref{eq:seely-digging-coherence}: to
summarize, equipped with the structure described above, $\FINV$ is a
Seely category, that is, a categorical model of classical \LL{}.

Applying the general recipe of Section~\ref{sec:modele-DILL}, we get the
contraction natural transformation $\Contr X:\Excl X\to\Tens{\Excl X}{\Excl X}$
and the weakening morphism $\Weak X:\Excl X\to\One$. We check that $(\Weak
X)_{m,*}=\Kronecker m{\Mset{}}$ and that $(\Contr X)_{m,(p,q)}=\Kronecker
m{p+q}$. We also get the cocontraction natural transformation $\Cocontr
X:\Tens{\Excl X}{\Excl X}\to\Excl X$ and the coweakening
morphism $\Coweak X:\One\to\Excl X$. And we check that $(\Coweak
X)_{*,m}=\Kronecker m{\Mset{}}$, and that $(\Cocontr
X)_{(p,q),m}=\Binom{p+q}{p}\Kronecker m{p+q}$ where 
\[
\Binom mp=\prod_{a\in\Web
  X}\frac{\Factor{m(a)}}{\Factor{p(a)}\Factor{(m(a)-p(a))}}\in\Natnz
\]
is a generalized binomial coefficient.

We also have a codereliction natural transformation $\Coder X:X\to\Excl X$
given by $(\Coder X)_{a,m}=\Kronecker m{\Mset a}$, which is easily seen to
satisfy the conditions of Section~\ref{sec:exp-struct}
and~\ref{sec:finctorial-exp}, so that $\FINV$ is a model of full differential
\LL{}.

\paragraph{An intrinsic presentation of function spaces.} 
We have seen that a morphism from $X$ to $Y$ of the linear category $\FINV$ can
be seen both as an element of $\Fmod\Field{\Limpl XY}$ and as a continuous
linear function from $\Fmod\Field X$ to $\Fmod\Field Y$. 

A morphism from $X$ to $Y$ in the Kleisli category $\FINVK$ is an element of
$\Fmod\Field{\Limpl{\Excl X}{Y}}$. Given $M\in\Fmod\Field{\Limpl{\Excl X}{Y}}$,
we have seen that we can define a function $\Fun M:\Fmod\Field X\to\Fmod\Field
Y$ by 
\[
\Funkofmat M(x)=M\Compl\Prom x=(\sum_{m\in\Web{\Excl X}}M_{m,b}x^m)_{b\in\Web
  Y}\,.
\]
Moreover, the correspondence $M\to\Funkofmat M$ is functorial. We provide here
an intrinsic characterization of these functions.

Let $E$ and $F$ be ltvs's. Let us say that a function $f:E\to F$ is
\emph{polynomial}
if there is $n\in\Nat$ and \emph{hypocontinuous} $i$-linear maps $f_i:E^i\to F$
(for $i=0,\dots,n$) such that
\begin{equation*}
  f(x)=f_0+f_1(x)+\cdots+f_n(x,\dots,x)\,.
\end{equation*}

A polynomial map $f$ of the form $f(x)=f_n(x,\dots,x)$, where $f_n$ is an
$n$-linear hypocontinuous function, is said to be \emph{homogeneous of degree
  $n$} (this condition implies of course $\forall t\in\Field\,f(tx)=t^nf(x)$,
and when $\Field$ is infinite, a polynomial function is homogeneous iff it
satisfies this latter condition).

Let $\Polyhom EF$ be the $\Field$-vector space of polynomial functions from $E$
to $F$. This space can be endowed with the linear topology of uniform
convergence on all linearly bounded subspaces, which admits the following
generating filter base of open neighborhoods of $0$: the basic opens are the
linear subspaces $\Ann(B,V)=\{f\in\Polyhom EF\St f(B)\subseteq V\}$, where $B$
is a linearly bounded subspace of $E$ and $V$ is a linear open subspace of $F$.
Let $\Anahom EF$ be the completion\footnote{A completion of an ltvs $E$ is a
  pair $(\tilde E,h)$ where $\tilde E$ is a complete ltvs and $h:E\to\tilde E$
  is a linear and continuous map such that, for any complete ltvs $F$
  and any linear continuous map $f:E\to F$, there is an unique linear and
  continuous map $\tilde f:\tilde E\to F$ such that $\tilde f\Comp h=f$. Using
  standard techniques, one can prove that any ltvs admits a completion, which
  is unique up to unique isomorphism.} of that ltvs.


\begin{theorem}
  Assume that $\Field$ is infinite.  For any finiteness spaces $X$ and $Y$, the
  ltvs $\Fmod\Field{\Limpl{\Excl X}Y}$ is linearly homeomorphic to
  $\Anahom{\Fmod\Field X}{\Fmod\Field Y}$.
\end{theorem}
\Beginproof
Let $h:\Fmod\Field X^n\to\Fmod\Field Y$ be an hypocontinuous $n$-linear
function of matrix $M\in\Fmod\Field{\Limpl{X\otimes\cdots\otimes X}{Y}}$, so
that $h(x_1,\dots,x_n)= M(x_1\otimes\cdots\otimes x_n)$. 

Remember that, using contraction and dereliction, we have defined
in~Section~\ref{sec:more-exp-struct} the morphism $\Derm
Xn\in\Fmod\Field{\Limpl{\Excl X}{X\otimes\cdots\otimes X}}$. Then we have
$N=M\Compl\Derm Xn\in\Fmod\Field{\Limpl{\Excl X}Y}$, and it is easy to see that
\[
\Funkofmat N(x)=h(x,\dots,x)\,.
\]
In that way, we see that any polynomial map from $\Fmod\Field X$ to
$\Fmod\Field Y$ is an element of $\Fmod\Field{\Limpl{\Excl X}{Y}}$; we have an
inclusion $\Polyhom{\Fmod\Field X}{\Fmod\Field
  Y}\subseteq\Fmod\Field{\Limpl{\Excl X}Y}$. Actually, the exponential
structure $\FINV$ is Taylor and this notion of polynomial map coincides with
the general notion of Section~\ref{sec:Taylor-structure}.

Conversely, let $(m,b)\in\Web{\Limpl{\Excl X}Y}$ with
$m=\Mset{a_1,\dots,a_n}$. The map $f:{\Fmod\Field X}^n\to\Field$ defined by
$f(x(1),\dots,x(n))=x(1)_{a_1}\dots x(n)_{a_n}$ is multilinear and
hypocontinuous. Hence the same holds for the map $x\mapsto f(x)e_b$ from
${\Fmod\Field X}^n$ to $\Fmod\Field Y$. Therefore we have
$\Field^{(\Web{\Limpl{\Excl X}Y})}\subseteq\Polyhom{\Fmod\Field X}{\Fmod\Field
  Y}$ (given a set $I$, remember that $\Field^{(I)}$ is the $\Field$-vector
space generated by $I$, that is, the space of all families $(a_i)_{i\in I}$ of
elements of $\Field$ such that $a_i=0$ for almost all $i$'s).

Hence $\Polyhom{\Fmod\Field X}{\Fmod\Field Y}$ is a dense subspace of
$\Fmod\Field{\Limpl{\Excl X}Y}$. To show that $\Fmod\Field{\Limpl{\Excl X}Y}$
is the completion of $\Polyhom{\Fmod\Field X}{\Fmod\Field Y}$ it suffices to
show that the above defined linear topology on that space (uniform convergence
on all linearly bounded subspaces) is the restriction of the topology of
$\Fmod\Field{\Limpl{\Excl X}Y}$.

Let $B\subseteq\Fmod\Field X$ be a linearly bounded subspace and let
$V\subseteq\Fmod\Field Y$ be linear open. Let $v'\in\Orth{\Fin Y}$ be such that
$\Neigh Y{v'}\subseteq V$.  By Proposition~\ref{prop:fspace-bounded-char},
$\Supp B\in\Fin X$, so $\Mfin{\Supp B}\in\Fin{\Excl X}$. Let 
\[
M\in \Neigh{\Limpl{\Excl X}Y}{\Mfin{\Supp B}\times
  v'}\subseteq\Fmod\Field{\Limpl{\Excl X}Y}\,,
\]
then $\Funofmat M(x)_b=0$ for each $x\in B$ and $b\in v'$. So we have
\[
\Neigh{\Limpl{\Excl X}Y}{\Mfin{\Supp B}\times v'}\cap\Polyhom{\Fmod\Field
  X}{\Fmod\Field Y}\subseteq\Ann(B,V)\,.
\]

Conversely let $U\in\Fin{\Excl X}$ and $v'\in\Orth{\Fin Y}$, then we have
$u=\bigcup_{m\in U}\Supp m\in\Fin X$ and hence the subspace
$B\subseteq\Fmod\Field X$ of all vectors which vanish outside $u$ is linearly
bounded. Let $M\in\Fmod\Field{\Limpl{\Excl X}Y}$ be such that the map
$\Funkofmat M:\Fmod\Field X\to\Fmod\Field Y$ is polynomial and belongs to
$\Ann(B,\Neigh Y{v'})$. Then for any $m=\Mset{a_1,\dots,a_n}\in\Mfin u$ and
$b\in v'$ we have $M_{m,b}=0$ because this scalar is the coefficient of the
monomial $\xi_1^{m(a_1)}\dots\xi_n^{m(a_n)}$ in the polynomial
$P\in\Field[\xi_1,\dots,\xi_n]$ such that $P(z_1,\dots,z_n)=\Funkofmat M(x)_b$
where $x\in\Fmod\Field X$ is such that $x_a=z_i$ if $a=a_i$ and $x_a=0$ if
$a\notin\Supp m$, and $P=0$ because $\Funkofmat M(B)\subseteq \Neigh Y{v'}$ by
assumption (we also use the fact that $\Field$ is infinite). Hence $M\in
\Neigh{\Limpl{\Excl X}{Y}}{U\times v'}$ and we have shown that 
\[
\Ann(B,\Neigh
Y{v'})\subseteq \Neigh{\Limpl{\Excl X}{Y}}{U\times v'}\cap\Polyhom{\Fmod\Field
  X}{\Fmod\Field Y}\,,
\]
showing that this latter set is a neighborhood of $0$ in the space of
polynomials.
\Endproof

The Taylor formula proved in~\cite{Ehrhard00b} for the morphisms of this
Kleisli category shows that actually any morphism is the sum of a converging
series whose $n$-th term is an homogeneous polynomial of degree $n$.

As an example, take $E=\Field[\xi]\Isom\Fmod\Field{\Limpl{\Excl\One}\One}$. The
corresponding topology on $E$ is the discrete topology. A typical example of
generalized polynomial map is the function $\phi:E\to\Field$ which maps a
polynomial $P$ to $P(P(0))$, in other words,
$\phi(x_0+x_1\xi+\cdots+x_n\xi^n)=x_0+x_1x_0+\cdots+x_nx_0^n$. Considered as a
generalized polynomial of infinitely many variables $x_0,x_1,\dots$, we see
that $\phi$ is not of bounded degree, and so it is not
polynomial. Nevertheless, it corresponds to a very simple and finite
computation on polynomials.

\paragraph{Antiderivatives.}
Just as $\REL$, the $\FINV$ exponential structure is Taylor in the sense of
Section~\ref{sec:Taylor-structure}. Moreover, if $\Field$ is of characteristic
$0$ (meaning that $\forall n\in\Nat\ n\,1=0\Implies n=0$) it has
antiderivatives in the sense of~\ref{sec:primitives}, because the morphism
$J_X=\Id_X+(\Coderc X\Compl\Derc X):\Excl X\to\Excl X$ satisfies
$(J_X)_{p,q}=(\Card p+1)\Kronecker pq$ for all $p,q\in\Web{\Excl X}$ and hence
is an isomorphism whose inverse $I_X$ is given by $(I_X)_{p,q}=\frac 1{\Card
  p+1}\Kronecker pq$.

\section*{Acknowledgment}
Part of the work reported in this article has been supported by the
French-Chinese project ANR-11-IS02-0002 and NSFC 61161130530 \emph{Locali}.

\bibliographystyle{alpha}
\bibliography{newbiblio}

\label{lastpage}

\end{document}

%% file: local.tex
\newcommand\CMLLPAR{
\usepackage{cmll}
\newcommand\IPar{\mathord{\parr}}
}

\CMLLPAR

%% file: notation.tex
\newtheorem{theorem}{Theorem}

\newtheorem{proposition}[theorem]{Proposition}
\newtheorem{lemma}[theorem]{Lemma}

\renewcommand\paragraph{\subsubsection}

\newcommand{\proofitem}[1]{\paragraph*{\mdseries\textit{#1}}}

\newcommand{\Beginproof}{\proofitem{Proof.}}
\newcommand{\Endproof}{
  \ifmmode 
  \else \leavevmode\unskip\penalty9999 \hbox{}\nobreak\hfill
  \fi
  \quad\hbox{$\Box$}
  \par\medskip}

\newenvironment{remark}%
{\smallbreak\noindent{\textit{Remark\/}: }\nobreak}%
{\smallbreak}

\renewcommand{\phi}{\varphi}
\renewcommand\epsilon{\varepsilon}

\newcommand{\Iff}{\quad\hbox{iff}\quad}

\newcommand{\Implies}{\Rightarrow}

\newcommand\Equiv{\Leftrightarrow}
\newcommand{\St}{\mid}

\renewcommand{\Bot}{{\mathord{\perp}}}
\newcommand{\Top}{\top}

\newcommand\cA{\mathcal{A}}
\newcommand\cB{\mathcal{B}}
\newcommand\cC{\mathcal{C}}
\newcommand\cD{\mathcal{D}}

\newcommand\cF{\mathcal{F}}
\newcommand\cG{\mathcal{G}}

\newcommand\cL{\mathcal{L}}

\newcommand\cT{\mathcal{T}}

\newcommand\cV{\mathcal{V}}

\newcommand\Fini{{\mathrm{fin}}}

\newcommand\Part[1]{{\mathcal P}({#1})}

\newcommand{\Linarrow}{\multimap}

\newcommand\Myleft{}
\newcommand\Myright{}

\newcommand\Web[1]{\Myleft|{#1}\Myright|}

\newcommand\Supp[1]{\operatorname{\mathsf{supp}}({#1})}

\newcommand\Mset[1]{[{#1}]}

\newcommand\Star{\star}

\newcommand\Par[2]{{#1}\IPar{#2}}
\newcommand\Parp[2]{({#1}\IPar{#2})}
\newcommand\ITens{\otimes}
\newcommand\Tens[2]{{#1}\ITens{#2}}
\newcommand\Tensp[2]{({#1}\ITens{#2})}
\newcommand\IWith{\mathrel{\&}}
\newcommand\With[2]{{#1}\IWith{#2}}

\newcommand\Orth[1]{#1^{\mathord\bot}}
\newcommand\Orthp[1]{(#1)^{\mathord\bot}}

\newcommand\Pair[2]{\langle{#1},{#2}\rangle}

\newcommand\Biorth[1]{#1^{\Bot\Bot}}

\newcommand\Triorth[1]{{#1}^{\Bot\Bot\Bot}}

\newcommand\One{1}

\newcommand\LL{\textsf{LL}}

\newcommand\Card[1]{\#{#1}}

\newcommand\Locun[1]{1^J}

\newcommand\Isom\simeq

\newcommand\Comp{\mathrel\circ}

\newcommand\Funinv[1]{{#1}^{-1}}

\newcommand\SET{\mathbf{Set}}

\newcommand\Limpl[2]{{#1}\Linarrow{#2}}

\newcommand\Nat{{\mathbb{N}}}
\newcommand\Natnz{{\Nat^+}}

\newcommand\Fin[1]{\mathsf{F}({#1})}

\newcommand\Diffsymb{\mathsf D}

\newcommand\App[2]{\left({#1}\right){#2}}
\newcommand\Abs[2]{\lambda{#1}\,{#2}}

\newcommand\Derp[3]{\frac{\partial{#1}}{\partial{#2}}\cdot{#3}}

\newcommand\List[3]{#1_{#2},\dots,#1_{#3}}

\newcommand\Kronecker[2]{\delta_{{#1},{#2}}}

\newcommand\Subst[3]{{#1}\left[{#2}/{#3}\right]}

\newcommand\Substbis[2]{{#1}\left[{#2}\right]}

\newcommand\Factor[1]{{#1}!}
\newcommand\Binom[2]{\genfrac{(}{)}{0pt}{}{#1}{#2}}
\newcommand\Multinom[2]{\left[#2\right]}

\newcommand\Real{\mathbb{R}}

\newcommand\Linapp[2]{\left\langle{#1}\right\rangle{#2}}

\newcommand\Fmod[2]{{#1}\langle{#2}\rangle}

\newcommand\Transp{\Orth}

\newcommand\Bcanon[1]{e_{#1}}

\newcommand\Mfin[1]{\mathcal M_\Fini({#1})}

\newcommand\Neigh[2]{\operatorname{\cV}_{#1}(#2)}

\newcommand\Evlin{\operatorname{\mathsf{ev}}}
\newcommand\Evdual{\operatorname{\mathsf{ev_\Bot}}}
\newcommand\Curdual{\Curlin_\Bot}
\newcommand\Curlin{\mathsf{cur}}
\newcommand\Curlinp[1]{\Curlin(#1)}
\newcommand\REL{\operatorname{\mathbf{Rel}}}

\newcommand\Diag[1]{\Delta^{#1}}
\newcommand\Codiag[1]{\operatorname{a}^{#1}}

\newcommand\Tpower[2]{{#1}^{\otimes{#2}}}

\newcommand\Sterms{\Delta}

\newcommand\Rel[1]{\mathrel{#1}}

\newcommand\Redst[1]{\mathop{\mathsf{Red}}}

\newcommand\Tay[1]{{#1}^*}

\newcommand\Deg[1]{\textsf{deg}_{#1}}

\newcommand\Dom[1]{\operatorname{\mathsf{D}}(#1)}

\newcommand\Symgrp[1]{\mathfrak S_{#1}}

\newcommand\Varset{\mathcal{V}}

\newcommand\Atoms{\cA}
\newcommand\Coatom[1]{\overline{#1}}

\newcommand\Dapp[2]{\Diffsymb{#1}\cdot{#2}}
\newcommand\Dappm[3]{\Diffsymb^{#1}{#2}\cdot{(#3)}}
\newcommand\Dsubst[3]{\frac{\partial #1}{\partial #3}\cdot{#2}}

\newcommand\Msetofsubst[1]{\bar F}

\newcommand\Retri\zeta
\newcommand\Retrp\rho

\newcommand\Impl[2]{{#1}\Rightarrow{#2}}

\newcommand\Funofmat[1]{\mathsf{fun}(#1)}
\newcommand\Funkofmat[1]{\mathsf{Fun}(#1)}

\newcommand\Tnat\iota

\newcommand\Loop\Omega

\newcommand\Timpl\Impl

\newcommand\Weak[1]{\WEAK_{#1}}
\newcommand\Coweak[1]{\COWEAK_{#1}}
\newcommand\Contr[1]{\CONTR_{#1}}
\newcommand\Cocontr[1]{\COCONTR_{#1}}
\newcommand\Contrm[2]{\CONTR_{#1}^{#2}}

\newcommand\Tensexp[2]{{#1}^{\mathord\otimes{#2}}}
\newcommand\Derm[2]{\mathsf d_{#1}^{#2}}

\newcommand\Der[1]{\DER_{#1}}
\newcommand\Coder[1]{\CODER_{#1}}
\newcommand\Digg[1]{\operatorname{\mathsf p}_{#1}}

\newcommand\Derc[1]{\partial_{#1}}
\newcommand\Dercm[2]{\partial_{#1}^{#2}}
\newcommand\Coderc[1]{\overline\partial_{#1}}
\newcommand\Codercm[2]{\overline\partial_{#1}^{#2}}
\newcommand\Partexcl[2]{\oc_{#1}#2}
\newcommand\Coderm[2]{{\overline{\mathsf d}}_{#1}^{#2}}

\newcommand\Fun{\operatorname{\mathsf{Fun}}}

\newcommand\Id{\operatorname{\mathsf{Id}}}

\newcommand\Proj[1]{\pi_{#1}}

\newcommand\Setpr[1]{\mathsf{pr}_{#1}}

\newcommand\Excl[1]{\oc{#1}}
\newcommand\Exclp[1]{\oc{(#1)}}
\newcommand\Int[1]{\wn{#1}}

\newcommand\Prom[1]{#1^!}

\newcommand\Mix{\operatorname{\mathsf{mix}}}
\newcommand\Mixvect[2]{\Mix(#1,#2)}
\newcommand\Permvect[2]{\operatorname{\mathsf{sym}}(#1,#2)}

\newcommand\Relincl\eta
\newcommand\Relrestr\rho

\newcommand\Tsym{\sigma}

\newcommand\Covar[1]{\overline{#1}}

\newcommand\Conset{\Sigma}
\newcommand\Arity{\mathsf{ar}}

\newcommand\Cut[2]{\langle#1\,\mathord\mid\,#2\rangle}
\newcommand\Net[2]{(#2\,;\,#1)}

\newcommand\Varnet{\mathsf{V}}
\newcommand\Fvarnet{\mathsf{FV}}

\newcommand\Typing[3]{#1\vdash_0{#2}:{#3}}
\newcommand\TypingCut[2]{#1\vdash_0{#2}}
\newcommand\Logic[3]{#1\vdash{#2}:{#3}}

\newcommand\Red{\mathord\leadsto}
\newcommand\Redcom{\mathord\leadsto_{\mathsf{com}}}
\newcommand\Redcut{\mathord\leadsto_{\mathsf{cut}}}

\newcommand\MLL{\textsf{MLL}}
\newcommand\DILLZ{\textsf{DiLL${}_0$}}
\newcommand\DILL{\textsf{DiLL}}

\newcommand\WEAK{\mathsf w}
\newcommand\COWEAK{\overline\WEAK}
\newcommand\DER{\mathsf d}
\newcommand\CODER{\overline\DER}
\newcommand\CONTR{\mathsf c}
\newcommand\COCONTR{\overline\CONTR}
\newcommand\PROM[2]{{#2}^{!(#1)}}
\newcommand\MIXZ{\xi_0}
\newcommand\MIXB{\xi_2}

\newcommand\Vect[1]{\overrightarrow{#1}}

\newcommand\TreeZ{\langle\rangle}
\newcommand\TreeO{*}
\newcommand\TreeB[2]{\langle#1,#2\rangle}
\newcommand\Nleaves{\mathsf L}
\newcommand\TreeExt[2]{{\mathord{#1}}^{#2}}
\newcommand\TreeExtp[3]{({\mathord{#1}}^{#2}(#3))}
\newcommand\Treeiso[3]{{{\mathord{#1}}^{#2}_{#3}}}
\newcommand\Symfunc[1]{\widehat{#1}}
\newcommand\Trees[1]{\cT_{#1}}
\newcommand\Symiso[3]{\widehat{\mathord{#1}}^{#2,#3}}

\newcommand\Complin{\,}
\newcommand\Compl{\Complin}
\newcommand\Vectors[2]{\vec{#1}(#2)}
\newcommand\Parvect[1]{\IPar(#1)}
\newcommand\Tensvect[2]{\mathord\otimes(#1,#2)}
\newcommand\Cutvect[2]{\mathsf{cut}(#1,#2)}

\newcommand\Tenscont[6]{\ITens_{#1,#2,#3,#4}(#5,#6)}

\newcommand\Op[1]{{#1}^{\textsf{op}}}

\newcommand\Biorthiso{\eta}

\newcommand\AXIOM{\textsf{axiom}}
\newcommand\CUTRULE{\textsf{cut rule}}
\newcommand\PERMRULE{\textsf{permutation rule}}
\newcommand\PARRULE{$\IPar$\textsf{-rule}}
\newcommand\TENSRULE{$\ITens$\textsf{-rule}}
\newcommand\MIXRULE{\textsf{mix rule}}
\newcommand\CONTEXT{\textsf{context}}
\newcommand\VARCUT{\textsf{ax-cut}}
\newcommand\WEAKENING{\textsf{weakening}}
\newcommand\COWEAKENING{\textsf{co-weakening}}
\newcommand\DERELICTION{\textsf{dereliction}}
\newcommand\CODERELICTION{\textsf{co-dereliction}}
\newcommand\CONTRACTION{\textsf{contraction}}
\newcommand\COCONTRACTION{\textsf{co-contraction}}
\newcommand\SUMRULE{\textsf{sum}}

\newcommand\Typesem[1]{[#1]}
\newcommand\Dersem[1]{[#1]}
\newcommand\Netsem[1]{[#1]}

\newcommand\Idpar{\mathsf{ax}}

\newcommand\Scalmult{\cdot}

\newcommand\ExpMonZ{\mu^0}
\newcommand\ExpMonB{\mu^2}
\newcommand\ExpMonT[1]{\mu^{#1}}

\newcommand\DiggT[2]{\Digg{#2}^{#1}}

\newcommand\Demorgan{\mathsf{dm}}
\newcommand\Groupoid[1]{{#1}_{\mathsf{iso}}}

\newcommand\Field{\mathbf k}

\newcommand\Beta{\beta}
\newcommand\Dbeta{\beta_{\mathsf d}}

\newcommand\Taym[2]{\mathrm T^{#2}_{#1}}
\newcommand\Primmor[1]{I_{#1}}

\newcommand\Primint[1]{I_{#1}}

\newcommand\Stermsh[2]{\Sterms_{#1}^{(#2)}}

\newcommand\Eqref[1]{(\ref{#1})}
\newcommand\SeelyZ{\mathsf m^0}
\newcommand\SeelyB{\mathsf m^2}

\newcommand\Ann{\mathsf{Ann}}
\newcommand\Bnd[2]{\cB_{#1}(#2)}
\newcommand\FINV{\mathbf{Fin}^\Field}
\newcommand\FINVK{\mathbf{Rel}^\Field_\oc}

\newcommand\Funofmatm{\mathsf{Fun}}
\newcommand\Compmat{\,}

\newcommand\Polyhom[2]{\mathsf{Pol}_\Field(#1,#2)}
\newcommand\Anahom[2]{\mathsf{Ana}_\Field(#1,#2)}

\newcommand\Mainport{\bullet}



%% file: tikznotation.tex
\usepackage{tikz-inet}

\tikzset{inetwire/.style={line width=0.15ex,rounded corners=2pt}}

\newcommand{\Prombox}[6][]{
  \node[#1,rectangle,draw,minimum width=#2cm,minimum height=#3cm](#4){#6};
  \inetcell[below=-0.1 of #4](#5){$\oc$}}

\newcommand{\Somenet}[5][]{
  \node[#1,fill=gray!30,rectangle,draw,thick,minimum width=#2cm,minimum height=#3cm,rounded
  corners=2pt](#4){#5};}

\newcommand{\Axioms}[3][]{
  \node[#1,rectangle,draw,thick,minimum width=#2cm,minimum height=0.5cm,rounded
  corners=2pt](#3){axiom links};}

\newcommand{\Sometree}[4][]{\inetcell[#1,fill=gray!30,minimum size=#2cm](#3){#4}}

\newcommand{\Inetcut}[3][-0.2]{\draw[inetwire](#2)|-++(0.3,#1)-|(#3);}
\newcommand{\Inetcutr}[3][-0.2]{\draw[inetwire](#2)|-++(-0.3,#1)-|(#3);}